\documentclass[preprints,review,accept,oneauthor,pdftex,a4paper]{Definitions/mdpi}

\usepackage{aas_macros}

\firstpage{1} 
\pubvolume{1}
\issuenum{1}
\articlenumber{0}
\pubyear{2021}
\copyrightyear{2020}
\datereceived{August 13, 2021} 
\dateaccepted{September 3, 2021} 
\datepublished{}
\hreflink{https://doi.org/} 

\pdfoutput=1

\Title{Phase-Space Correlations Among Systems of Satellite Galaxies}

\Author{Marcel S. Pawlowski$^{1,\dagger}$\orcidA{}}

\AuthorNames{Marcel S. Pawlowski}

\address[1]{%
$^{1}$ \quad Leibniz-Institut f\"ur Astrophysik Potsdam (AIP), An der Sternwarte 16, D-14482 Potsdam, Germany}

\corres{Correspondence: mpawlowski@aip.de}

\firstnote{Email: mpawlowski@aip.de} 

\abstract{
Driven by increasingly complete observational knowledge of systems of satellite galaxies, mutual spatial alignments and relations in velocities among satellites belonging to a common host have become a productive field of research. Numerous studies have investigated different types of such phase-space correlations, and were met with varying degrees of attention by the community. The Planes of Satellite Galaxies issue is maybe the best-known example, with a rich field of research literature and an ongoing, controversial debate on how much of a challenge it poses to the $\Lambda$CDM model of cosmology. Another type of correlation, the apparent excess of close pairs of dwarf galaxies, has received considerably less attention despite its reported tension with $\Lambda$CDM expectations. With the fast expansion of proper motion measurements in recent years, largely driven by the Gaia mission, other peculiar phase-space correlations have been uncovered among the satellites of the Milky Way. Examples are the apparent tangential velocity excess of satellites compared to cosmological expectations, and the unexpected preference of satellites to be close to their pericenters. At the same time, other kinds of correlations have been found to be more in line with cosmological expectations, specifically lopsided satellite galaxy systems and the accretion of groups of satellite galaxies. The latter has mostly been studied in cosmological simulations thus far, but offers the potential to address some of the other issues by providing a way to produce correlations among the orbits of a group's satellite galaxy members.
This review is the first to provide an introduction to the highly active field of phase-space correlations among satellite galaxy systems. The emphasis is on summarizing existing, recent research and highlighting interdependencies between the different, currently almost exclusively individually considered types of correlations. Future prospects in light of upcoming observational facilities and an ever-expanding knowledge of satellite galaxy systems beyond the Local Group are also briefly discussed.
}


\begin{document}



\section{Introduction}
 \label{sect:intro}

Advances in observational facilities and successful observational surveys detecting increasingly fainter dwarf galaxies around the Milky Way (MW) and other nearby galaxies, coupled with improvements in numerical modelling of the cosmic evolution of matter and the formation of galaxies, have opened opportunities to investigate the highly non-linear regime of structure- and galaxy formation. The resulting research subject of near-field cosmology involves testing predictions of the cosmological model, the nature of dark matter, and the processes of galaxy formation and evolution on the scale of nearby galaxies and their satellite galaxy systems.

The comparison of expectations derived from cosmological simulations with observed galaxies and their satellites has revealed several ‘small-scale problems’ for the $\Lambda$\ Cold Dark Matter ($\Lambda$CDM) model of cosmology. Here, small refers to galaxy stellar masses of $M_\star \leq 10^9\,M_\odot$\ and distribution scales below 1 Mpc (see \cite{2017ARA&A..55..343B} for a review). These include the \textit{Missing Satellites Problem} \citep{1999ApJ...522...82K, 1999ApJ...524L..19M}, i.e. that cosmological simulations predict that there are thousands of dark matter sub-halos of mass $\geq 10^7\,M_\odot$\ around a galaxy such as the MW, but only 40–50 satellites with stellar masses $\geq 300\,M_\odot$\ have been discovered.
The \textit{Core-Cusp Problem} \citep{1991ApJ...378..496D, 2011ApJ...742...20W}, i.e. that the density of dark matter halos in dark-matter-only simulations increases towards their center as $\rho(r) \propto 1/r$, resulting in a central density peak (a cusp), whereas in contrast the observed dynamics of some satellite galaxies indicate a shallow profile (a core).
And the \textit{Too Big To Fail Problem} (TBTF, \cite{2011MNRAS.415L..40B}), i.e. that the dark matter halo density deduced from the dynamics of the most-luminous satellite galaxies is lower than that of the most massive dark matter sub-halos in simulated MW analogs, which should be too big to have failed to form stars.

While they are deemed catastrophic for $\Lambda$CDM by some (e.g. \cite{2012PASA...29..395K}), a less extreme approach has been claimed to be adequate to address some of these problems: they do not necessarily indicate fundamental issues with the underlying cosmological model, but could also be caused by lacking modeling of baryonic effects or by observational shortcomings that were not fully considered. The missing satellites problem is nowadays widely understood to be solved by considering baryonic effects that influence which sub-halos host luminous galaxies and which do not. In particular, galaxy formation in halos with virial velocities below $20\,\mathrm{km\,s}^{-1}$\ is expected to be suppressed because their virial temperatures are below the $\sim 10^4\,$K to which cosmic reionisation heats the intergalactic medium. Gas in these halos thus can not cool and form stars, such that the halos remain dark. Baryonic effects can also account for cores. Simulations show that feedback from supernova explosions can drive the gas in a dark matter halo to larger distances. The resulting change in the overall potential can also drive out dark matter from the center, transforming a cusp into a core (see e.g. \cite{2017ARA&A..55..343B} and references therein). However, both the observational and the theoretical situation is somewhat unclear: some studies argue that cores in dwarf galaxies might just be an observational artifact since the inferred central density slope varies with the viewing direction for mock-observed simulated dwarf galaxies \citep{2018MNRAS.474.1398G}. Additionally, some simulations disagree with the more common finding that cores only form in galaxies of $M_\star > 10^6\,M_\odot$: they either find no core formation even at those masses \citep{2016MNRAS.457.1931S} or core formation in dwarf galaxies of all masses \citep{2016MNRAS.459.2573R}. The TBTF problem has similarly been claimed to be solved by baryonic processes \citep{2016MNRAS.457.1931S}, but the situation is again complex and unclear. While interactions with a host galaxy (tidal stripping, disk or tidal shocking, and ram pressure stripping) can reduce the central masses of satellites \cite{2014ApJ...786...87B}, the TBTF problem also seems to be present in field galaxies that do no experience these effects \cite{2015A&A...574A.113P}, yet observational evidence for the TBTF problem for field galaxies has also been questioned \cite{ 2017ApJ...850...97B,  2019MNRAS.482.5606D}.

Others advocate for alternatives to the $\Lambda$CDM model rather than only baryonic physics to solve the small-scale problems. These alternatives aim to preserve the successes of $\Lambda$CDM on large scales. Examples are Warm Dark Matter (WDM) and Self-Interacting Dark Matter (SIDM). WDM particles have higher velocities in the early Universe, giving them a larger free streaming length than CDM. This washes out smaller-scale density fluctuations, which results in fewer small-scale dark matter halos, alleviating the missing satellites problem. The TBTF problem is addressed because WDM halos form later than their CDM counterparts, resulting in lower central densities \citep{2005PhRvD..71f3534V, 2013JCAP...03..014A}. SIDM postulates that dark matter particles have strong self-interactions \citep{2000PhRvL..84.3760S}. SIDM preserves the large-scale structure and the distribution of sub-halos, but the self-interaction affects the internal structure of dark matter halos, forming isothermal cores which alleviate the core-cusp and TBTF problems.

In summary, the investigation of smaller-scale structures can provide constraints on our cosmological model, the fundamental nature of the Universe, and the processes and feedback effects governing galaxy formation and evolution. However, the small-scale problems are currently subject to considerable uncertainties in interpretation. Many proposed solutions either rely on modeling baryonic effects and feedback processes, or on alternatives to the CDM particle. At the same time the observational evidence for the problems has in part been questioned. It is therefore difficult to judge how serious the problems really are, which possible solution (or combination thereof) is correct, and thus what specific constraints the problems provide for our understanding of cosmology and galaxy formation. It would therefore be favorable to have tests of the $\Lambda$CDM model, its alternatives, and the overall galaxy formation paradigm on the scale of galaxies and their satellites that are more robust against these effects. Such tests must be independent of the internal structure of the involved galaxies to avoid being dominated by choices in modeling sub-grid physics.

\begin{figure}[H]
\centering
\includegraphics[width=0.7\textwidth]{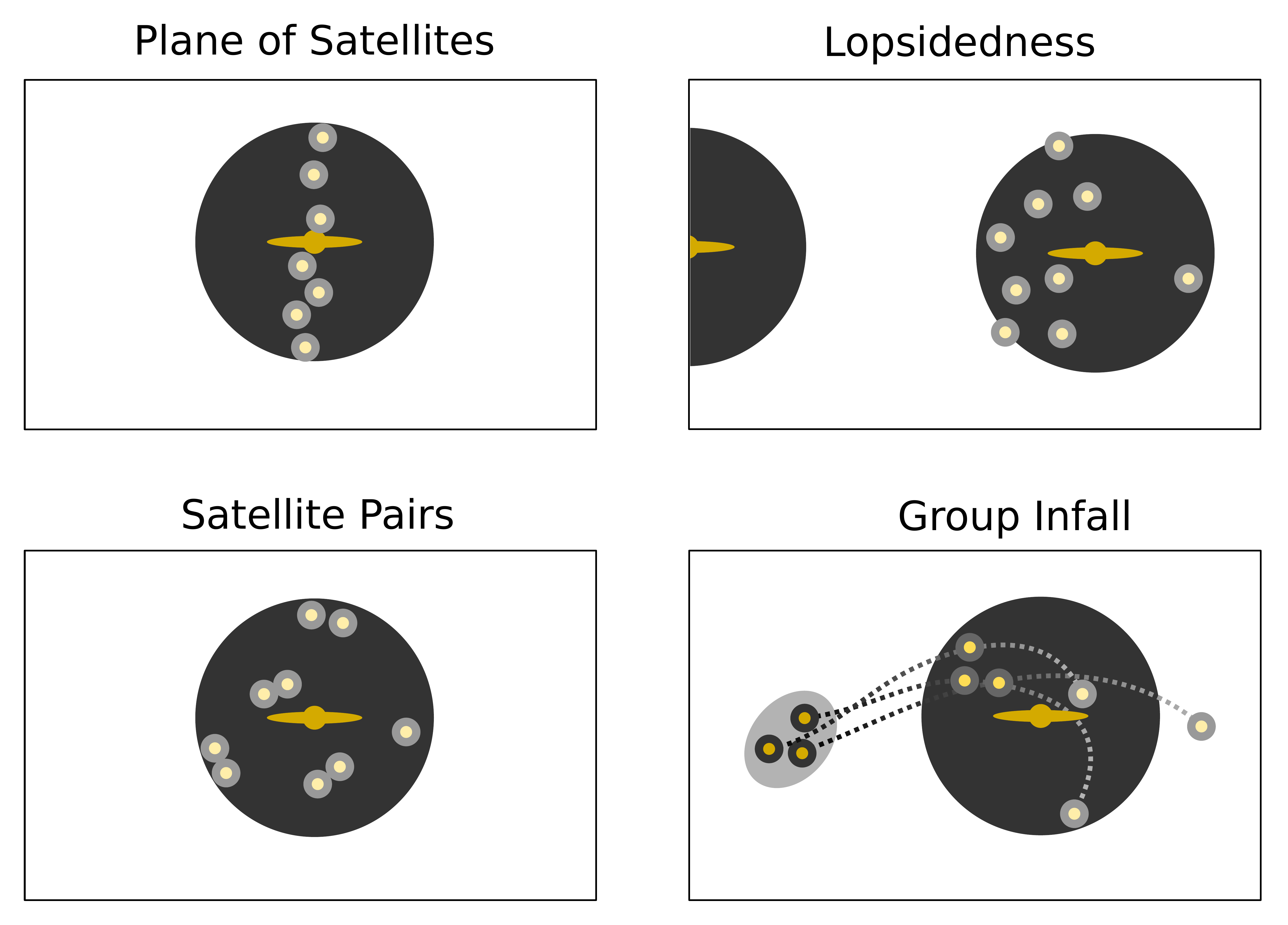}
\caption{Schematic representations of different types of phase-space correlations that have been studied among systems of satellite galaxies. Clockwise starting from the top-left: Planes of satellite galaxies (Sect. \ref{sect:planesofsatellites}), lopsided satellite galaxy systems (Sect. \ref{sect:lopsided}), the infall of satellites in groups (Sect. \ref{sect:groups}), and close pairs of satellite galaxies (Sect. \ref{sect:pairs}). \label{fig:sketch}}
\end{figure}

The phase-space distribution of satellite galaxies around their hosts fulfill these requirements. The positions and motions of satellite galaxies within their host galaxy's halo, on scales of 100s of kpc, are largely independent of the detailed physics internal to them, be it sub-grid modelling of baryonic physics, or the detailed nature of dark matter that can affect the dark matter density distribution in the centers of halos. As our knowledge of systems of satellite galaxies around hosts has grown over the recent decade -- starting with the Milky Way, but expanding to M31 and more distant hosts -- considerable research has already been performed on these topics. Numerous different types of phase-space correlations have been investigated (see Fig. \ref{fig:sketch}). The maybe most popular aspect in this regard is the Planes of Satellite Galaxies issue, but other approaches and questions such as lopsided satellite galaxy systems, or the influence of satellite accretion in groups, have also lead to interesting results and both controversial challenges as well as reassuring confirmations of the $\Lambda$CDM paradigm. However, a systematic overview of different approaches to study phase-space correlations among satellite galaxy systems, and in particular the potential interplay between different such features, has thus far been lacking. This review provides such an overview and introduction, and an assessment of the current state of research. Future observational facilities and upcoming research projects promise to continue to develop, implement, and apply tests of satellite phase-space correlations to ever-expanding datasets.

This review focusses on the large-scale phase-space distribution of dwarf galaxies. Multiple additional dimensions of properties for dwarf galaxies are subject of extensive research and hold a plethora of additonal information. These include many, largely spectroscopically determined, internal properties such as ages, star formation histories, metallicities, and internal dynamics of dwarf galaxies, but also aspects such as total stellar and dynamical mass, spatial extend or shape and orientation. These should be considered for a fully comprehensive picture of the origin and evolution of individual dwarf galaxies and the structures they are part of \cite{2015ApJ...799L..13C}. Yet, to keep this review sufficiently brief and accessible, these aspects will not be covered in detail in the following.

\section{Phase-Space Correlations}
\label{sect:correlations}

The distribution and motion of satellite galaxies within the halo of their host galaxy on scales of $\sim100$\,kpc is expected to be largely unaffected by the internal dynamics of the involved galaxies (with sizes $\leq 10\,$kpc). The dynamics on these scales is to first order also independent of the exact type of dark matter (CDM, WDM, or SIDM, see e.g. fig. 15 in \citealt{2017ARA&A..55..343B}). Thus, the phase-space distribution of satellite galaxies provides a unique opportunity to test galaxy formation simulations independent of the exact details of baryonic and dark matter modeling. Consequently, it identifies issues that need to be addressed to fully understand galaxy formation and evolution in any model.

\subsection{An Impending Observational Revolution: Satellite Galaxy Systems and their Phase-Space Distribution}
\label{sect:observationalrevolution}

Observational access to satellite galaxy systems was severely limited for a long time. Satellites are very faint compared to their hosts. Typical $L_\star$\ galaxies like the MW and M31 have $M_\mathrm{V} \approx -21$, whereas the twelveth brightest MW satellite Canes Venatici has $M_\mathrm{V} = -8.6$, and the ultra-faint dSph satellites of the MW reach down to $M_\mathrm{V} > -4$. Similarly, the surface brightness of satellites can be as faint as $\mu_\mathrm{V} \geq 26 \mathrm{mag\,arcse}c^{-2}$, orders of magnitude fainter than their host. However, in the past $\sim$20 years, our knowledge of the faintest dwarf galaxies has increasingly grown. A major driver was the Sloan Digital Sky Survey (SDSS, \cite{2000AJ....120.1579Y}), which observed 35\% of the sky in five filters. It more than doubled the number of known MW satellite galaxies (from 11 to 27, listed in \cite{2016MNRAS.456..448P}). SDSS has also provided a large sample of more distant hosts with 1-3 bright satellites. More MW satellites have been discovered by the Dark Energy Survey (DES, e.g. \cite{2015ApJ...807...50B}) and other programs (compiled in e.g. \cite{2015MNRAS.453.1047P, 2019ARA&A..57..375S}). For M31, the PAndAS survey had a transformatory effect by providing homogeneous coverage of a large volume around that galaxy. It discovered at least 15 previously unknown satellites \citep{2013ApJ...776...80M}.

Progress on satellite systems beyond the Local Group was initially slower, but is picking up speed, especially in the Local Volume (within $\approx 10$\,Mpc).
Satellite galaxies were discovered and studied in the inner regions of Centaurus\,A at $\sim4$\,Mpc distance by the PISCeS survey \cite{2016ApJ...823...19C}, but also in the wider field of the Centaurus group \citep{2017A&A...597A...7M}, in the M81 group at $\sim4$\,Mpc distance \citep{2013AJ....146..126C}, and the M101 group at $\sim7$\,Mpc distance \citep{2014ApJ...787L..37M, 2017A&A...602A.119M}. The Exploration of Local VolumE Satellites (ELVES) Survey \cite{2021arXiv210503435C} has recently compiled a sample of 223 satellite galaxies around 80\% of host galaxies within 12\,Mpc. An example of an ongoing, ambitious project targeting hosts at distances between 20 and 40\,Mpc is the SAGA survey \citep{2017ApJ...847....4G}. SAGA aims to spectroscopically confirm satellites in the halos ($\leq 300$\,kpc) of 100 MW analogs down to the brightness of Leo\,I ($M_\mathrm{r} < -12.3$). So far, 36 hosts have been covered, resulting in the confirmation of 69 additional satellite galaxies for a complete sample of 126 satellites \cite{2021ApJ...907...85M}. 
In contrast, the MATLAS survey \cite{2020MNRAS.491.1901H} has targeted the surroundings of isolated early-type galaxies and has identified 2210 dwarf galaxy candidates, though currently spectroscopic velocities are available for only a small fraction of those.
The Dwarf Galaxy Survey with Amateur Telescopes (DGSAT) follows a less traditional approach by using small telescopes to identify satellite galaxies in the Local Volume as extended low surface brightness structures. So far, DGSAT discovered 13 satellite candidates \citep{2016A&A...588A..89J, 2017A&A...603A..18H}. 

These ongoing efforts will revolutionize our understanding of satellite galaxies by building a larger sample of known systems of satellite galaxies around a common host (in contrast to studies of individual satellite galaxies). Future telescopes and surveys will further advance our knowledge in this domain. The Vera C. Rubin Observatory (previously known as the Large Synoptic Survey Telescope) plans to start its 10 year Legacy Survey of Space and Time (LSST) in 2022. It will provide a much more complete picture of the nearby satellite systems by discovering faint dwarf galaxies all over the Local Group \cite{2019ARA&A..57..375S}. This will soon be followed by the next revolution in optical astronomy: the commissioning of extremely large telescopes with $\sim 30$m diameters. Their unique capabilities enable deep photometric observations of the halos of distant galaxies, and with their high light collecting ability they can measure spectroscopic velocities of fainter satellites around more distant hosts than currently feasible. Such velocities are necessary for significant tests of a satellite galaxy system's phase-space structures, because spatial alignments alone are not nearly as constraining \citep{2014MNRAS.442.2362P}. This especially applies for distant satellite galaxy systems where only projected positions can be reliably determined, but distance measurement uncertainties exceed the physical extend of the systems\footnote{Assuming a distance uncertaintiy of 5\% for tip-of-the-red-giant-branch (TRGB) distances, the distribution of satellites around a MW-like host with virial radius of 250\,kpc can not be resolved anymore beyond $\sim 5$\,Mpc.}.

In the Local Group, especially around the MW, full 3D velocity information can be obtained by measuring the proper motions of satellite galaxies. After being dominated by ground- and Hubble-Space-Telescope (HST) based measurements for a long time \cite{2006AJ....131.1445P, 2013ApJ...764..161K, 2013ApJ...768..139S, 2015AJ....149...42P, 2017ApJ...849...93S}, recently the Gaia mission revolutionized the field \cite{2016A&A...595A...1G}. Using data release 2 (DR2), the proper motions for 39 MW satellites were measured, many of them for the first time \cite{2018ApJ...863...89S, 2018A&A...619A.103F}. Additional satellite proper motions have been obtained since \cite{ 2018ApJ...867...19K,  2018A&A...620A.155M,  2019ApJ...875...77P,  2019A&A...623A.129F} and the recent Gaia EDR3 has allowed to update the previous measurements, resulting in a typical improvement in accuracy by a factor of 2.5 \cite{2021ApJ...916....8L,  2020RNAAS...4..229M,  2021arXiv210608819B}. The future end-of-mission data release 4 is expected to further improve this accuracy.

We are thus in the midst of an observational revolution in the study of not only satellite galaxies, but whole systems of satellite galaxies, both nearby and afar. 
We can thus now use the already increased observational information (while simultaneously preparing for the coming influx of data) to test our cosmological model expectations regarding the distribution and dynamics of satellite systems. This approach is made even more pressing by current analyses of well studied satellite systems that reveal a serious small-scale problem of cosmology: The Planes of Satellite Galaxies Problem.

\subsection{Planes of Satellite Galaxies}
\label{sect:planesofsatellites}

\subsubsection{The Satellite Planes Around the MW and M31}

\begin{figure}
\centering
\includegraphics[width=0.7\textwidth]{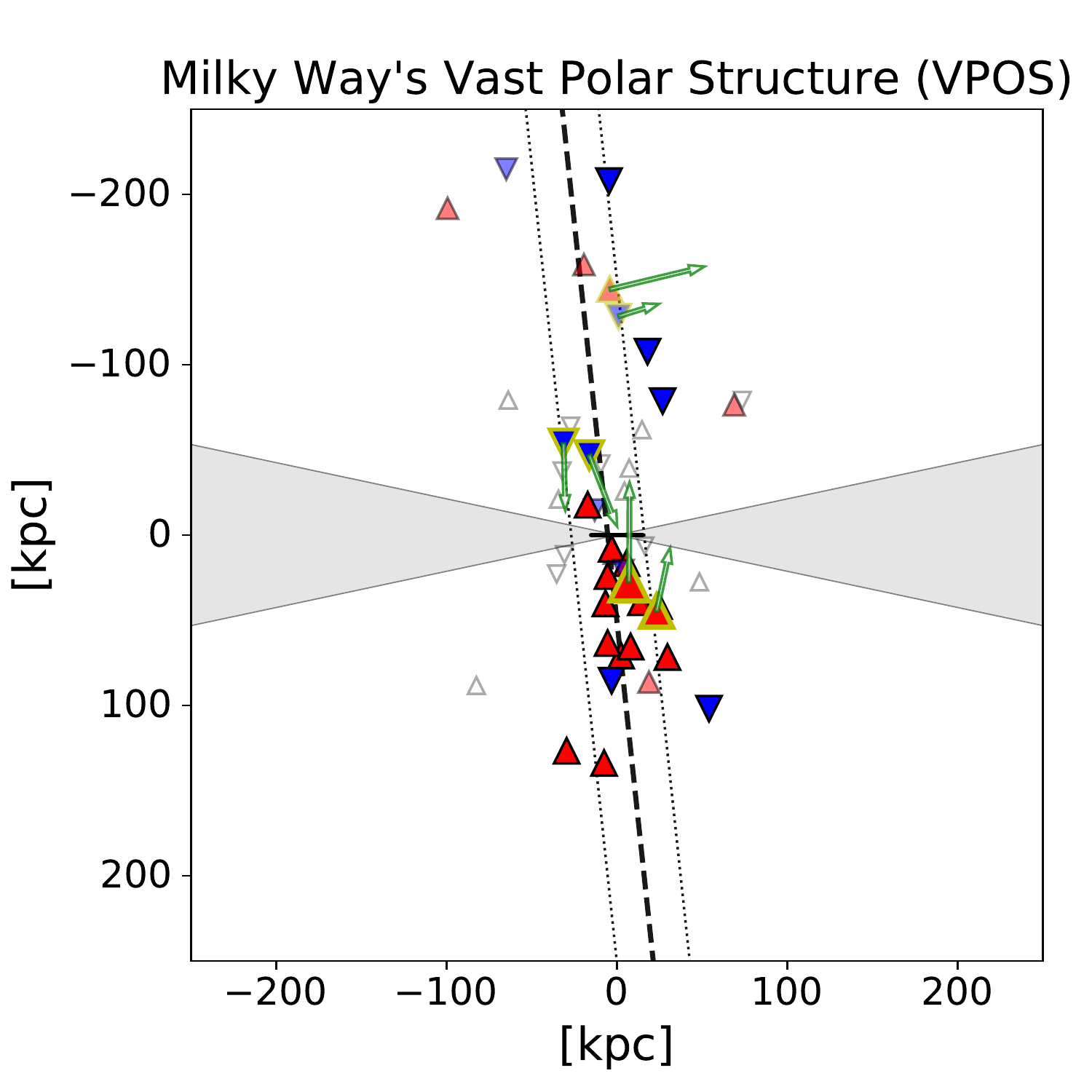}
\caption{Distribution of dwarf galaxies around the Milky Way. The Milky Way is seen edge-on in the center (black line), with grey shaded areas indicating a $\pm 12^\circ$\ region around the disk plane that is obscured and where dwarf galaxy searches are expected to be most incomplete. The view is oriented to show the Vast Polar Structure (VPOS) edge-on, which is illustrated with a black dashed line and its root-mean-quare width is indicated with the dotted lines.
The dwarf galaxy positions are indicated with upward (downward) pointed triangles, indicating wheter a dwarf is moving into (out of) the plot. The dwarfs that are likely or possibly orbiting along the VPOS are colored according to their most-likely motion (red is receding from the viewer into the plot, blue is approaching out of the plot). Dwarfs which are consistent within their proper motion uncertainties are shown in light colors, dwarf which are clearly not part of the VPOS are shown as empty symbols.
Three proposed pairs of satellite galaxies are highlighted with yellow contours. These are, from top to bottom: Leo\,IV and Leo\,V, Draco and Ursa Minor, and the LMC and SMC. Their most-likely velocities in this projection (from \cite{2021ApJ...916....8L}) are indicated with green arrows. They are consistent with following similar orbital directions. Draco, Ursa Minor, and both Magellanic Couds are also well consistent with co-orbiting along the VPOS; Leo\,IV and Leo\,V have only weakly constrained proper motions but are consistent with sharing a very similar orbit.
\label{fig:MW}}
\end{figure}

\begin{figure}
\centering
\includegraphics[width=0.7\textwidth]{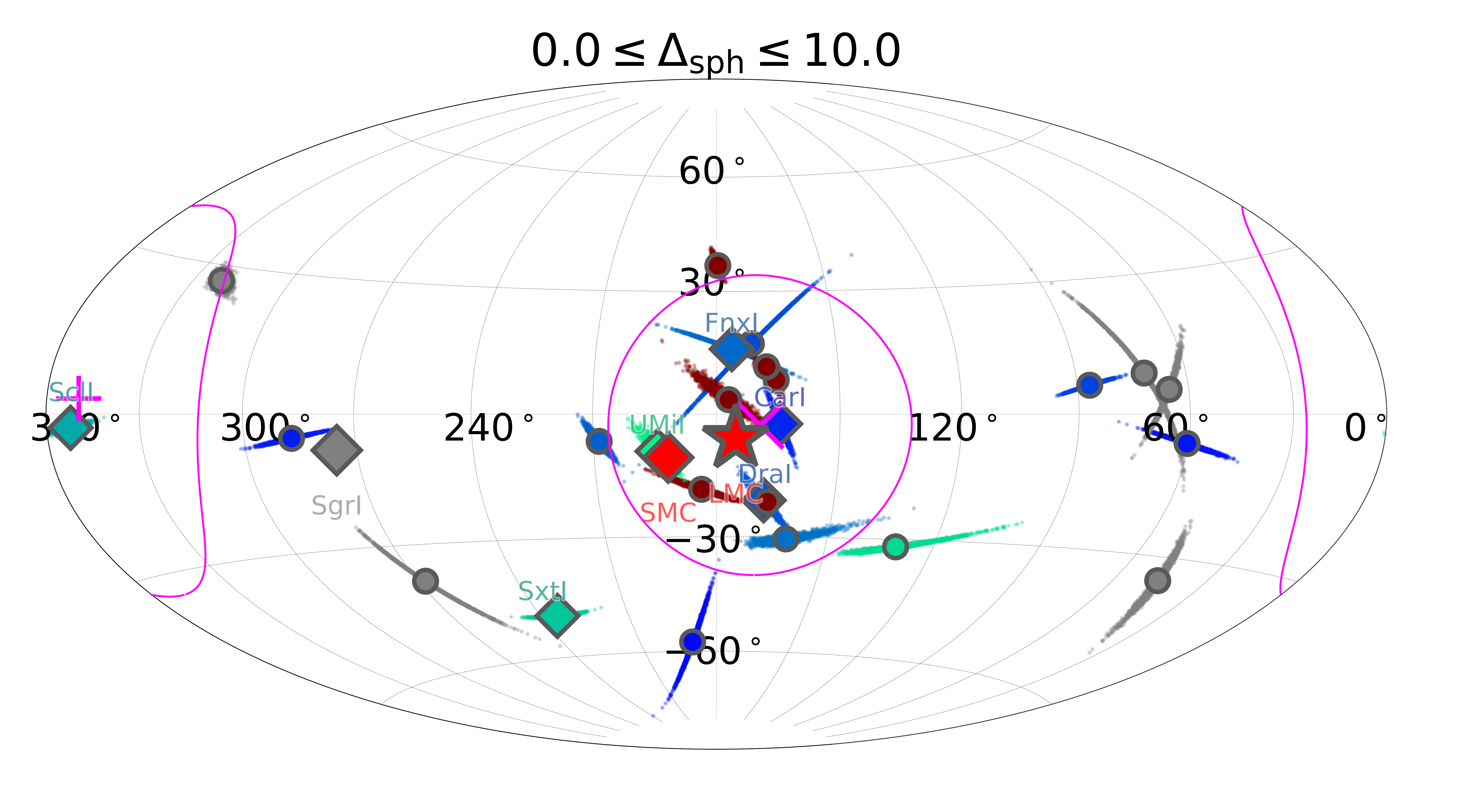}
\includegraphics[width=0.7\textwidth]{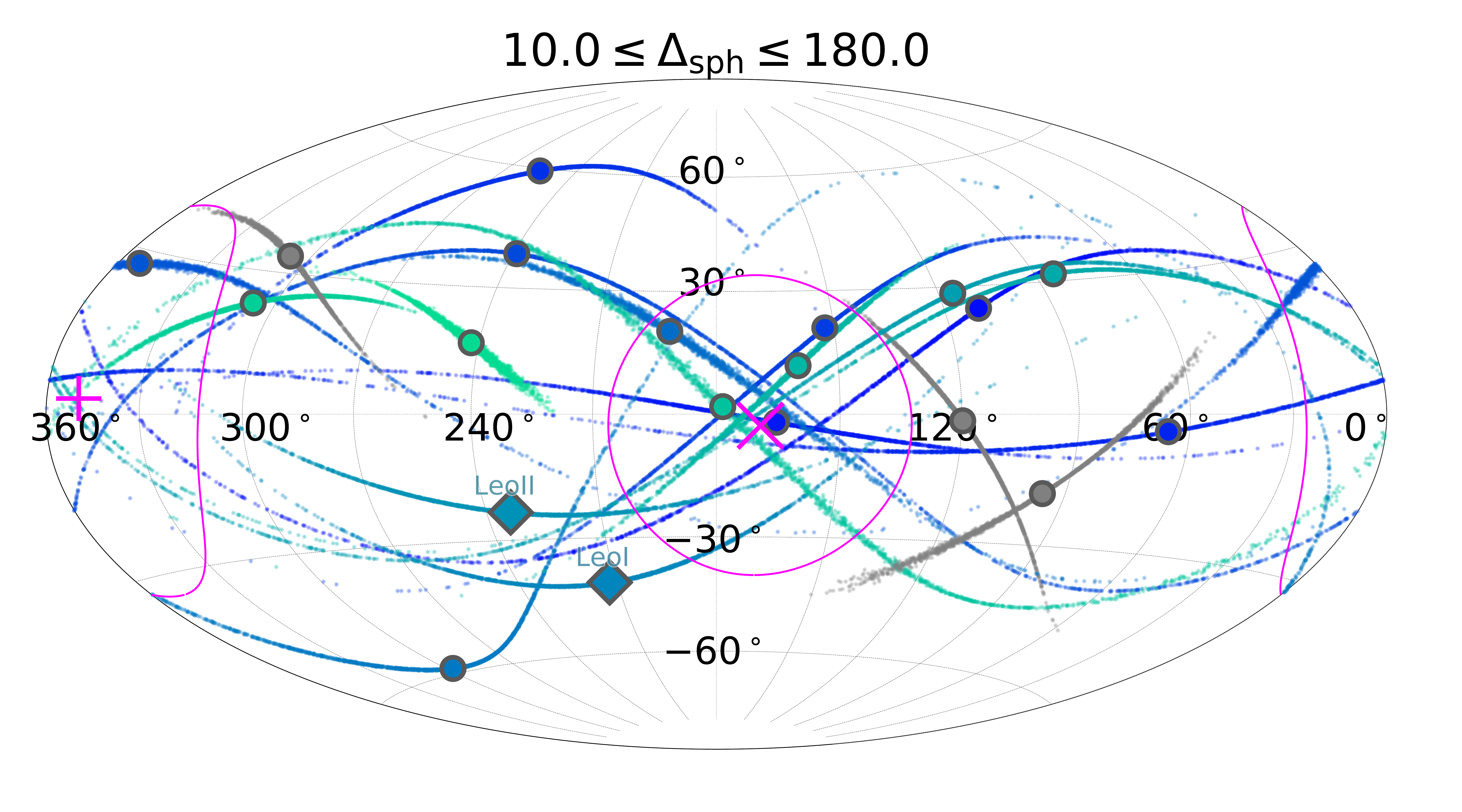}
\caption{ 
Most-likely orbital poles (directions of angular momenta, big symbols) and their uncertainties (colored bands) for the MW satellite galaxies, based on their Gaia eDR3 proper motions from \cite{2021ApJ...916....8L}. Also included are the orbital poles for the LMC, SMC, and Sagittarius from \cite{2020MNRAS.491.3042P}. The classical satellites are shown as diamonds and labelled, the fainter satellites are shown as dots. The uncertainties in orbital pole directions are obtained by Monte-Carlo sampling the proper motion and distance uncertainties.
The upper panel shows the well-constrained orbital poles (1$\sigma$ uncertainty in direction $\Delta_\mathrm{sph} < 10^\circ$), while the lower panel shows those with only weakly constrained or unconstrained orbital poles ($\Delta_\mathrm{sph} > 10^\circ$).
Satellites whose position alone already places them outside of the VPOS are plotted in grey, their orbital poles have no chance to align with the VPOS normal vector. Likely satellites of the LMC are plotted in dark red \cite{2020ApJ...893..121P}.
The pink cross and circle indicate the direction of the normal vector of the VPOS and the 10\% area around it. Among the satellites with well-constrained orbital poles (upper panel), there is a pronounced preference to cluster in this direction, indicating that satellites preferentially co-orbit along this plane. This is not only driven by the likely LMC satellites, but also by other MW satellites, including more massive classical satellites at large distances from the LMC. Among the satellites with only weakly constrained orbital poles (lower panel), many are consistent with orbiting along the VPOS within their uncertainties.
 \label{fig:poles}}
\end{figure}

\begin{figure}
\centering
\includegraphics[width=0.7\textwidth]{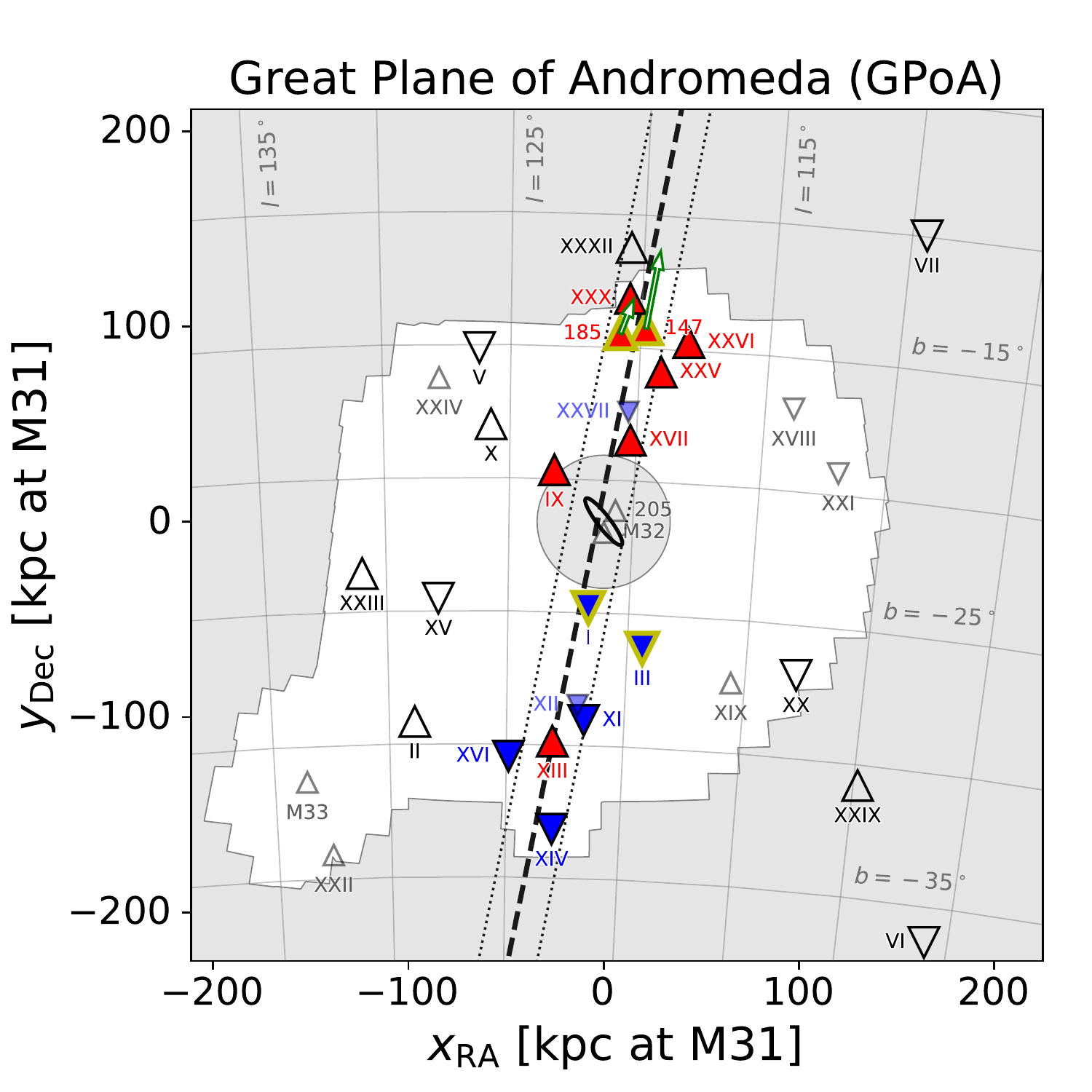}
\caption{Distribution of dwarf galaxies around M31, as seen from our position. The shaded regions indicate the PAndAS survey footprint, within which most M31 satellite galaxies have been discovered. The position and orientation of M31 itself is indicated with the black ellipse in the center. The GPoA is seen edge-on, and is indicated with a dashed line, while the dotted lines give it's root-mean-square width. The symbols indicate whether a dwarf galaxy is approaching (downward pointing triangle) or receding (upward pointing triangle) from us relative to M31's systemic velocity. The on-plane satellites are also color coded as in Fig. \ref{fig:MW}. Smaller, fainter symbols indicate that a dwarf lies behind M31, while bigger, brighter symbols indicate the dwarf is in front of M31. This illustrates the strong lopsidedness of the M31 dwarf galaxy system, with a majority of satellites -- and almost all GPoA members -- residing in front of the galaxy as seen from the Milky Way. The two proposed pairs of satellites, NGC\,147 and NGC\,185 in the North, and Andromeda\,I and Andromeda\,III in the South, are highlighted by yellow outlines. Both pairs are part of the GPoA. The green arrows indicate the most-likely velocity of NGC\,147 and NGC\,185, based on their proper motion, and show that these two dwarfs are consistent with co-orbiting along the GPoA in the sense indicated by the GPoA's line-of-sight velocity trend, but differ enough in their velocities to not constitute a bound pair \cite{2020ApJ...901...43S}. \label{fig:M31}}
\end{figure}

It has already been noticed 45 years ago \citep{1976RGOB..182..241K,1976MNRAS.174..695L} that the then-known MW satellites, some distant globular clusters, and the Magellanic Stream lie along a common great-circle. Over the last years, supported by the discovery of additional satellite galaxies, this suspicion was corroborated (see Figure \ref{fig:MW}). It was shown that the Milky Way is surrounded by a Vast Polar Structure (VPOS), a rotating plane of satellite galaxies, which also contains several globular clusters and stellar and gaseous streams \cite{2012MNRAS.423.1109P}, though the preferential alignment of streams has been questioned with an expanded sample \cite{ 2020MNRAS.494..983R}. The VPOS has a root-mean-square (rms) height of only 20-30 kpc, but a radius of ~250 kpc. As fainter satellite galaxies around the MW were discovered, they were also found to align with the VPOS \cite{2009MNRAS.394.2223M, 2015MNRAS.453.1047P}. A major concern was that since many discoveries were made by SDSS, the survey’s footprint might cause the alignment with the VPOS. Yet, a statistical analysis that carefully reproduced the survey’s sky coverage showed this worry to be unsubstantiated: the SDSS satellites do add significance to the structure (raising it from $4\sigma$\ to $5\sigma$) \cite{2016MNRAS.456..448P}. 

The VPOS is not only a spatial alignment, but also displays substantial kinematic correlation. The orbital poles (directions of angular momentum vectors) of 8 of the 11 classical MW satellites with measured proper motions are aligned with the satellite plane normal vector. This indicates that they are orbiting within the VPOS \cite{2013MNRAS.435.2116P}. Gaia DR2 proper motions confirm these findings for the 11 classical satellite galaxies \citep{2020MNRAS.491.3042P}, but were initially less clear for the fainter systems due to the in part very large proper motion uncertainties (see fig. 3 in \cite{2018A&A...619A.103F}). 
With the recent updated proper motions based on Gaia EDR3 \cite{2021ApJ...916....8L}, it became apparent that $1/2$\ to $3/4$\ of those MW satellite galaxies whose spatial positions place them within the VPOS are indeed orbiting along this structure, with the vast majority co-orbiting in the same sense (Fig. \ref{fig:poles}). Future Gaia data releases and ongoing HST projects will help to clarify the situation further, as many orbital poles of fainter and more distant satellite galaxies are still only weakly constrained, indicating the need for better proper motions to more reliably measure the degree of orbital coherence among the full set of known MW satellites.

A similar satellite plane, the Great Plane of Andromeda (GPoA), has then been discovered around M31 (see Figure \ref{fig:M31}). It consists of 15 out of 27 satellites within the PAndAS footprint, with an rms height of ~12.6 kpc, and is seen edge-on from the MW \cite{2013Natur.493...62I, 2013ApJ...766..120C}. The line-of-sight velocities of 13 of the satellites indicate that they have the same orbital sense around M31, such that the GPoA appears to be co-rotating, too. However, structures with correlated line-of-sight velocities in simulated satellite systems are not necessarily kinematically coherent structures, but can also be transitory features as often found in cosmological simulations \citep{2015ApJ...800...34G, 2016MNRAS.460.4348B}. In this regard it it intriguing that the first HST proper motion measurements for the two M31 satellites NGC \,147 and NGC\,185, which are part of the GPoA, demonstrate that these two objects are indeed likely orbiting along the satellite plane \cite{2020ApJ...901...43S}. This finding strongly supports the possibility that the GPoA is a coherent structure akin to the VPOS, but it will need additional proper motion measurements to fully assess the orbital alignment of the on-plane satellites. The dynamical stability of satellite planes similar to the GPoA was tested by \citet{2017MNRAS.465..641F, 2018MNRAS.473.2212F}, who find a strong dependency on the orientation and shape of the underlying potential (e.g. more spherical halo shapes are beneficial for long-term survival).

\subsubsection{Satellite Planes Outside of the Local Group}

The satellite galaxy planes in the Local Group have helped to motivate searches for satellite structures around more distant hosts (see Table \ref{tab:planes} ). Examples are a flattened distribution of the dSph satellites of M81 \cite{2013AJ....146..126C}, a flattened satellite distribution around M101 that is potentially connected to a surrounding filamentary structure on 3\,Mpc scales \cite{2017A&A...602A.119M}, and a flattened distribution of dwarf galaxies on 300 to 600\,kpc scales around NGC\,253 which additionally shows a potential kinematic correlation in line-of-sight velocities, and could be an extended satellite plane or again be related to a cosmic filament with embedded dwarf galaxies \cite{2021arXiv210608868M}. Motivated by the subsequent discovery of two additional dwarf satellite galaxies around NGC\,253, it was concluded that this spatial alignment indeed appears to extend to the inner satellite galaxies of the system \cite{arXiv:2108.09312}. In addition, six of the seven satellite galaxies within projected 150\,kpc around NGC\,2750 show a pronounced kinematic correlation reminiscent of a rotating satellite plane, a feature which has been argued to make this an analog of the Centaurus\,A satellite system around a lower-mass host \cite{2021ApJ...917L..18P}.

The to-date best-studied case of a satellite galaxy system external to the Local Group, however, is that surrounding the nearby giant elliptical galaxy Centaurus A. Originally, two parallel planes of satellite galaxies were identified around Centaurus A \cite{2015ApJ...802L..25T}. Using new dwarf galaxy candidates around Centaurus A it was later found that the additional objects reduce the significance of this particular double-planar feature, but the additional data support the notion of a single satellite plane \cite{2016A&A...595A.119M}. Since the Centaurus A Satellite Plane (CASP) is again seen close to edge-on, line-of sight velocities allow to test for potential rotation, similar to what was done for M31. Using archival data, it was shown that 14 out of 16 Cen A satellite galaxies exhibit a clear signal of co-rotation in their line-of-sight velocities, and that this signal is aligned along the major axis direction of the Centauris A satellite plane as expected for a rotating struture \cite{2018Sci...359..534M}. Additional velocity measurements have since been obtained for another 12 satellites with VLT MUSE \cite{2021A&A...645A..92M}. These confirm a strong kinematic coherence, with a total of 21 out of 28 satellites following a coherent velocity trend \cite{2021A&A...645L...5M}.
A recent major merger was proposed to explain numerous features of Centaurus A and its stellar halo \cite{2020MNRAS.498.2766W}. Intriguingly, the preferred orbital orientation of this major merger also matches the orientation of the observed, flattened distribution of dwarf satellite galaxies.

Beyond in-depth studies of individual satellite systems, some early statistical approaches have been carried out. These generally suffer from very small numbers of satellite galaxies per host. Since the aim is to study mutual correlation between satellites, a simple stacking of systems is not possible. Instead, at least two satellites need to be known per host. Under the assumptions that a majority of satellite galaxies are embedded in co-rotating satellite planes, it would then be expected that they show opposing velocity vectors on opposite sides of their host. By selecting satellites that are observed to be on diametrically opposed sides of the host, the chance to observe their common satellite plane in an edge-on orientation is increased. If the satellites indeed co-orbit in a common direction and are observed along their common orbital plane, then one should approach and one should recede from the observer in the host's velocity rest frame if they are on opposite sides, i.e. one expecte an anti-correlation in the line-of-sight velocities for diametrically opposed satellites.

\begin{figure}
\centering
\includegraphics[width=0.7\textwidth]{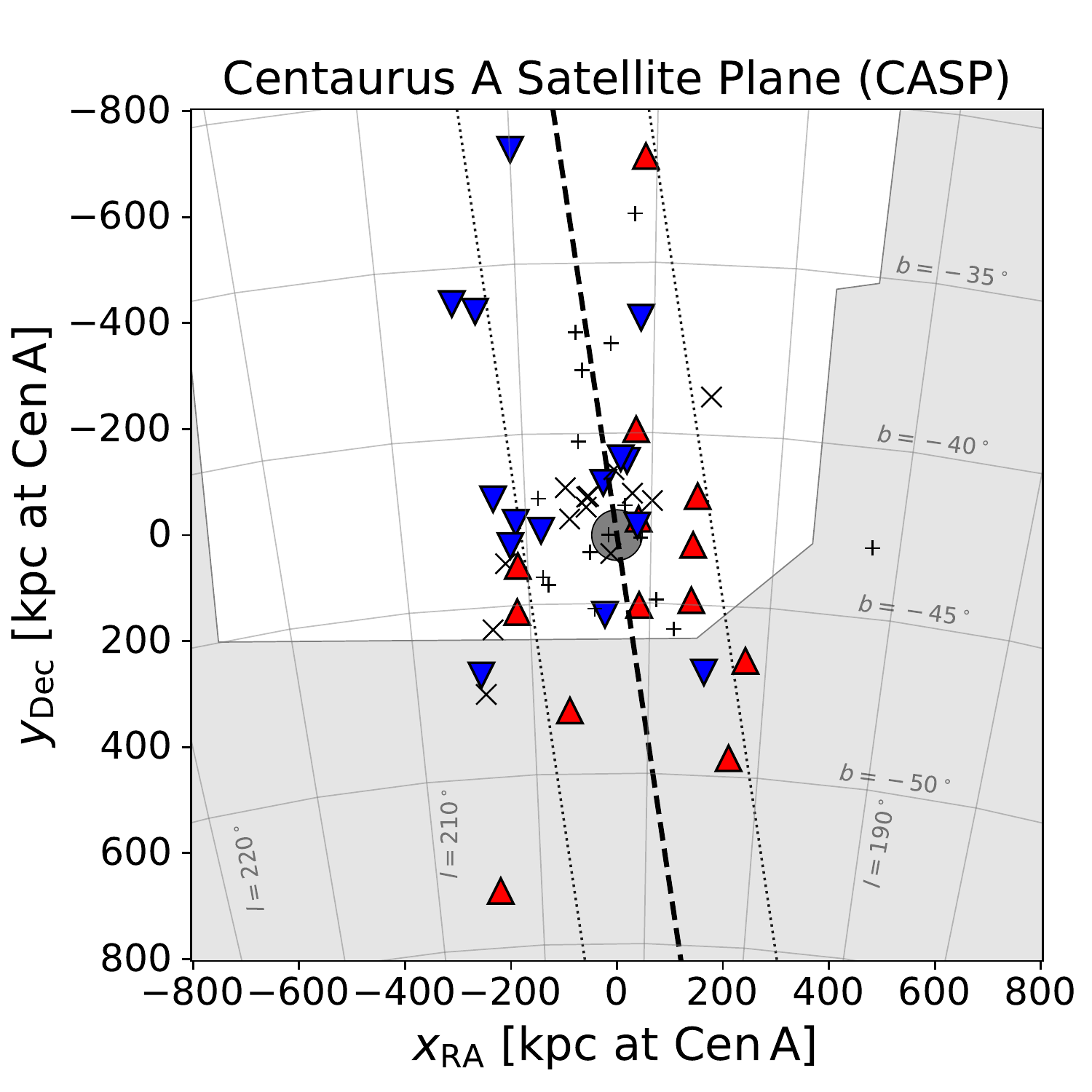}
\caption{Distribution of dwarf galaxies around Centaurus \,A (dark grey circle in the center) as seen from the Milky Way, plotted similar to Figs. \ref{fig:MW} and \ref{fig:M31}. Dwarf galaxies with known line-of-sight velocities are colored, other dwarf galaxies with consistent distance measurements (big crosses) or satellite galaxy candidates (black plus signs) are also shown. The dashed line indicates the major axis of the on-sky satellite distribution, and the dotted lines its root-mean-square heigth. \label{fig:CenA}}
\end{figure}

\startlandscape
\begin{table}
\caption{Proposed Planes of Satellite Galaxies in the Local Volume. \label{tab:planes}
Flattening is either given as absolute rms height $\Delta_\mathrm{rms}$, or as relative flattening as $c/a$\ (minor-to-major axis ratio) or $b/a$ (on-sky minor-to-major axis ratio).
LSS gives the alignment with the surrounding large-scale structure or super-galactic plane (SGP). 
}
\centering
\tablesize{\scriptsize}
\begin{tabular}{lccccccl}
\toprule
\textbf{Host}	& $\mathbf{N_\mathrm{member}}$ 	& \textbf{Extent} & \textbf{Spatial Flattening} & \textbf{Kinematic Coherence} & \textbf{LSS} & \textbf{Ref.} & \textbf{Comments}  \\
\midrule
MW & 11 & 250\,kpc & $c/a = 0.182$, $\Delta_\mathrm{rms} = 19.6$\,kpc & 8 aligned orbital poles & $40^\circ$ & \cite{2012MNRAS.423.1109P,2020MNRAS.491.3042P} & classical satellites only \\
                      & 40-50 & & $c/a = 0.2$-$0.3$, $\Delta_\mathrm{rms} = 20$-$30$\,kpc & 50-75\% aligned orbital poles & & \cite{2015MNRAS.453.1047P, 2021ApJ...916....8L} & 3 counter-orbit \\
M31		& 15-19 & 400\,kpc & $c/a = 0.107$, $\Delta_\mathrm{rms} = 12.6$\,kpc & 13/15 have $v_\mathrm{los}$\ coherence & $5^\circ$ & \cite{2013Natur.493...62I,2013MNRAS.435.1928P} & \\
                 & &  & &  2/2 with PMs co-orbit  & &  \cite{2020ApJ...901...43S} & \\
Cen\,A & 28 & 800\,kpc & $b/a = 0.52$ & 21/28 have $v_\mathrm{los}$\ coherence & $10^\circ$ & \cite{2021AnA...645L...5M} &   \\
M\,81   & 19 & 600\,kpc  & $b/a = 0.5$, $\Delta_\mathrm{rms} = 61$\,kpc & unknown & SGP aligned & \cite{2013AJ....146..126C} & gas-poor satellites only \\
M\,83   & 6 & 210\,kpc   & $\Delta_\mathrm{rms} = 20$\,kpc & unknown & $58^\circ$ & \cite{2018AnA...615A..96M} &  \\
M\,101  & 11 & 1.5\,Mpc  & $\Delta_\mathrm{rms} = 46$\,kpc & unknown & $32^\circ$ & \cite{2017AnA...602A.119M} & filament? \\
NGC\,253  & 7 & 600\,kpc & $c/a = 0.14$, $\Delta_\mathrm{rms} = 31$\,kpc & 4/5 have $v_\mathrm{los}$\ coherence & $17^\circ$ & \cite{2021arXiv210608868M} &  filament? \\
NGC\,2750  & 7 & 150\,kpc & N/A & 6 with $v_\mathrm{los}$\ show strong coherence & N/A & \cite{2021ApJ...917L..18P} &  lower-mass Cen\,A analog? \\
\bottomrule
\end{tabular}
\end{table}
\finishlandscape

This hypothesis has been tested with SDSS data by analysing systems containing at least two known satellites within a projected radius of 150\,kpc that have measured line-of-sight velocities \cite{2014Natur.511..563I}. It was found that there is indeed a strong velocity anti-correlation of line-of-sight velocity directions for satellites with a positional offset of $\approx 180 \pm 8^\circ$\ around their host. This is consistent with ~60 per cent of all satellites living in co-rotating planes. A similar approach was applied to the preliminary satellite sample of the SAGA survey, but did not reveal significant evidence for the presence of co-orbiting planes of satellites \cite{2021ApJ...907...85M}.
The small sample size of only ~20 suitable satellite systems, the limit to within 150\,kpc of the host, and the focus on diametrically opposed satellites of the original study have since been criticized \citep{2015MNRAS.453.3839P,2015MNRAS.449.2576C}. However, it was later shown that there is indeed a significant overabundance of satellites on the opposite side of spectroscopically confirmed satellites around giant host in SDSS, further supporting the notion that some causal link gives rise to a correlation in the spatial and kinematic distribution of satellite galaxies observed in the Universe, yet which is absent in mock-observed cosmological simulations \cite{2015ApJ...805...67I}.

Most recently, a systematic search for flattened arrangements of dwarf galaxies was carried out using the more than 2000 dwarf candidates identified by the MATLAS survey \cite{2021arXiv210810189H}. The study reports the discovery of 31 significant flattened dwarf galaxy arrangements, out of 119 investigated systems. The identified structures have striking similarities in their extent and relative flattening with the known satellite planes in the Local Group and around Centaurus A. If these were indeed edge-on satellite planes, the total incidence of such structures would be even larger because face-on arrangements can not be identified by the adopted analysis. The reason is that for most dwarf candidates only their two-dimensional spatial positions projected on the sky are currently available. It will thus be of interest to investigate the kinematics of the identified candidate satellite planes, once spectroscopy observations confirm the association of their dwarf candidates as satellites  of the proposed hosts, and measure their relative systemic velocities.

\subsubsection{Comparison to Cosmological Expectations: The Planes of Satellite Galaxies Problem}

The planes of satellites have been intensively discussed in the context of $\Lambda$CDM, which predicts typical satellite sub-halo systems to be substantially more isotropic and uncorrelated. \citet{2005A&A...431..517K} first pointed out that a flattened structure as narrow as the 11 then-known, brightest (classical) satellite galaxies of the MW is extremely unlikely to occur by chance, and therefore constitutes a problem for $\Lambda$CDM. However, it has since been cautioned that the null hypothesis of an isotropic distribution of sub-halos in $\Lambda$CDM is incorrect \cite{2005ApJ...629..219Z, 2011MNRAS.411.1525L, 2018MNRAS.473.2234G}: sub-halos are preferentially accreted along filaments, which causes some anisotropy. However, while this anisotropy is significant, it is not necessarily strong enough to explain the observed satellite planes \cite{2012MNRAS.424...80P}.

  Figure \ref{fig:SatPlaneSims} compiles comparisons to the three best-studied planes of satellite galaxies with one common, current state-of-the art hydrodynamical cosmological simulation, Illustris TNG-100 \cite{2019ComAC...6....2N}. Hosts are selected to reproduce the mass, isolation, and absence of an ongoing major merger in the three hosts, and satellites were selected to mimick observational constraints, see the respective publications this data is based on for details \cite{2020MNRAS.491.3042P, 2021A&A...645L...5M, PawlowskiSohn2021}. The figures illustrate that there is strong tension between cosmological expectations for typical satellite galaxy systems and the observed flattening and kinematic coherence of the planes of satellites. Only of order 1 in 1000 simulated systems can simultaneously reproduce or exceed the respective observational flattening and kinematic coherence measures. Similarly small frequencies of as extreme analogs were reported for other cosmological simulations \cite{2014ApJ...784L...6I, 2014MNRAS.442.2362P, 2014ApJ...789L..24P, 2019ApJ...875..105P, 2021MNRAS.504.1379S, 2019MNRAS.488.1166S}.

\begin{figure}
\centering
\includegraphics[width=0.45\textwidth]{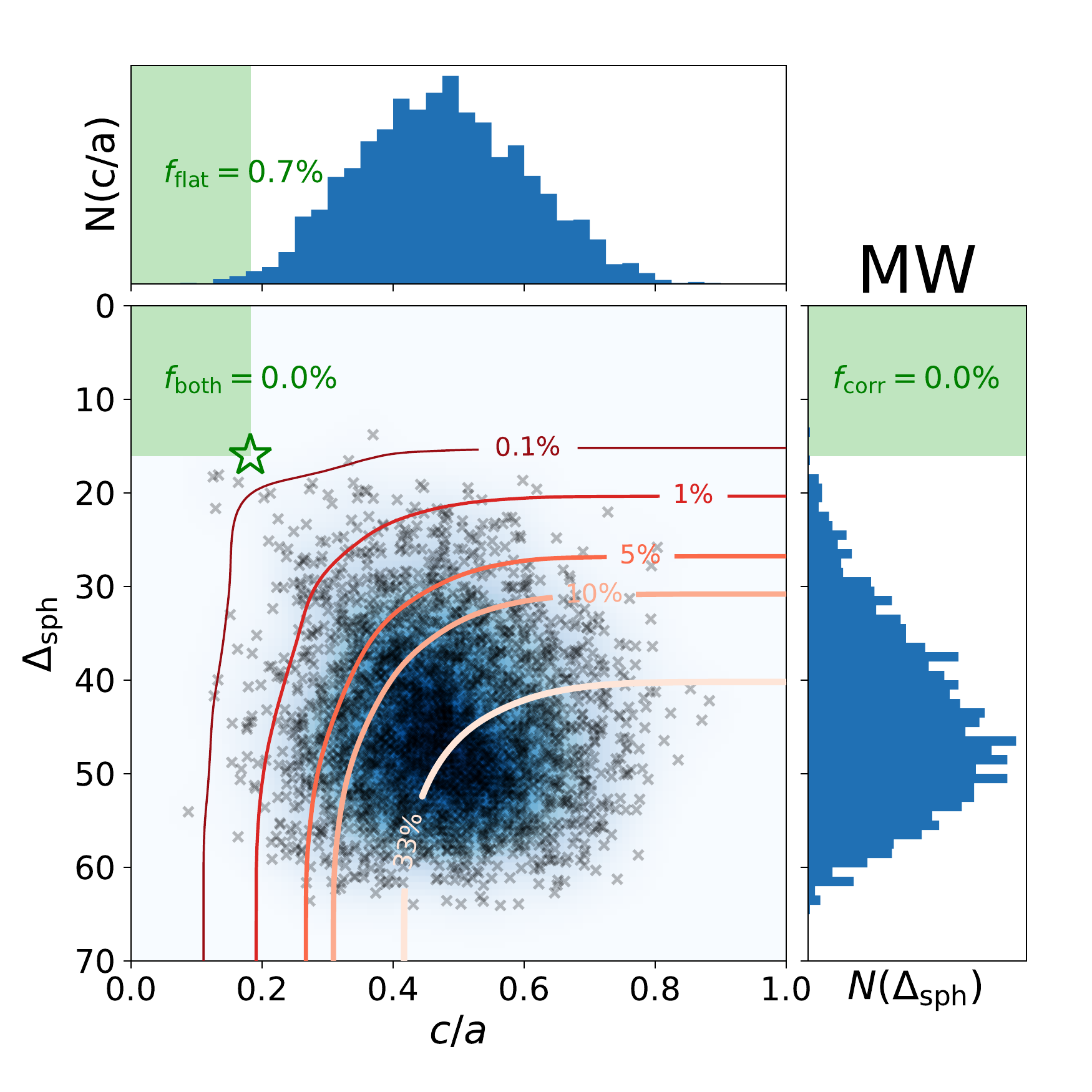}
\includegraphics[width=0.45\textwidth]{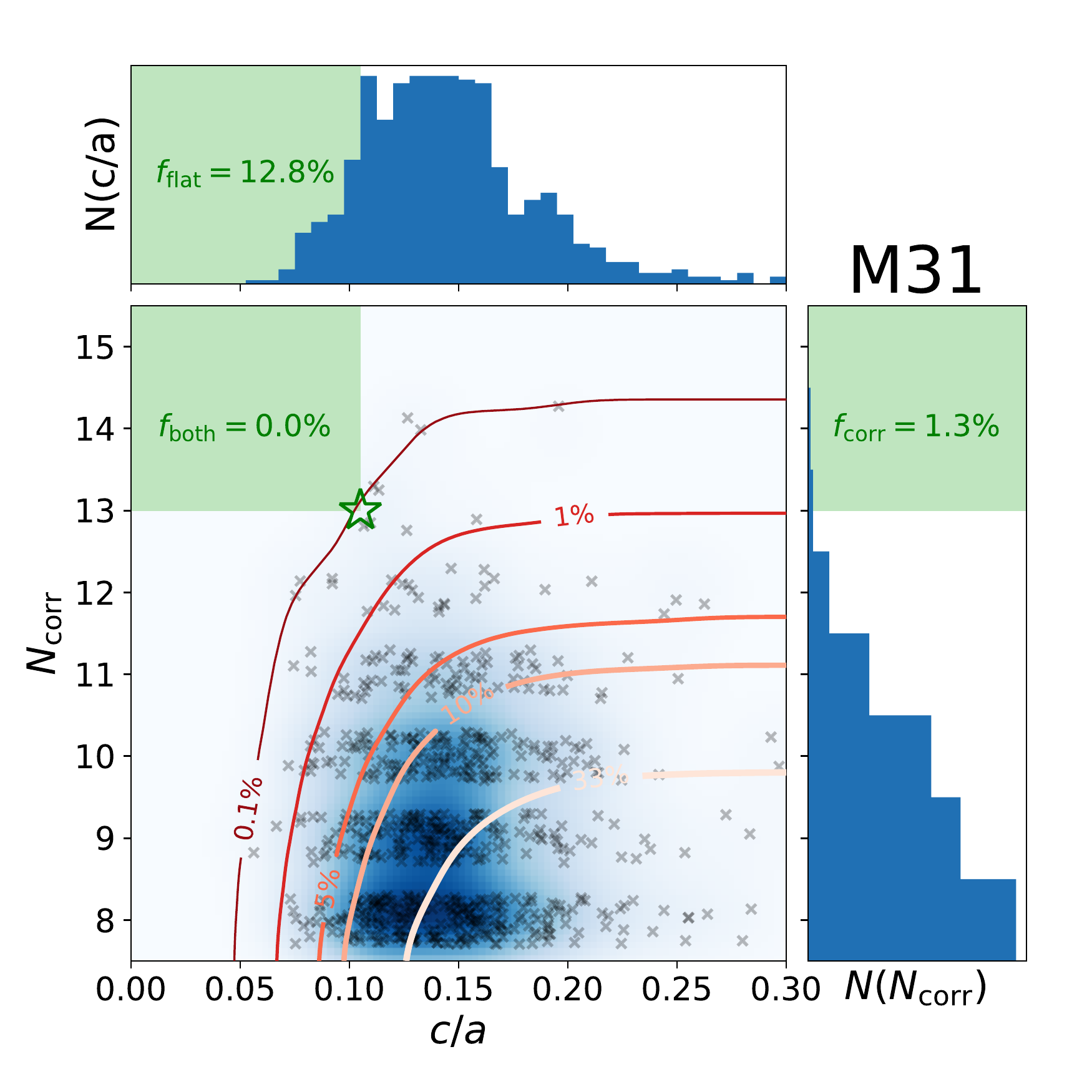}
\includegraphics[width=0.45\textwidth]{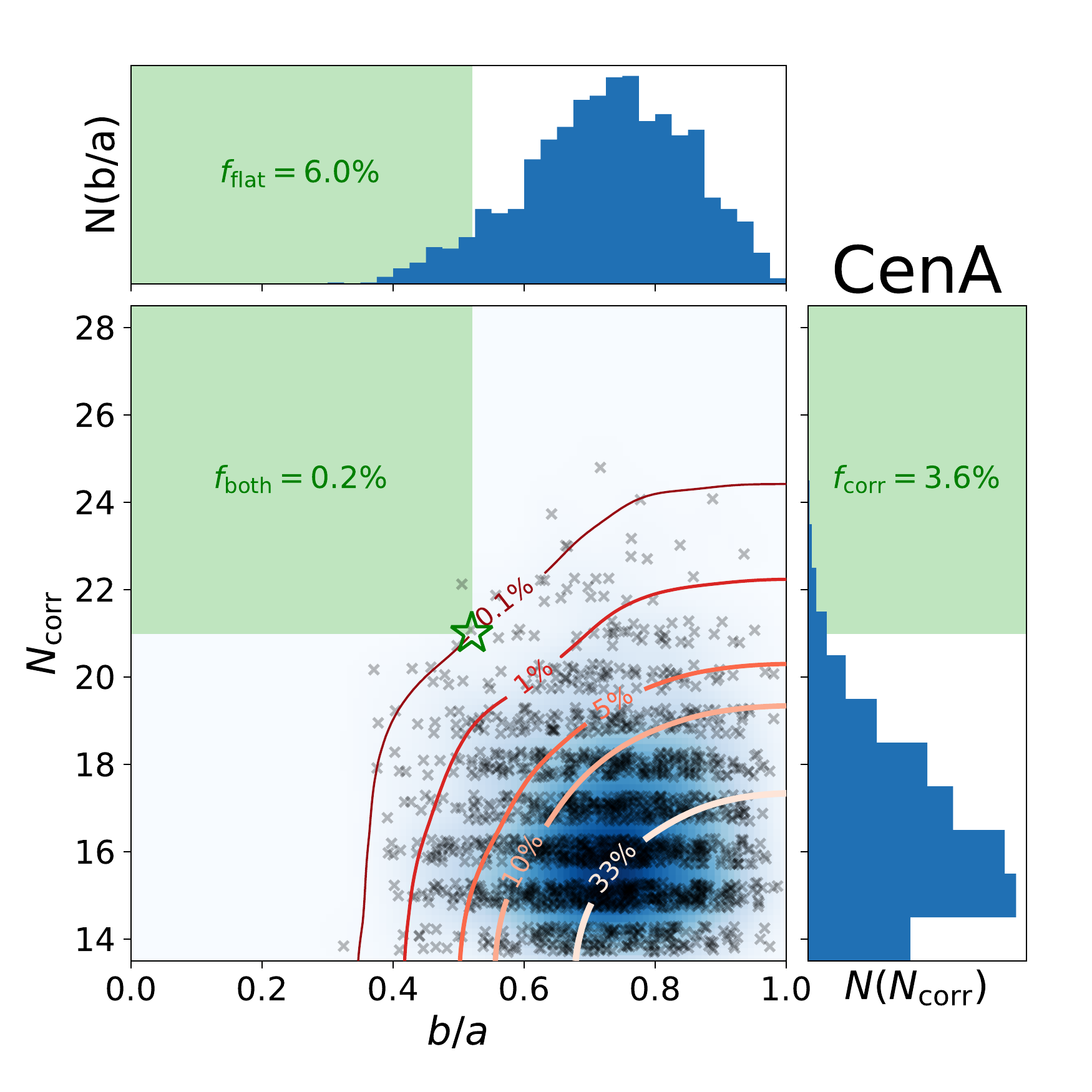}
\end{figure}
\clearpage
\begin{figure}
\caption{
Comparison of flattening (horizontal axes) and kinematic coherence (vertical axes) of observed (green stars) satellite galaxy systems and those found around hosts of similar mass in cosmological simulations. 
For the Milky Way (top panel), flattening is measured as the minor-to-major axis ratio $c/a$\ of the 11 brightest (classical) satellites, while the kinematic coherence is measured using the spherical standard distance of the seven most-clustered orbital poles $\Delta_\mathrm{sph}$. The data is based on the analysis presented in \citet{2020MNRAS.491.3042P}, who report overall results to be very similar for other numbers of combined orbital poles, too.
For Andromeda/M31 (middle panel), the minor-to-major axis ratio flattening $c/a$\ of the 15 most-flattened out of 27 satellites within a mock PAndAS footprint are compared with the $N_\mathrm{corr}$\ number of satellites showing a coherent trend in the sign of their line-of-sight velocities relative to the host, as presented in Pawlowski \& Sohn [{
\it submitted}].
For Centaurus\,A (bottom panel), only the on-sky minor-to-major axis ratio $b/a$\ is used to determine the flattening of the obseved system (to ensure independence of the more considerable distance uncertainties), while the kinematic coherence is quantified as in the case of Andromeda with the number of satellites following a coherent line-of-sight velocity trend. This comparison is based on the analysis presented in \citet{2021A&A...645L...5M}.
All three comparisons are based on the same cosmological simulation, the highest-resolution, hydrodynamical Illustris TNG-100 run \cite{2019ComAC...6....2N}. The blue shaded region indicates the density of individual satellite system realizations in the simulated analogs (grey crosses). The red contours indicate the fraction of systems lying to the top-left of each point, i.e. the systems that are more flattened and more coherent in their velocities.
The green-shaded areas delinate the regions where systems would be as- or more-extreme than the observed ones, and the fractions $f$\ indicate how many simulated systems fulfil these criteria. Clearly, none to only a few simulated systems are simultaneously as flattened and as kinematically coherent as the observed satellite planes. Only of order 0.1\% of the simulated systems reproduce these two quentities of the observed satellite galaxy planes.
\label{fig:SatPlaneSims}}
\end{figure}

  Claims of both tension and consistency of the observed satellite planes (around the MW, and later M31 and Centaurus A) with $\Lambda$CDM simulations have resulted in an active dialogue within the community ( e.g. \cite{2011MNRAS.413.3013L, 2012MNRAS.424...80P, 2013MNRAS.429.1502W, 2013MNRAS.435.2116P, 2013MNRAS.435.1928P, 2014MNRAS.438.2916B, 2014ApJ...784L...6I, 2016MNRAS.457.1931S, 2015ApJ...815...19P}). This has resulted in an extensive body of literature on this topic. Consequently, a full discussion of the different arguments, claims, and contributions would vastly exceed the scope of this review, which aims to not only focus on the planes of satellite galaxies issue. Therefore, the reader is referred to the more in-depth review focussed solely on the Planes of Satellite galaxies problem by \citet{2018MPLA...3330004P}, as well as the review on small-scale challenges of $\Lambda$CDM by \citet{2017ARA&A..55..343B}.

  This ongoing debate is highly productive for the community, and has advanced the level of analyses substantially. It has constrained important properties of the observed satellite planes further, especially in light of what to consider when searching for simulated analogs. The debate has also identified biases which needs to be avoided when measuring the frequency of satellite planes in simulations. Examples are inaccurately modelled survey shapes or Galactic obscuration, or the effect that a more radially concentrated satellite distribution biases towards more narrow satellite planes in absolute height, independent of relative flattening. Another important aspect that was highlighted is the influence of low-number statistics on measures of satellite plane flattening and correlation. This prohibits directly comparing systems of different numbers of satellite galaxies, because the less numerous ones are strongly biased toward more extreme flattening measures. Comparison to purely isotropic systems are necessary to help identify and account for such intrinsic effects.

  Another important aspect to consider it the look-elsewhere effect, which applies particularly to the M31 plane: the choice to assign 15 out of 27 satellites to one planar structure appears arbitrary. For example, a cosmological simulation might not contain an as narrow plane of 15 satellites, but an only marginally wider plane of 17 satellites that is in fact more significant. Yet, considering all such exceptional satellite alignments, still only ~5\% and ~9\% of systems in the Millennium II simulation contain a structure at least as prominent as the VPOS and GPoA, respectively \cite{2015MNRAS.452.3838C}. The Local Group is thus a 0.45\% exception if the two systems can be considered independent. Note, however, that this must be treated as an upper limit, since no observational uncertainties were considered in the analysis, but uncertainties in position would result in wider, less significant satellite planes, while uncertainties in velocity would result in a lower degree of inferred kinematic correlation especially for intrinsically highly correlated structures \cite{2017AN....338..854P}.

\subsubsection{Proposed Origins of Planes of Satellite Galaxies}

Three major scenarios to explain the origin of satellite planes have been suggested, but none of these currently offer a satisfactory solution \cite{2018MPLA...3330004P}. The first two are based on features identified in large-scale simulations of cosmological structure formation. 
One is the preference of matter to be accreted onto host galaxies along cosmic filaments, thereby showing a preference in the accretion direction of satellite galaxies \cite{2011MNRAS.411.1525L,  2014MNRAS.443.1274L}. However, the effect of satellite accretion along filaments is already self-consistently included in cosmological simulations. Since those simulations are in tension with the observed structures, filaments appear to be insufficient to explain the strong correlation in satellite galaxy planes \cite{2012MNRAS.424...80P, 2018MNRAS.476.1796S}.
The other proposed mechanism at play in cosmological simulations it the accretion of satellite galaxies in small groups, the members of which thus share similar orbits. Such group infall will be discussed in more detail in the following section. Here it should only be pointed out that this effect is already present in cosmological simulations, too. It is thus also self-consistently considered in the previously discussed determinations of the very low frequencies of analogs to the observed satellite planes in cosmological simulations -- unless one wishes to pre-select only those systems from the simulations that show very pronounced, recent group infall events, possibly to mimick the presence of the LMC around the MW \cite{2021MNRAS.504.1379S}.

The third proposed scenario for the formation of planes of satellite galaxies is based on tidal dwarf galaxies (TDGs). These are second generation galaxies formed in the debris of interacting galaxies. Such debris populate a phase-space distribution similar to co-rotating satellite planes \citep{2011A&A...532A.118P}. However, TDGs are not embedded in dark matter halos, and should thus have low mass-to-light ratios of $M/L \approx 3$, which is inconsistent with the observed MW and M31 satellites which have $M/L$ of up to 1000. This discrepancy can, however, be avoided if one is willing to adopt other far-reaching changes.
One is abandoning the popular dark-matter-based cosmology in favor of a modified gravity approach such as Modified Newtonian Dynamics (MOND, \cite{1983ApJ...270..365M}). In this case, the timing argument mandates a past encounter between the MW and M31 due to their known baryonic masses, separation, and mutual velocity and driven by the enhanced gravitational acceleration in MOND \cite{2013A&A...557L...3Z}. This encounter in turn could produce tidal debris with distributions and motions broadly consistent with the observed planes of satellites around the MW and M31 \cite{2018A&A...614A..59B,  2018MNRAS.477.4768B}.
Alternatively, one could abandon the notion that the majority of observed satellite galaxies in the Local Group is in dynamical equilibrium. This could easily be the case if they are TDGs, because their lack of a stabilizing potential sourced by a massive dark matter halo would make them much more susceptible to tidal disruption and tidal shocks close to pericenter. This disruption would inflate their velocity dispersions relative to their decreased surface brightness, thereby resulting in overestimated inferred dynamical $M/L$\ ratios \cite{1997NewA....2..139K, 1998ApJ...498..143K}. Since newly-formed TDGs are gas-rich, their potential is dominated by the gas component. Consequently, ram-pressure stripping of the gas could also result in inferring inflated dynamical $M/L$ ratios \cite{2013MNRAS.431.3543H, 2014MNRAS.442.2419Y, 2019ApJ...883..171H}. Intriguingly, predictions of this latter hypothesis \cite{2014MNRAS.442.2419Y} are now within reach of being tested by measuring the internal, tangential velocity dispersions of MW satellites, or also radial expansion motion of their stars. This will be feasible via accurate proper motion studies, once current pioneering approaches based on combining HST first-epochs with Gaia second epochs \cite{2018NatAs...2..156M,2020A&A...633A..36M} have been supplanted by more accurate HST proper motion measurements that will boost the accessible number of stars from the current 15-45 per galaxy to several thousand.

  In summary, there is compelling evidence for the existence of satellite planes around the MW, M31, and Cen A, and promising indications for the presence of similar structures around a wider number of hosts. Despite ongoing research, we currently do not known what causes planes of satellite galaxies. This indicates that we need a more systematic approach to studying correlations in the phase-space distributions of satellite galaxies, which moves beyond focussing on individual structures and adresses this need by comprehensively investigating and interpreting satellite planes in combination with other phase-space correlation both observationally and in numerical simulations.

\subsection{Group Infall}
\label{sect:groups}

\begin{figure}
\centering
\includegraphics[width=0.35\textwidth]{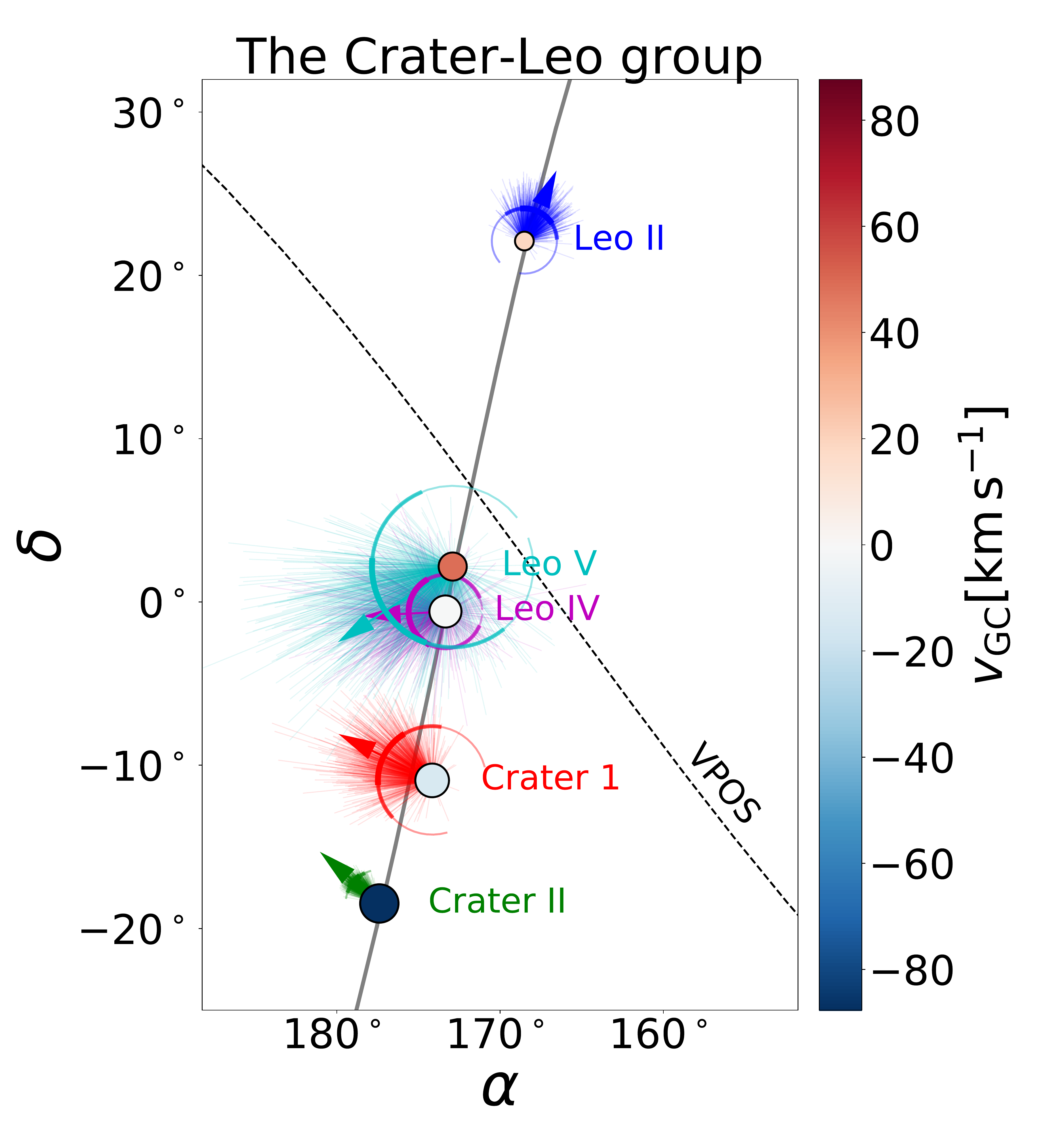}
\includegraphics[width=0.37\textwidth]{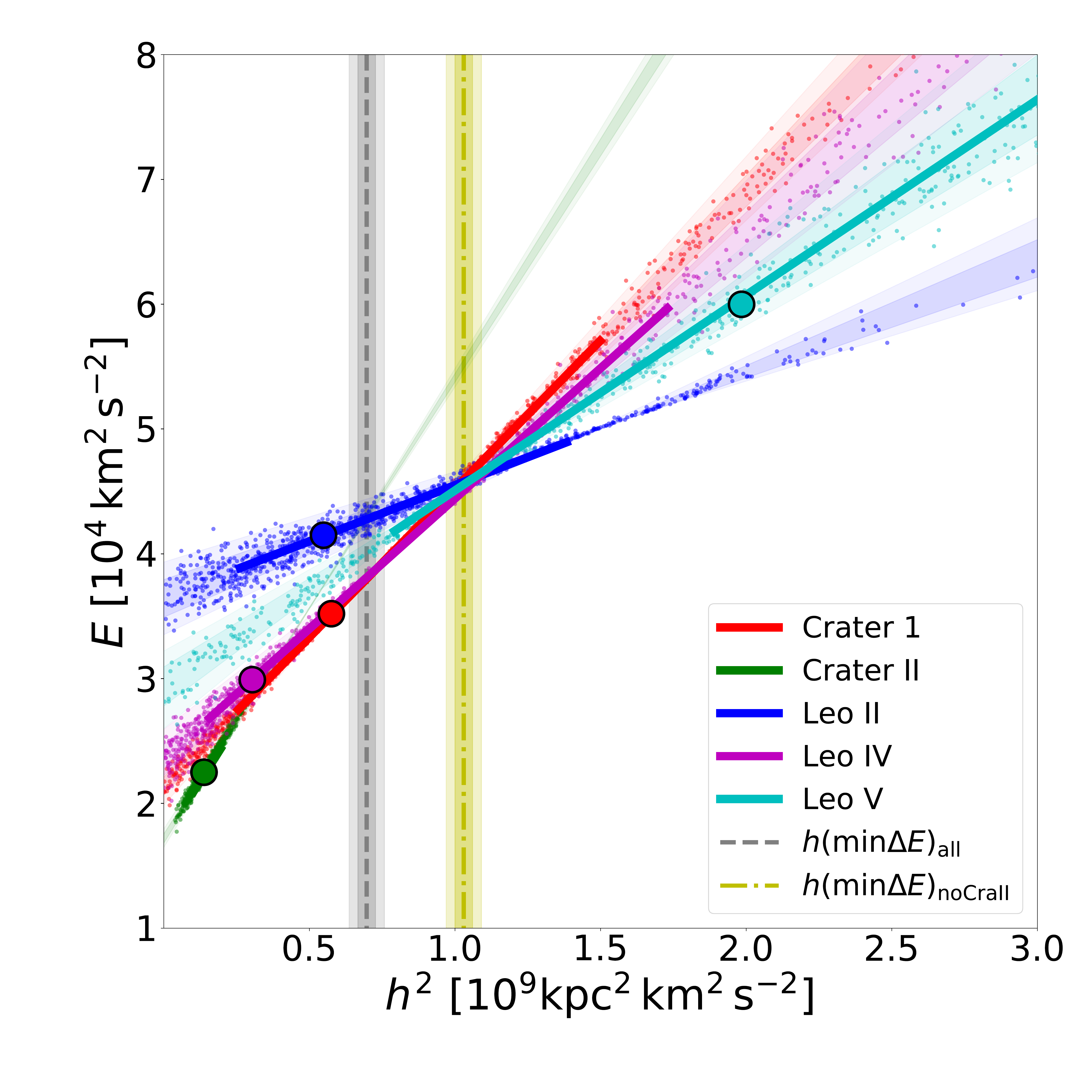}
\end{figure}
\begin{figure}
\caption{The proposed Crater-Leo group of MW satellite galaxies. 
The left panel plots the on-sky positions of the five proposed group members in equatorial coordinates. They align along one great circle (solid grey line), which is inclined relative to the VPOS (dashed line). All satellites have relatively small galactocentric velocities $v_\mathrm{GC}$\ (color-coding of symbols). Smaller symbols correspond to more distant satellites, highlighting the smooth distance gradient among the members from the $\sim 110$\,kpc distant Crater\,II in the South to Leo\,II at $\sim 210$\,kpc in the North. The thick arrows indicate the most-likely direction of motion based on current proper motion estimates, the thinner arrows indicate Monte-Carlo samplings from the proper motion uncertainties, and the circle segments indicate the 1, 2, and 3 $\sigma$\ uncertainties in the velocity directions of the satellites. The velocities are currently not sufficiently constrained to clearly judge a possible association. All but Crater\,II are currently consistent with moving along the direction defined by the proposed group, but all could also could also move along the VPOS like Crater\,II does.
The right panel plots the specific total energy $E$\ and square of the specific angular momentum $h$\ for the five proposed objects, assuming a MW potential contain a mass of $1.5 \times 10^{12}\,M_\odot$\ within 250\,kpc . It is expected that satellites accreted as one group share similar energy and angular momentum. Within the current proper motion uncertainties (small dots indicate individual Monte-Carlo sampled realizations of the observed data), Crater\,1, Leo\,II, Leo\,IV, and Leo\,V are consistent with sharing very similar energy and angular momentum (the point of closest crossing of all four is indicated by the vertical yellow band). Crater\,II appears to be unassociated due to its much smaller energy and angular momentum. \label{fig:CraterLeo}}
\end{figure}

  An important question to address when investigating phase-space correlations of satellite galaxies is whether there is evidence that they are accreted onto a host galaxy in groups. Group infall is one suggested origin of satellite planes \cite{2008MNRAS.385.1365L, 2008ApJ...686L..61D} and could explain the high incidence of satellite galaxy pairs in the Local Group (see Sect. \ref{sect:pairs}). It could also provide constraints on the environmental influences on dwarf galaxy formation and evolution if some dwarfs are conclusively identified to have been accreted together. The best system to search for such evidence is the MW, since we can obtain full 6D phase-space information: positions, distances, and line-of-sight velocities are easily measured, and proper motions can be obtained with reasonable temporal baselines. 

  Recent research, largely driven by Gaia proper motions, has shown that the LMC has brought in a significant number of satellites of their own (e.g. \cite{2017MNRAS.465.1879S, 2018ApJ...867...19K, 2020MNRAS.495.2554E, 2020ApJ...893..121P}). However, this is a highly hierarchical system and an exceptional event unable to explain the existence of the VPOS in its entirety with the LMC on first infall \citep{2011ApJ...742..110N}. In a study covering a wide range of MW and LMC masses and even accounting for the influence of the SMC, none of the other classical MW satellites were found to be dynamically associated to the Magellanic Clouds \cite{2020ApJ...893..121P}. Notably, this includes Carina and Fornax, for which such an association had previously been suggested. In the absence of a dynamical association, group infall does not seem to be a suitable explanation for their similarities in orbital planes. Note, however, that other recent works -- while also concluding that an association with the LMC of these two dwarf galaxies, in particular of Fornax, is unlikely -- can not firmly rule out such an association using current data \cite{2021MNRAS.504.4551S, 2021arXiv210608819B}.

  The effect of infall of cosmologically more common groups has received surprisingly little attention. $\Lambda$CDM simulations predict that some satellites are accreted in groups, albeit there is a lack of evidence supporting the notion that it can cause sufficiently strongly correlated satellite planes. This is in part because only a fraction of ~ 30\% all satellites were part of groups, and the typical number of 2-5 satellites per group is much smaller than the number of $\geq 10$\ satellites making up the satellite planes \cite{2008MNRAS.385.1365L, 2013MNRAS.429.1502W, 2015ApJ...807...49W}. The infall of satellites in groups thus appears to be self-consistently included in cosmological simulations, but insufficient to explain the extreme observed structures. However, a potential way out would be finding evidence for a strong environmental dependence of dwarf galaxy luminosity, number of dwarf galaxies belonging to one group, and the group’s compactness. If the groups with the largest number of members also form the brightest dwarf galaxies, their infall into the Milky Way could dominate the observed satellite population. The effect of group infall thus must be studied in more detail in cosmological simulations, in particular in hydrodynamic simulations including the effects of star formation and inhomogenious reionization. 

  Satellites that are accreted as a common group are expected to share similar total energy and specific angular momentum \citep{1995MNRAS.275..429L}, providing a path to identification. The dwarf galaxies Leo II, IV, and V, Crater II, and the star cluster Crater 1 are good candidate case of group infall (see Fig. \ref{fig:CraterLeo}). They align along one great circle on the sky, follow a common distance gradient (from the more nearby Crater\,II at 110\,kpc distance, to the most distant Leo\,II at 210\,kpc), and have similar Galactocentric velocities \cite{2016MNRAS.459.2370T}. Gaia DR3 proper motions (arrows with arcs indicating $1-3\sigma$\ intervals on direction of motion in Fig. \ref{fig:CraterLeo}) are unfortunately still inconclusive due to the large distance of the putative group. However, HST observations are a promising route to measure proper motions of the distant Crater-Leo objects, which not only promise to conclusively confirm or refute whether these objects share a common origin, but also offers a possibility to constrain the MW potential out to $>100$\,kpc.

  For M31, it had been suggested that the dwarf elliptical galaxy NGC\,205 might have been the most massive member of an infalling group, and that this could potentially explain the observed, apparently co-orbiting GPoA \cite{2016MNRAS.462.3221A}. However, the highest probability of reproducing the observed thinness of the GPoA in this scenario was 1\%, and additionally requiring a match of the line-of-sight kinematics reduced this to around 0.01\%. The reason reported by the authors was that the detailed satellite phase-space distribution made it difficult to find a convincing match. Excluding a number of particularly problematic satellites from the observed sample can increase the probability of a match to around 0.1\%, but this remains comparable to the frequency of GPoA analogs around hosts drawn from cosmological simulations, and thus does not appear to offer strong additional explanatory power regarding the observed phase-space arrangement. It would, however, be worth to revisit this scenario once a more complete understanding of the full 3D velocities of M31's satellite system via proper motion measurements has been established.

\subsection{Lopsidedness}
\label{sect:lopsided}

\begin{figure}
\centering
\includegraphics[width=0.35\textwidth]{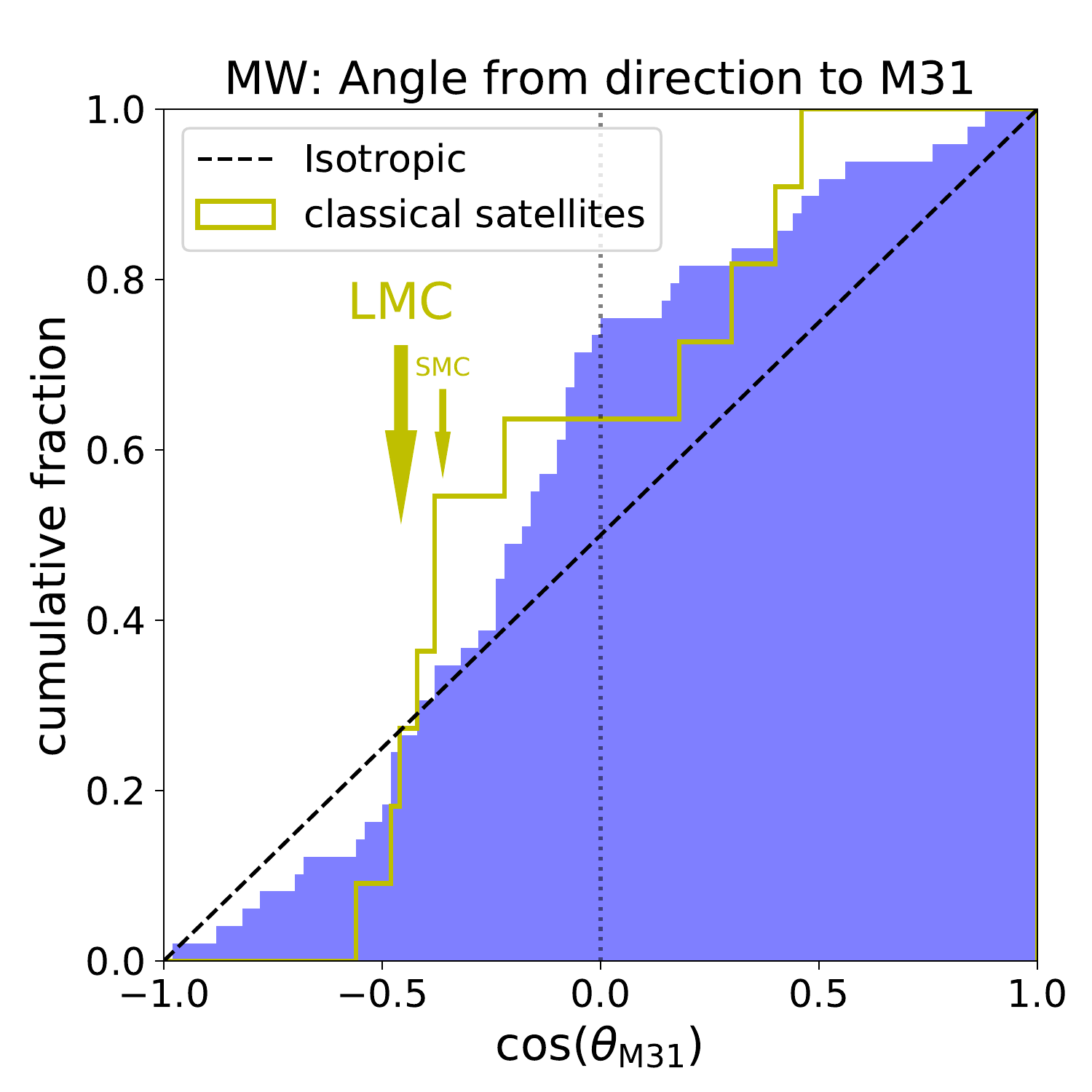}
\includegraphics[width=0.35\textwidth]{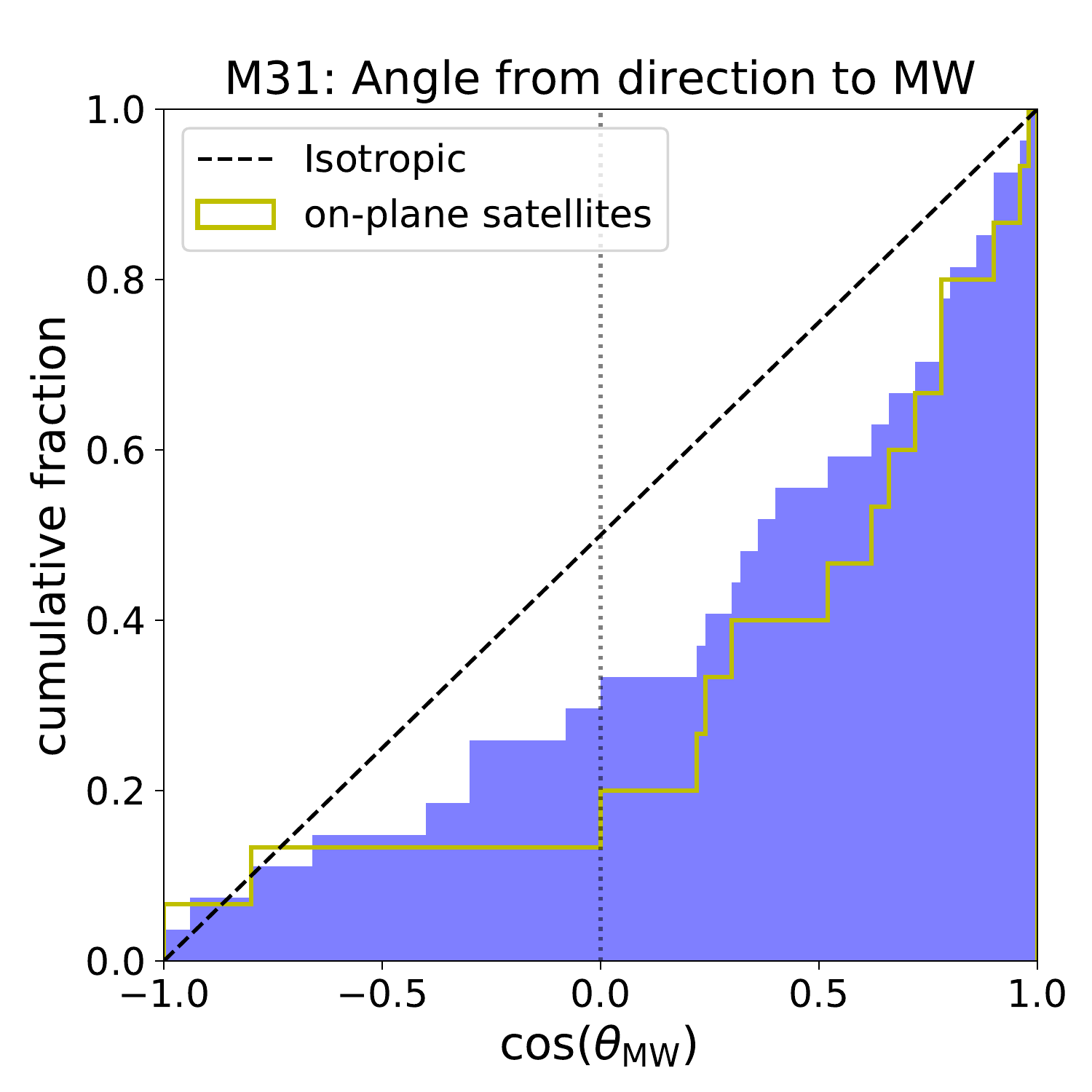}
\caption{Cumulative distribution of the Cosine of the angle $\theta$\ between the position of satellites around the MW (left panel) and M31 (right panel), and, respectively, the direction from the MW to M31, and from M31 to the MW.
The dashed black line indicates the expectation for an isotropic distribution. The dotted vertical line separates the hemispheres facing away from (left half, $\cos(\theta) < 0$), or towards (right half, $\cos(\theta) > 0$) the other major galaxy. The yellow lines give the cumulative distribution for only the 11 classical satellites of the MW, and for only the on-plane satellites of M31, respectively. For M31, there is a clear lopsidedness towards the MW, which is even more pronounced for the satellites that are members of its satellite plane. The MW satellites tend to preferentially face away from M31 (i.e. globally in the same direction as the M31 satellites), but uneven survey coverage areas and the presence of the LMC (corresponding angle indicated with a yellow arrow) with an entourage of its own satellites makes this difficult to interpret.
  \label{fig:MWM31lopsided}}
\end{figure}   

Another observed phase-space correlation is the lopsidedness of satellite galaxy systems. The most nearby example is M31, for which it was found most likely that 20 out of 27 of its satellites in the PAndAS survey lie on the side towards the MW \cite{2013ApJ...766..120C}. A highly asymmetric distribution of satellite galaxy candidates was also reported around M101, with seven low-surface-brightness satellite candidates all residing on one side of the host \cite{2014ApJ...787L..37M}.

Figure \ref{fig:MWM31lopsided} shows the cumulative angular distribution of the MW and M31 satellite galaxies, relative to the line connecting the MW and M31. The observed satellite distributions differ from that expected for isotropy. While the lopsidedness around M31 can be considered to be well established, for the Milky Way, the situation is more difficult to reliably evaluate. This is due to the substantial obscuration by the MW disk and the uneven coverage of the sky by surveys searching for satellite galaxies. A further complication is the LMC, which brings along an entourage of its own satellites that likely result in an overdensity of faint dwarf galaxies in that direction. Of the presently known $\sim 50$\ satellite galaxies, 75\% are found within the hemisphere away from M31 (seven of the 11 classical satellites). Globally, for the Local Group both the MW and M31 thus appear to have a lopsidedness towards the same direction, defined as the direction from M31 towards the MW. 

Motivated by the M31 satellite system, \citet{2016ApJ...830..121L} made a first attempt to statistically study lopsidedness in external satellite systems. They identified pairs of host galaxies in the SDSS, and stacked their surrounding satellite candidates by measuring their position angle relative to the line connecting the two galaxies. This revealed a similar feature as seen for M31: a significant excess of ~8\% of satellites in the direction towards a partner host compared to a random distribution.

Such a lopsidedness was not a priori predicted by cosmological simulations, thus offering another test and potential challenge for $\Lambda$CDM. An exploration of lopsidedness of satellite systems using the Millennium I \& II, Illustris, and ELVIS cosmological simulations, following the selections made in the observational study \cite{2016ApJ...830..121L}, has demonstrated that a lopsided signal very similar to the observed one is present and highly significant for the two Millennium simulations (5.9 and 4.4$\sigma$, respectively) \cite{2017ApJ...850..132P}. There are also some indications that lopsidedness is stronger among more massive satellites, but a theoretical understanding of the origin of the signal, as well as its possible connection to the filamentary structure of the cosmic web, was still missing.

This was rectified in a follow-up study using the Extremely Small Multidark simulation \cite{2019MNRAS.488.3100G}. By tracing back the accretion history of satellite galaxies of paired hosts, which show a lopsided satellite distribution at $z=0$, it became clear that the lopsidedness of $z=0$ satellites was stronger in the past. The signal was also driven mainly by recently accreted objects which have not yet completed a full orbit around their hosts. Taken together, this appears to indicate that recent accretion events are a major driver in producing lopsided satellite distributions. In contrast to other phase-space correlations, the observed degree of lopsidedness of distant satellite galaxy systems thus appears not to be problematic for the $\Lambda$CDM model of cosmology and current galaxy formation models.

While the previous studies have focussed on pairs of host galaxies, motivated by the Local Group, lopsided satellite distributions have now also been reported around isolated host galaxies \citep{2020ApJ...898L..15B}, with pairs of satellites having a strong tendency to be located on the same side of a host. The signal appears to be stronger for blue than for red host galaxies, and for satellites at the outskirts compared to the inner regions of a host. Similar trends have since been identified in the hydrodynamical cosmological simulation Illustris TNG-300 \cite{2021ApJ...914...78W}.

\section{Other Studies of Satellite Galaxy Phase-Space Distributions}

Due to a lack of sufficiently deep observations and spectroscopic follow ups, the typical number of satellites (e.g. in the SDSS) known around a distant host is $\sim 1 - 2$. Satellite alignments are therefore often studied by stacking many systems and comparing them to properties of the host galaxy, e.g. its major axis \citep{1969ArA.....5..305H, 2006MNRAS.369.1293Y}. These studies are therefore not directly applicable for the current context: they do not consider mutual alignments and correlations between satellites. However, a number of other phase-space pecularities have been studied in the Local Group, and even though these have received less attention than the previously discussed correlations, they deserve to be included for completeness and as encoragement to further pursue these avenues of research as better data become available.

\subsection{Pairs of Satellite Galaxies}
\label{sect:pairs}

  Among the Milky Way satellites, the Large and Small Magellanic Clouds are certainly the best-known pair of galaxies. They are the brightest nearby satellites, visible by naked eye and thus known to humanity since ancient times. With an angular separation of only $20^\circ$, they are sufficiently close on the sky to suggest an association, and this intuition was since confirmed by their similar distance, velocity, and proper motions, indicating that they share an (inter)active recent history \cite{ 2013ApJ...764..161K, 2012MNRAS.421.2109B}.

  A number of other dwarf galaxies in the Local Group have also been suggested to constitute pairs, or binary satellite galaxies, due to their spatial proximity and similarities in line-of-sight velocities (see Table \ref{tab:pairs}). These are the M31 satellites NGC\,147 and NGC\,185 \cite{1998AJ....116.1688V, 2010ApJ...711..361G}, and the MW satellites Leo\,IV and Leo\,V \cite{2008ApJ...686L..83B}, which were suggested to be a possibly bound pair of satellite galaxies \cite{2010ApJ...710.1664D,2012A&A...542A..61B}. In addition, another two potential pairs, the Milky Way satellites Draco and Ursa Minor and the M31 satellites Andromeda\,I and Andromeda\,III, were subsequently identified in a systematic analysis of known satellite galaxies \cite{2013MNRAS.431L..73F}. Looking at the spatial and velocity distributions of dwarf galaxies in the Local Group, the study uncovered a high frequency of pairs of satellite galaxies of comparable magnitude. About 30\% of all observed satellites brighter than $M_\mathrm{V} = -8$\ are found in likely pairs, whereas a frequency of only 4\% is expected from cosmological simulations. However, for NGC\,147 and NGC\,185, more recent distance measurements place them at a 3D separation of $\sim 90$\,kpc \cite{2015ApJ...811..114G}, which is more in line with a chance alignment in projection than spatially compact binary satellites.

No clear explanations is available yet for this apparent over-abundance of pairs compared to simulations. The situation is further complicated by the finding that, with the possible exception of Leo\,IV and Leo\,V, none of the four identified paired satellites appear to currently be gravitationally bound \cite{2014MNRAS.440.1225E}. This is because, even if they had sufficient mass to have negative total energy, the pairs' estimated tidal radii are still smaller than their separations.

Proper motion measurements have the potential to claify the situation. Specifically, if the members of proposed pairs have very different motions on the sky, any associations would have to be ruled chance alignments. Unfortunately, nature isn't this forthcoming, and proper motion measurements have more complicated than resolved the situation. While their likely large separation in 3D makes an association unlikely, the proposed pair NGC\,147 and NGC\,185 do constitute the first two M31 satellites for which proper motions have been measured with the HST \cite{2020ApJ...901...43S}. The puzzling finding of this study is that the two object are in fact consistent with moving in a similar direction around M31, but their typical orbits are too distinct to be in line with a bound satellite pair. However, within the substantial uncertainties in distance and proper motions, it cannot yet be firmly ruled out that they used to be a bound pair some time in the past.

For Leo\,IV and Leo\,V, their large distances result in rather uncertain Gaia proper motions. Even with EDR3, all that can be concluded is that they are consistent with sharing a similar orbit \cite{2021ApJ...916....8L}. It will probably require accurate HST proper motions to develop a better understanding of their possible phase-space association.

\begin{figure}
\centering
\includegraphics[width=0.245\textwidth]{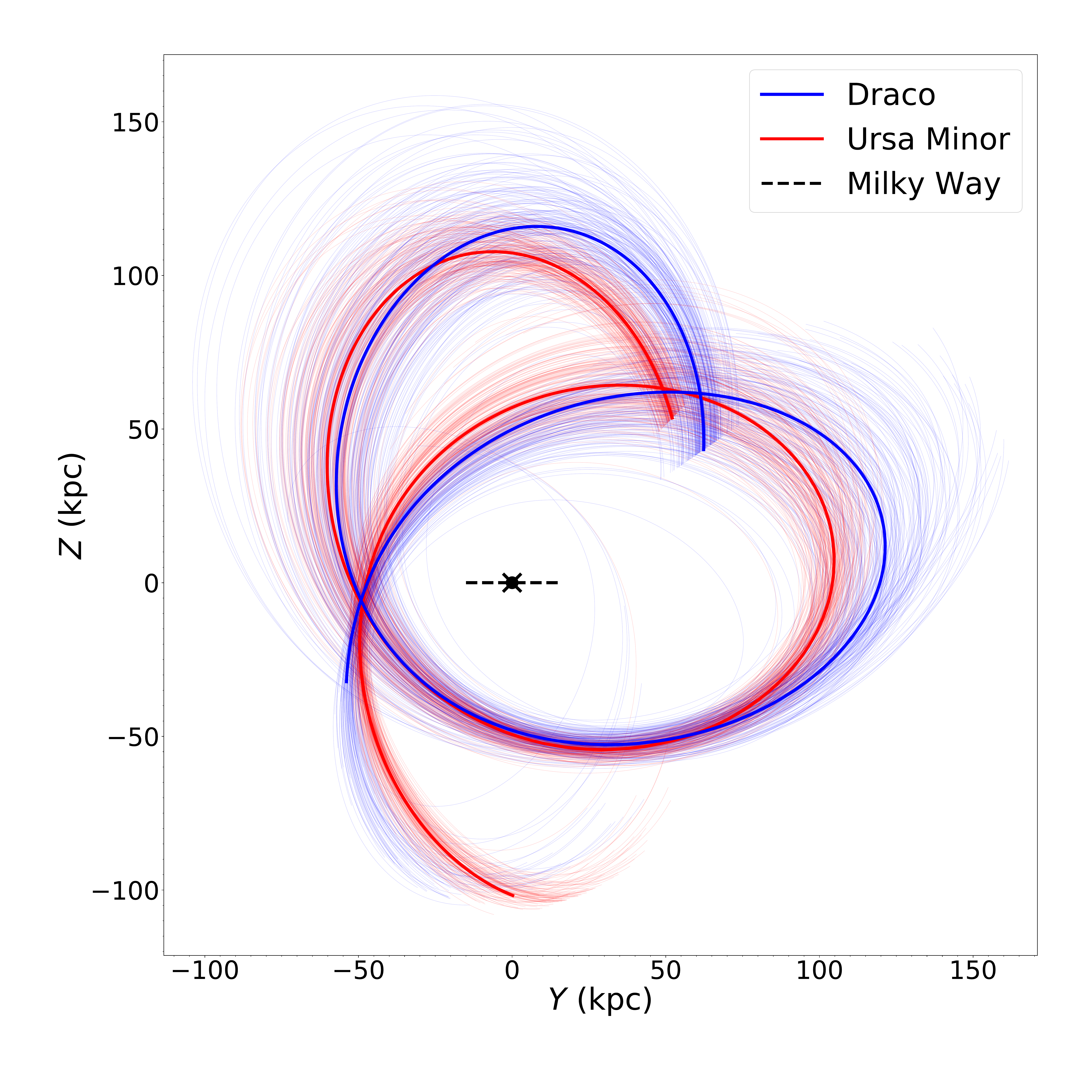}
\includegraphics[width=0.47\textwidth]{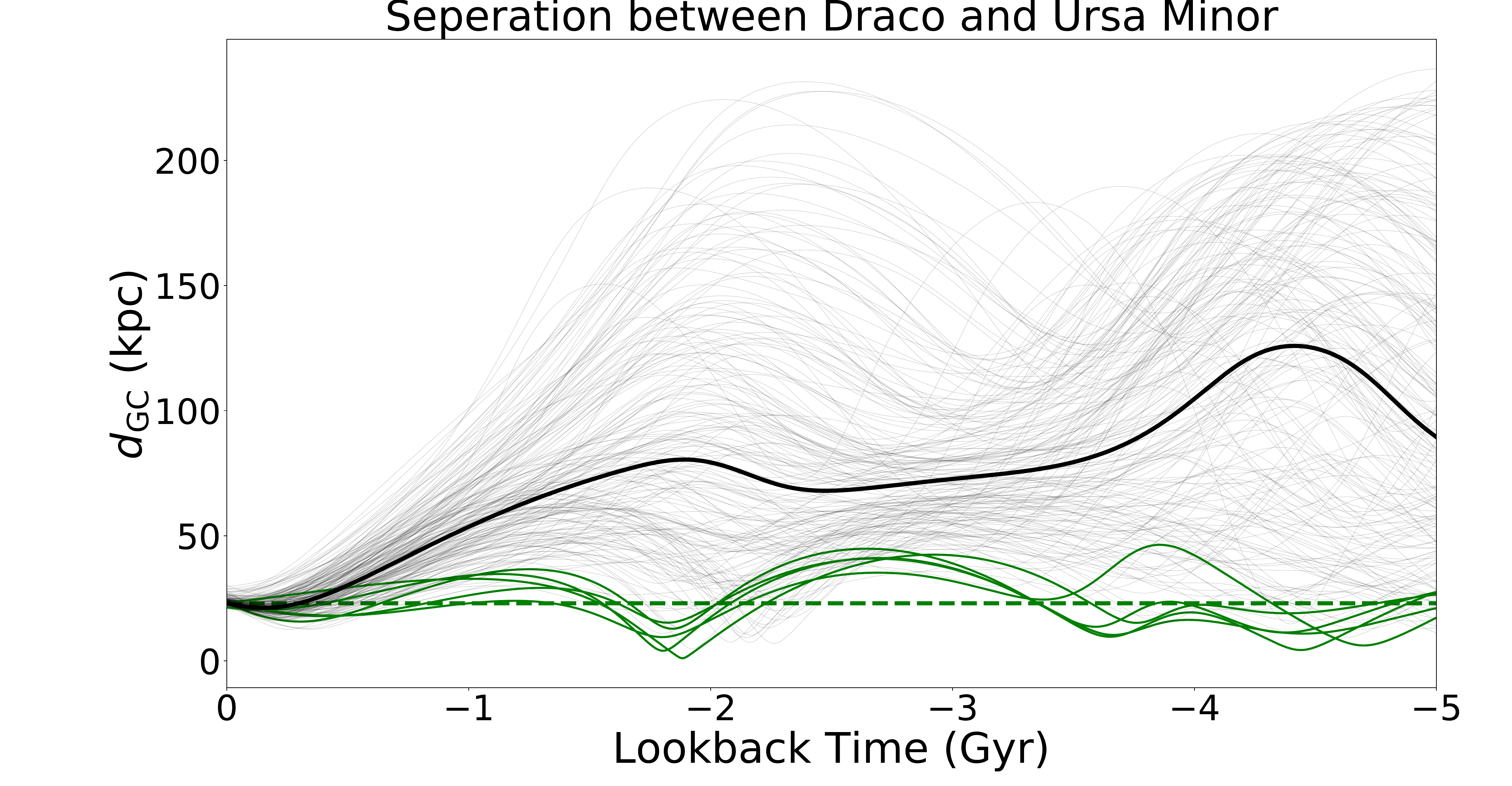}
\caption{
Possible orbits of the dSphs Draco and Ursa Minor, a proposed pair of MW satellite galaxies.
The left panel shows the orbits in the y-z plane of a Galactocentric Cartesian coordinate system. This orientation is a close-to face-on view of the orbital planes of both dwarfs (and the VPOS). Each line corresponds to one possible orbit realized by drawing from the observational uncertainties (most notably the distance and proper motion) in a Monte-Carlo fashion. Draco and Ursa Minor are consistent with sharing the same orbital plane, orbital direction, orbital eccentricity, and similar apo- and pericenters. The right panel shows the separation between Draco and Ursa Minor resulting from different Monte-Carlo realizations, integrated backwards for 5\,Gyr. While the proper motion uncertainties remain too high to conclusively confirm that the two satellites share a similar orbit, they are consistent with co-orbiting close to each other, because numerous realizations result in an average mutual separation below 30\,kpc (highlighted in green).
The orbits were integrated using the galpy package, adopting its MWpotential14 with a dark halo mass of $0.8 \times 10^{12}\,M_\odot$\ as the potential of the MW \cite{2015ApJS..216...29B}.
  \label{fig:DraUMiPair}}
\end{figure}   

\startlandscape
\begin{table}
\caption{Proposed pairs of satellite galaxies in the Local Group. \label{tab:pairs}}
\centering
\begin{tabular}{lccl}
\toprule
\textbf{Pair}	& \textbf{Host}	 & \textbf{Satellite Plane Alignment} & \textbf{Comments}\\
\midrule
LMC + SMC		& MW  & yes (in pos + 3D vel) & group bringing in a number of smaller satellites, too \\
Draco + Ursa Minor		& MW  & yes (in pos + 3D vel) & within PM uncertainties small separation possible for long time\\
Leo\,IV + Leo\,V		& MW  & yes (in pos, 3D vel unclear) & possible group with Crater\,1, Crater\,II, Leo\,II? \\
And\,I + And\,III & M31 & yes (in pos, los vel.) & no PMs available yet \\
NGC\,147 + NGC\,185 & M31 & yes (pos + 3D vel) & PMs indicate similar orbital plane but different orbits\\
\bottomrule
\end{tabular}
\end{table}
\finishlandscape

Draco and Ursa Minor, in constrast, share almost exactly the same orbital properties in pericenter, apocenter, and eccentriciy based on recent Gaia EDR3 proper motions \cite{2021ApJ...916....8L}. They also orbit in the same direction in a common orbital plane, which furthermore aligns well with the VPOS. This makes these two objects the currently most promising candidate for a longer-term pair of satellites in the Local Group. The striking similarity of their orbits is highlighted in Fig. \ref{fig:DraUMiPair}. While available proper motions are not sufficiently constrained to conclusively prove that the two objects have remained close to each other for a longer period, within the current measurement uncertainties such orbits are possible and not particularly unlikely (green lines in the right panel of Fig. \ref{fig:DraUMiPair}). 
Furthermore, the two satellites currently reside on the opposite side of the MW from the LMC, which should minimize the latter's influence on the potential pair. Given their similarity in position, distance, velocity, and proper motion, one can also expect that any effect the LMC (or the MW's resulting reflex motion) has on their orbits would influence both Draco and Ursa Minor to a very similarly degree.

\subsection{Tangential Velocity Excess}
\label{sect:tangentialvelocity}

\citet{2017MNRAS.468L..41C} investigated the relation between radial and tangential velocity of MW satellites with measured proper motions. They discovered a significant tangential velocity excess: 9 of 10 satellites have $\geq 80\%$\ of their orbital kinetic energy in tangential motion. This constitutes another surprising tension with $\Lambda$CDM simulations, in which a similar situation occurs in only 1.5\% of satellite systems. This finding is independent of, and complementary to, the issue of satellite galaxy planes because it considers only the radial and tangential velocity components but not their direction, whereas most studies of the satellite plane only consider the three  position coordinates and the orbital direction, but not the speed. However, it is interesting that the tangential velocity excess results in a similar degree of tension between the observed MW satellite system and $\Lambda$CDM expectations based on simulations. One might hope for a joint solution to both.

While the original study relied largely on HST proper motions and considered only the most-massive MW satellites, the measurement of proper motions for a larger number of satellites with Gaia made a re-assessment possible. Analysis of a total of 38 MW satellites largely confirmed the previously found tangential velocity excess, though inclusion of fainter satellites has resulted in a somewhat reduced fraction of $\sim 80$\% of all satellites having a tangentially biased motion\cite{2019MNRAS.486.2679R}. This study also reports a radial dependency, with the satellites close to the MW having tangentially biased and the more distant ones radially biased velocities. This is in line with the general trend in cosmological simulations that contain massive baryonic host galaxies in their centers, indicating that tidal disruption in the central regions might play a role by preferentially destroying satellites on radial orbits. The now available Gaia EDR3 proper motion measurements \cite{2020RNAAS...4..229M,2021ApJ...916....8L,2021arXiv210608819B}, as well as future data releases, will help clarify the situation. However, the issue might also be related to another pecularity in the orbital distribution of satellite galaxies: their preference to lie close to pericenter.

\subsection{Closeness to Pericenter}
\label{sect:pericenter}

The orbital velocity of an object such as a satellite galaxy orbiting a host is slowest at apocenter. Consequently, a satellite galaxy orbiting the MW on an eccentric orbit spends more of its time around its apo- than its pericenter. While any one satellite galaxy observed at a random time can be at any phase along its orbit, for an ensemble of independent satellites it is to be expected that they are preferentially found close to their apocenters.

Yet, when proper motions first became available for many of the fainter MW satellites via Gaia DR2, a pronounced preference of satellites to be found close to their orbital pericenter was reported \cite{2018A&A...619A.103F, 2018ApJ...863...89S}. This remains valid for a variety of different MW potential models, though more massive models ($1.6 \times 10^{12}\,M_\odot$\ in \cite{2018A&A...619A.103F}) somewhat reduce the exceess at pericenter. Nevertheless, despite pronounced orbital eccentricities of the observed satellites, no preference to lie at apocenter could be achived even for such a high-mass MW model.

\begin{figure}
\centering
\includegraphics[width=0.35\textwidth]{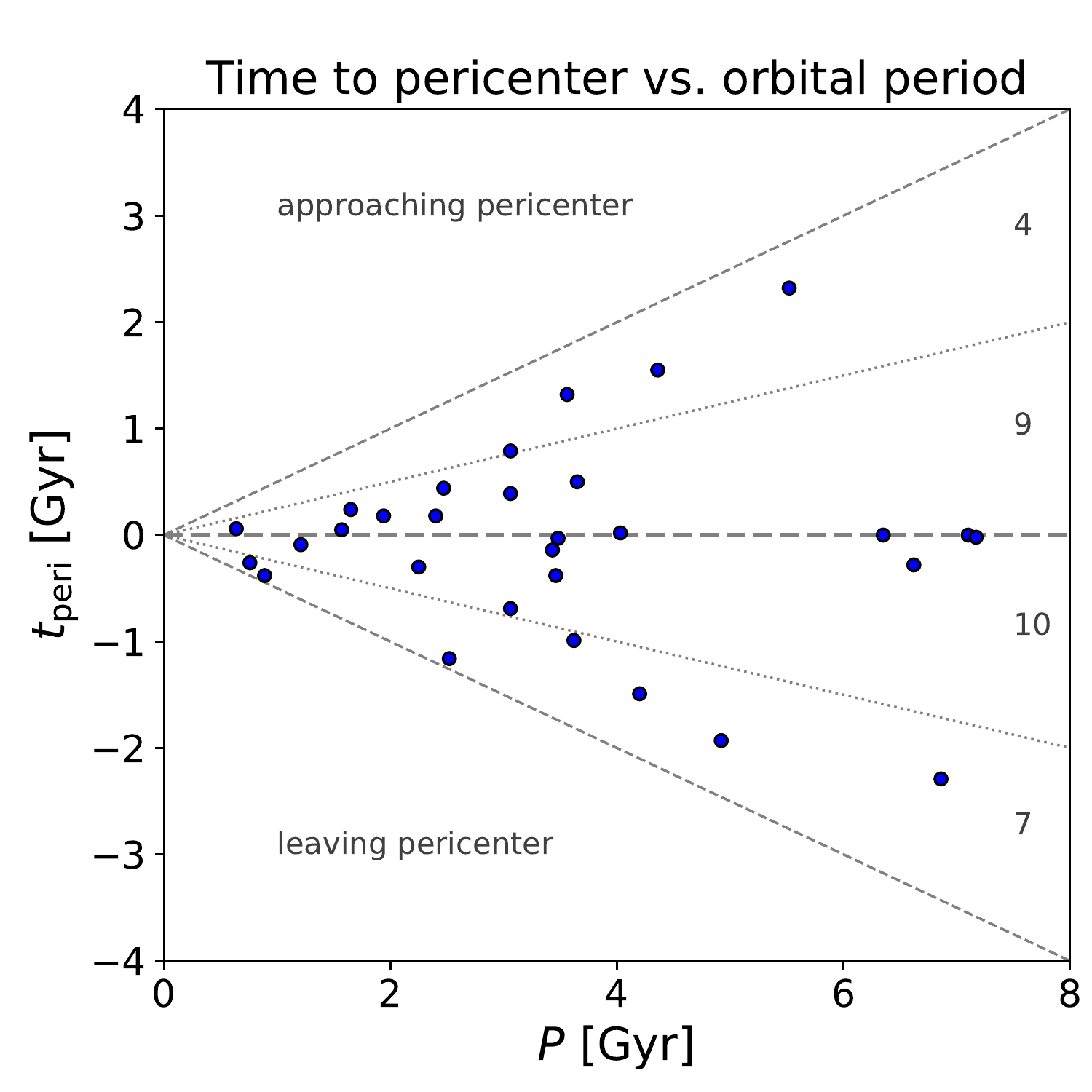}
\includegraphics[width=0.35\textwidth]{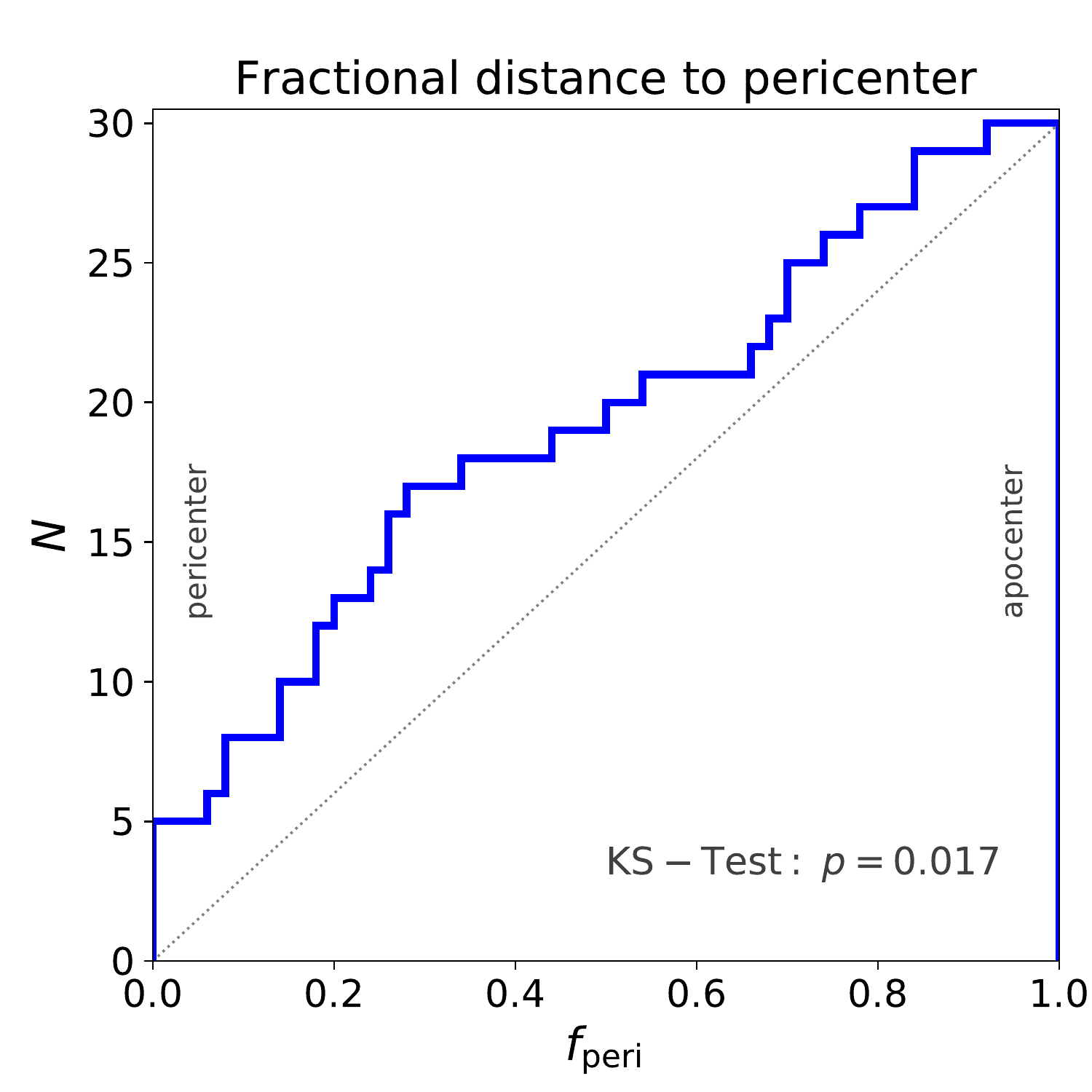}
\caption{
Phase along their orbits of the observed MW satellites with apocenters within 300\,kpc, adpoting the orbital parameters for the lower-mass MW model from \citet{2021arXiv210608819B}.
The left panel plots $P$,  the period of the satellite's orbit, against $t_\mathrm{peri}$, the time to the next pericenter (if positive) or the time since the last pericenter passage (if negative). The thin dashed lines indicate the possible limits (i.e. satellites at apocenter), the thick dashed line indicates where satellites are at their pericenter. The dotted lines split these regions in half, which for satellites in random phases should be populated equally. However, the observed satellites (blue points) lie preferentially closer to their peri- than their apocenters, as indicated by the counts on the right. 
The right panel shows the cumulative distribution of $f_\mathrm{peri}$,  the fractional position in time of the satellite between pericenter ($f_\mathrm{peri} = 0$) and apocenter ($f_\mathrm{peri} = 1.0$). For satellites at random phase, the latter should result in a uniform distribution (dashed line in right panel), yet the observed satellites appear to be preferentially found close to their pericenters. A Kolmogorov-Smirnov test provides a probability of only 1.7\% that the observed distribution was drawn from this expectation.
  \label{fig:pericenters}}
\end{figure}

One potential systematic that could have produced this finding would be a bias to over-estimated tangential velocities, which can easily happen if proper motions are uncertain and not well constrained. This would result in excessive inferred orbital energies, artificially increased apocenter distances, and would thus make one falsely believe that more satellites are close to their pericenter than actually realized. However, given that the improved accuracy of proper motion measurements with Gaia EDR3 has confirmed the preference of satellites to be found close to the pericenters of their orbits \cite{2021ApJ...916....8L}, this bias does not seem to be the dominant reason for the finding. Figure \ref{fig:pericenters} confirms this finding for another set of proper motions obtained from Gaia EDR3 \cite{2021arXiv210608819B}.

An alternative explanation would be that we have only discovered the most nearby satellite galaxies, which are naturally close to their pericenters. This predicts that a large number of satellite galaxies are still to be discvovered at larger distances ($> 100$\,kpc), which would then be closer to their orbital apocenters. This predictions will be tested with upcoming observational surveys such as LSST, though it appears at odds with some estimated satellite discovery limits \cite{2019ARA&A..57..375S, 2020ApJ...893...47D}, and an analysis of a volume-complete sample of satellites and their Gaia EDR3 proper motions that also reports a pronounced excess of satellites close to their pericenters \cite{2021ApJ...916....8L}.

The preferred closeness to pericenter could also possibly be related to a large number of considered satellites being in fact satellites brought in as a group with the LMC. They would then not constitute independent satellites, and a clustering close to a common orbital phase (specifically the pericenter, because the LMC is close to its pericenter) would be expected, in particular in a first-infall scenario. However, current estimates of the number of satellites that are associated with the LMC appear to be insufficient to fully account to the observed preference of satellites to be close to their pericenters \cite{2018A&A...619A.103F}.

Another possible explanation assumes that the observed satellite galaxies are in fact dark matter free dSphs, possibly formed as TDGs during a major merger in Andromeda \cite{2013MNRAS.431.3543H}. The elevated observed velocity dispersions for their surface brightness of dwarf galaxies surrounding the MW would in this scenario be caused by tidal shocks when they pass their pericenters. Since this will effectively destroy the dwarfs shortly after their pericenter passage, it could explain why an excess of objects is found at pericenter instead of at apocenter: many dwarfs in this scenario would not reach the latter as an identifiable object \cite{2020ApJ...892....3H}.

\section{Interdependencies Between Different Types of Satellite Galaxy Phase-Space Correlations}
\label{sect:interdependencies}

Studies thus far have mostly focussed on individual types of phase-space correlations discussed above. Yet, the different types show interesting interconnections which could help to ultimately understand their origins and possibly explain some of the pecularities of specific observed satellite galaxy systems.

One connection that has received some attention in the literature is the strong relation between the M31 satellite plane and the lopsidedness of the M31 satellite system: while the off-plane satellites do not show a pronounced lopsidedness, of the 15 on-plane satellites 13 are in the hemisphere pointing towards the MW (see yellow line in Fig. \ref{fig:MWM31lopsided}). The lopsided distribution of M31 satellites thus appears to be entirely driven by the on-plane satellite galaxies \cite{2013ApJ...766..120C}.

Another interesting but thus far not discussed connection is that between planes of satellite galaxies and pairs of satellites. Of the five possible satellite pairs in the Local Group listed in Table \ref{tab:pairs}, all are part of the two spatial satellite planes. This can be seen in Figures \ref{fig:MW} and \ref{fig:M31}, where the proposed pairs are highlighted in yellow. This appears to not only be restricted to a spatial alignment, but also is present in their orbital motions: for three of these five pairs, proper motions indicate that both pair members co-orbit along the respective satellite plane: the LMC and SMC as well as Draco and Ursa Minor around the MW, and NGC\,147 and NGC\,185 around M31. The proper motions of the other two pairs are either not sufficiently constrained (Leo\,IV and Leo\,V are consistent with sharing similar orbits and with orbiting along the VPOS), or have not been measured yet (Andromeda\,I and III). It is intriguing that these pairs, which with exception of the LMC-SMC pair are neither interacting nor appear to presently be gravitationally bound, nevertheless move in similar directions along much more extended satellite structures. This speaks against a chance alignmend, and suggests some sort of causal connection. Yet, identifying the nature of this connection is complicated, because numerous possible explanations can be envisioned.

One such potential explanation is that the presence of flattened distribution of satellite galaxies simply boosts the likelyhood of chance alignments. By being preferentially confined to a common, flattened structure instead of a more spherical, isotropic distribution, the chance to find close satellites and identify them as apparent, spatially close pairs might be boosted. If satellites preferentially co-orbit along a common, flattened structure, such apparent pairs might also be expected to share similar orbital directions without having a direct connection.

Pairs of satellite galaxies can affect many other phase-space correlations, and appear directly related to group infall. After all, a pair, if indeed sharing a common origin and thus orbit, would just be a special case of a group of satellites. Identifying a pair might even give hints to identify additional satellites that once were associated with those two in a more numerous group. This connection between pairs and groups is exemplified by the potential Leo\,IV and Leo\,V pair, both of which are also proposed members of a Crater-Leo group of satellites. Interestingly, while group infall is expected in $\Lambda$CDM simulations \cite{ 2008MNRAS.385.1365L, 2013MNRAS.429.1502W, 2015ApJ...807...49W, 2018MNRAS.476.1796S}, the apparent excess of close satellite pairs observed in the Local Group \cite{2013MNRAS.431L..73F} indicates that such groups in simulations are not compact or not frequent enough to reproduce the observed close clustering of satellite galaxies.

A different possible explanation would turn things around and postulate that satellites preferentially arrange in pairs and groups, likely before even being accreted onto a host halo. This is the underlying scenario for attempts to explain the presence of planes of satellite galaxies via the accretion of satellites in groups \cite{2008ApJ...686L..61D, 2021MNRAS.504.1379S}. The accretion of a single group of satellites constituting a substantial fraction of the overall satellite population of their host would initially introduce a degree of lopsidedness if the infalling group is sufficiently compact, too \citep{2020MNRAS.492..456W}. Only once the group has dispersed sufficiently along its orbit would such an asymmetry vanish. This is a possible explanation for the asymmetry in the distribution of MW satellites as displayed in Fig. \ref{fig:MWM31lopsided}, though uneven sky coverage and differences in survey depths also contribute to this and preclude a robust interpretation at this point. This idea could possibly be tested observationally, by studying the grouping of observed dwarf galaxies in the field (i.e. away from hosts) against predictions of the incidence and properties of such groups in cosmological simulations.

The tangential velocity excess might be directly related to the closeness of satellites to their pericenters, which after all is the one of two points along their orbit where the radial velocity component approaches zero. A possible connection between the accretion of groups of satellite and the excess of satellites close to their pericenter was also suggested in the context of the LMC bringing in an entourage of its own satellites. With the LMC close to its pericenter, its satellites would also be found in a similar orbital phase, yet this appears to be insufficient to account for the full pericenter excess \cite{2018A&A...619A.103F}.

As discusses in Sect. \ref{sect:pericenter}, the pericenter excess was also suggested to point at an alternative nature of the observed dwarf satellite galaxies as dark-matter-free dSphs that possibly originated as gas-rich tidal dwarf galaxies (TDGs). Interestingly, before the pericenter excess was even known TDGs had already been proposed as a possible explanation of the observed planes of satellite galaxies. Galaxies formed out of tidal debris are expected to preferentially co-orbit in a common plane set by the interplay of the orbital angular momentum and internal spins of their colliding parent galaxies. However, adopting such a scenario has substantial further implications, as it would either worsen the missing-satellites problem if it is to be realized in a $\Lambda$CDM framework (because many observed satellites would then not be dark-matter dominated, primordial dwarfs), or by requiring to adopt a modified gravity paradigm such as MOND (to ensure that also TDGs show enhanced gravitational acceleration). Either way, this possibility alone already illustrates that understanding the mutual connections between different types of phase-space correlations can have far-reaching implications for our interpretation of the origin, evolution, and nature of dwarf and satellite galaxies, the history of the Local Group, the nature of Dark Matter, and thus (near-field) cosmology.

\section{Conclusions}
\label{sect:conclusion}

Several different types of phase-space correlations among systems of satellite galaxies have been studied. Some of these, such as the planes of satellite galaxies, are recognized as either a problem, or at least a challenge, for the $\Lambda$CDM model of cosmology. Other such challenges include the apparent overabundance of close pairs of satellites, which interestingly are found to be members of the planes of satellites, the excess of tangential velocities for the observed MW satellites compared to satellites in cosmological simulations, and potentially the closeness of MW satellites to their pericenters. Other phase-space correlations, lopsided distributions of satellites around their hosts and possibly the accretion of satellite galaxies in groups, appear to be in better agreement with cosmological expectations. However, the latter are both also connected to the other types of phase-space correlations, which makes a judgement on the source and status of the mismatches as well as the agreements between the observed satellite galaxy systems and $\Lambda$CDM predictions from cosmological simulations difficult. 

Yet, this can be seen as a sign that the study of phase-space correlations among satellite galaxies is a still relatively young field of research within the wider context of near-field cosmology. Consequently, numerous, in part fundamental, questions regarding the origins and interconnections between different types of phase-space correlations of satellite galaxies remain open at this point. What is the origin of planes of satellite galaxies? Are they a feature unique to the most nearby systems, or are similar structures common throughout the Universe?  Do planes of satellites artificially boost the incidence of close pairs? Or does a clustering of satellites in pairs or compact groups, possibly to a degree exceeding that expected from current cosmological simulations, result in the identification of apparent satellite alignments and correlated orbital directions? Are lopsided satellite systems indicative of such group accretion events, or could they be a natural consequence of preferentially co-orbiting satellite systems? What is the impact of planes of satellite galaxies on estimations of the total number of dwarf galaxy satellites around the MW, which is often extrapolated from the observed systems under the assumption of an underlying isotropic distribution? How do phase-space correlated dwarf galaxies compare in their internal properties and star formation histories, and could this help shed light on similarities in their history and evolution, or reveal hints that they shared a similar environment before being accreted? 

Addressing and answering these and other related questions in the coming years will require a multi-pronged approach. It will involve additional studies using cosmological simulations and in particular the dynamical evolution of satellite system in them, dedicated orbit modelling informed by increasingly accurate and complete data for the most nearby satellite galaxy systems, as well as an expansion to additional, more distant systems of satellite galaxies around hosts beyond the Local Group. Current progress in numerical modelling, a wide interest in the topic triggered by the wealth of information provided by Gaia's accurate 6D phase-space information for the MW satellite galaxies, its promise to deliver even better measurements towards the mission's end, and substantial observational progress in the investigation of more distant satellite galaxy systems which will further benefit from numerous upcoming observational facilities, all promise rich potential for future research.

\vspace{6pt} 



\funding{
This work was supported by a Leibniz-Junior Research Group grant (project number J94/2020) via the Leibniz Competition, and a Klaus Tschira Boost Fund provided by the Klaus Tschira Stiftung and the German Scholars Organization.
}

\acknowledgments{
I thank Salvatore Taibi, Yanbin Yang, and Oliver M\"uller for helpful discussions and comments.
This research has made use of NASA's Astrophysics Data System
and adstex (\url{https://github.com/yymao/adstex}).
Software packages used in the preparation of this work include 
astropy \citep{2018AJ....156..123A, 2013A&A...558A..33A},  
galpy \citep{2015ApJS..216...29B}, 
IPython \citep{PER-GRA:2007},
matplotlib \citep{Hunter:2007}, 
NumPy \citep{harris2020array}, and
SciPy \citep{Virtanen_2020}.
}

\conflictsofinterest{The author declares no conflict of interest.}

%


\end{paracol}
\reftitle{References}

\externalbibliography{yes}
\bibliography{phasespacereviewrefs}

\begin{thebibliography}{999}

\bibitem[{Bullock} and {Boylan-Kolchin}(2017)]{2017ARA&A..55..343B}
{Bullock}, J.S.; {Boylan-Kolchin}, M.
\newblock {Small-Scale Challenges to the {\ensuremath{\Lambda}}CDM Paradigm}.
\newblock {\em \araa} {\bf 2017}, {\em 55},~343--387,
  \href{http://xxx.lanl.gov/abs/1707.04256}{{\normalfont
  [arXiv:astro-ph.CO/1707.04256]}}.
\newblock
  doi:{\changeurlcolor{black}\href{https://doi.org/10.1146/annurev-astro-091916-055313}{\detokenize{10.1146/annurev-astro-091916-055313}}}.

\bibitem[{Klypin} \em{et~al.}(1999){Klypin}, {Kravtsov}, {Valenzuela}, and
  {Prada}]{1999ApJ...522...82K}
{Klypin}, A.; {Kravtsov}, A.V.; {Valenzuela}, O.; {Prada}, F.
\newblock {Where Are the Missing Galactic Satellites?}
\newblock {\em \apj} {\bf 1999}, {\em 522},~82--92,
  \href{http://xxx.lanl.gov/abs/astro-ph/9901240}{{\normalfont
  [arXiv:astro-ph/astro-ph/9901240]}}.
\newblock
  doi:{\changeurlcolor{black}\href{https://doi.org/10.1086/307643}{\detokenize{10.1086/307643}}}.

\bibitem[{Moore} \em{et~al.}(1999){Moore}, {Ghigna}, {Governato}, {Lake},
  {Quinn}, {Stadel}, and {Tozzi}]{1999ApJ...524L..19M}
{Moore}, B.; {Ghigna}, S.; {Governato}, F.; {Lake}, G.; {Quinn}, T.; {Stadel},
  J.; {Tozzi}, P.
\newblock {Dark Matter Substructure within Galactic Halos}.
\newblock {\em \apjl} {\bf 1999}, {\em 524},~L19--L22,
  \href{http://xxx.lanl.gov/abs/astro-ph/9907411}{{\normalfont
  [arXiv:astro-ph/astro-ph/9907411]}}.
\newblock
  doi:{\changeurlcolor{black}\href{https://doi.org/10.1086/312287}{\detokenize{10.1086/312287}}}.

\bibitem[{Dubinski} and {Carlberg}(1991)]{1991ApJ...378..496D}
{Dubinski}, J.; {Carlberg}, R.G.
\newblock {The Structure of Cold Dark Matter Halos}.
\newblock {\em \apj} {\bf 1991}, {\em 378},~496.
\newblock
  doi:{\changeurlcolor{black}\href{https://doi.org/10.1086/170451}{\detokenize{10.1086/170451}}}.

\bibitem[{Walker} and {Pe{\~n}arrubia}(2011)]{2011ApJ...742...20W}
{Walker}, M.G.; {Pe{\~n}arrubia}, J.
\newblock {A Method for Measuring (Slopes of) the Mass Profiles of Dwarf
  Spheroidal Galaxies}.
\newblock {\em \apj} {\bf 2011}, {\em 742},~20,
  \href{http://xxx.lanl.gov/abs/1108.2404}{{\normalfont
  [arXiv:astro-ph.CO/1108.2404]}}.
\newblock
  doi:{\changeurlcolor{black}\href{https://doi.org/10.1088/0004-637X/742/1/20}{\detokenize{10.1088/0004-637X/742/1/20}}}.

\bibitem[{Boylan-Kolchin} \em{et~al.}(2011){Boylan-Kolchin}, {Bullock}, and
  {Kaplinghat}]{2011MNRAS.415L..40B}
{Boylan-Kolchin}, M.; {Bullock}, J.S.; {Kaplinghat}, M.
\newblock {Too big to fail? The puzzling darkness of massive Milky Way
  subhaloes}.
\newblock {\em \mnras} {\bf 2011}, {\em 415},~L40--L44,
  \href{http://xxx.lanl.gov/abs/1103.0007}{{\normalfont
  [arXiv:astro-ph.CO/1103.0007]}}.
\newblock
  doi:{\changeurlcolor{black}\href{https://doi.org/10.1111/j.1745-3933.2011.01074.x}{\detokenize{10.1111/j.1745-3933.2011.01074.x}}}.

\bibitem[{Kroupa}(2012)]{2012PASA...29..395K}
{Kroupa}, P.
\newblock {The Dark Matter Crisis: Falsification of the Current Standard Model
  of Cosmology}.
\newblock {\em \pasa} {\bf 2012}, {\em 29},~395--433,
  \href{http://xxx.lanl.gov/abs/1204.2546}{{\normalfont
  [arXiv:astro-ph.CO/1204.2546]}}.
\newblock
  doi:{\changeurlcolor{black}\href{https://doi.org/10.1071/AS12005}{\detokenize{10.1071/AS12005}}}.

\bibitem[{Genina} \em{et~al.}(2018){Genina}, {Ben{\'\i}tez-Llambay}, {Frenk},
  {Cole}, {Fattahi}, {Navarro}, {Oman}, {Sawala}, and
  {Theuns}]{2018MNRAS.474.1398G}
{Genina}, A.; {Ben{\'\i}tez-Llambay}, A.; {Frenk}, C.S.; {Cole}, S.; {Fattahi},
  A.; {Navarro}, J.F.; {Oman}, K.A.; {Sawala}, T.; {Theuns}, T.
\newblock {The core-cusp problem: a matter of perspective}.
\newblock {\em \mnras} {\bf 2018}, {\em 474},~1398--1411,
  \href{http://xxx.lanl.gov/abs/1707.06303}{{\normalfont
  [arXiv:astro-ph.GA/1707.06303]}}.
\newblock
  doi:{\changeurlcolor{black}\href{https://doi.org/10.1093/mnras/stx2855}{\detokenize{10.1093/mnras/stx2855}}}.

\bibitem[{Sawala} \em{et~al.}(2016){Sawala}, {Frenk}, {Fattahi}, {Navarro},
  {Bower}, {Crain}, {Dalla Vecchia}, {Furlong}, {Helly}, {Jenkins}, {Oman},
  {Schaller}, {Schaye}, {Theuns}, {Trayford}, and {White}]{2016MNRAS.457.1931S}
{Sawala}, T.; {Frenk}, C.S.; {Fattahi}, A.; {Navarro}, J.F.; {Bower}, R.G.;
  {Crain}, R.A.; {Dalla Vecchia}, C.; {Furlong}, M.; {Helly}, J.C.; {Jenkins},
  A.; {Oman}, K.A.; {Schaller}, M.; {Schaye}, J.; {Theuns}, T.; {Trayford}, J.;
  {White}, S.D.M.
\newblock {The APOSTLE simulations: solutions to the Local Group's cosmic
  puzzles}.
\newblock {\em \mnras} {\bf 2016}, {\em 457},~1931--1943,
  \href{http://xxx.lanl.gov/abs/1511.01098}{{\normalfont
  [arXiv:astro-ph.GA/1511.01098]}}.
\newblock
  doi:{\changeurlcolor{black}\href{https://doi.org/10.1093/mnras/stw145}{\detokenize{10.1093/mnras/stw145}}}.

\bibitem[{Read} \em{et~al.}(2016){Read}, {Agertz}, and
  {Collins}]{2016MNRAS.459.2573R}
{Read}, J.I.; {Agertz}, O.; {Collins}, M.L.M.
\newblock {Dark matter cores all the way down}.
\newblock {\em \mnras} {\bf 2016}, {\em 459},~2573--2590,
  \href{http://xxx.lanl.gov/abs/1508.04143}{{\normalfont
  [arXiv:astro-ph.GA/1508.04143]}}.
\newblock
  doi:{\changeurlcolor{black}\href{https://doi.org/10.1093/mnras/stw713}{\detokenize{10.1093/mnras/stw713}}}.

\bibitem[{Brooks} and {Zolotov}(2014)]{2014ApJ...786...87B}
{Brooks}, A.M.; {Zolotov}, A.
\newblock {Why Baryons Matter: The Kinematics of Dwarf Spheroidal Satellites}.
\newblock {\em \apj} {\bf 2014}, {\em 786},~87,
  \href{http://xxx.lanl.gov/abs/1207.2468}{{\normalfont
  [arXiv:astro-ph.CO/1207.2468]}}.
\newblock
  doi:{\changeurlcolor{black}\href{https://doi.org/10.1088/0004-637X/786/2/87}{\detokenize{10.1088/0004-637X/786/2/87}}}.

\bibitem[{Papastergis} \em{et~al.}(2015){Papastergis}, {Giovanelli}, {Haynes},
  and {Shankar}]{2015A&A...574A.113P}
{Papastergis}, E.; {Giovanelli}, R.; {Haynes}, M.P.; {Shankar}, F.
\newblock {Is there a ``too big to fail'' problem in the field?}
\newblock {\em \aap} {\bf 2015}, {\em 574},~A113,
  \href{http://xxx.lanl.gov/abs/1407.4665}{{\normalfont
  [arXiv:astro-ph.GA/1407.4665]}}.
\newblock
  doi:{\changeurlcolor{black}\href{https://doi.org/10.1051/0004-6361/201424909}{\detokenize{10.1051/0004-6361/201424909}}}.

\bibitem[{Brooks} \em{et~al.}(2017){Brooks}, {Papastergis}, {Christensen},
  {Governato}, {Stilp}, {Quinn}, and {Wadsley}]{2017ApJ...850...97B}
{Brooks}, A.M.; {Papastergis}, E.; {Christensen}, C.R.; {Governato}, F.;
  {Stilp}, A.; {Quinn}, T.R.; {Wadsley}, J.
\newblock {How to Reconcile the Observed Velocity Function of Galaxies with
  Theory}.
\newblock {\em \apj} {\bf 2017}, {\em 850},~97,
  \href{http://xxx.lanl.gov/abs/1701.07835}{{\normalfont
  [arXiv:astro-ph.GA/1701.07835]}}.
\newblock
  doi:{\changeurlcolor{black}\href{https://doi.org/10.3847/1538-4357/aa9576}{\detokenize{10.3847/1538-4357/aa9576}}}.

\bibitem[{Dutton} \em{et~al.}(2019){Dutton}, {Obreja}, and
  {Macci{\`o}}]{2019MNRAS.482.5606D}
{Dutton}, A.A.; {Obreja}, A.; {Macci{\`o}}, A.V.
\newblock {NIHAO - XVII. The diversity of dwarf galaxy kinematics and
  implications for the H I velocity function}.
\newblock {\em \mnras} {\bf 2019}, {\em 482},~5606--5624,
  \href{http://xxx.lanl.gov/abs/1807.10518}{{\normalfont
  [arXiv:astro-ph.GA/1807.10518]}}.
\newblock
  doi:{\changeurlcolor{black}\href{https://doi.org/10.1093/mnras/sty3064}{\detokenize{10.1093/mnras/sty3064}}}.

\bibitem[{Viel} \em{et~al.}(2005){Viel}, {Lesgourgues}, {Haehnelt},
  {Matarrese}, and {Riotto}]{2005PhRvD..71f3534V}
{Viel}, M.; {Lesgourgues}, J.; {Haehnelt}, M.G.; {Matarrese}, S.; {Riotto}, A.
\newblock {Constraining warm dark matter candidates including sterile neutrinos
  and light gravitinos with WMAP and the Lyman-{\ensuremath{\alpha}} forest}.
\newblock {\em \prd} {\bf 2005}, {\em 71},~063534,
  \href{http://xxx.lanl.gov/abs/astro-ph/0501562}{{\normalfont
  [arXiv:astro-ph/astro-ph/0501562]}}.
\newblock
  doi:{\changeurlcolor{black}\href{https://doi.org/10.1103/PhysRevD.71.063534}{\detokenize{10.1103/PhysRevD.71.063534}}}.

\bibitem[{Anderhalden} \em{et~al.}(2013){Anderhalden}, {Schneider},
  {Macci{\`o}}, {Diemand}, and {Bertone}]{2013JCAP...03..014A}
{Anderhalden}, D.; {Schneider}, A.; {Macci{\`o}}, A.V.; {Diemand}, J.;
  {Bertone}, G.
\newblock {Hints on the nature of dark matter from the properties of Milky Way
  satellites}.
\newblock {\em \jcap} {\bf 2013}, {\em 2013},~014,
  \href{http://xxx.lanl.gov/abs/1212.2967}{{\normalfont
  [arXiv:astro-ph.CO/1212.2967]}}.
\newblock
  doi:{\changeurlcolor{black}\href{https://doi.org/10.1088/1475-7516/2013/03/014}{\detokenize{10.1088/1475-7516/2013/03/014}}}.

\bibitem[{Spergel} and {Steinhardt}(2000)]{2000PhRvL..84.3760S}
{Spergel}, D.N.; {Steinhardt}, P.J.
\newblock {Observational Evidence for Self-Interacting Cold Dark Matter}.
\newblock {\em \prl} {\bf 2000}, {\em 84},~3760--3763,
  \href{http://xxx.lanl.gov/abs/astro-ph/9909386}{{\normalfont
  [arXiv:astro-ph/astro-ph/9909386]}}.
\newblock
  doi:{\changeurlcolor{black}\href{https://doi.org/10.1103/PhysRevLett.84.3760}{\detokenize{10.1103/PhysRevLett.84.3760}}}.

\bibitem[{Collins} \em{et~al.}(2015){Collins}, {Martin}, {Rich}, {Ibata},
  {Chapman}, {McConnachie}, {Ferguson}, {Irwin}, and
  {Lewis}]{2015ApJ...799L..13C}
{Collins}, M.L.M.; {Martin}, N.F.; {Rich}, R.M.; {Ibata}, R.A.; {Chapman},
  S.C.; {McConnachie}, A.W.; {Ferguson}, A.M.; {Irwin}, M.J.; {Lewis}, G.F.
\newblock {Comparing the Observable Properties of Dwarf Galaxies on and off the
  Andromeda Plane}.
\newblock {\em \apjl} {\bf 2015}, {\em 799},~L13,
  \href{http://xxx.lanl.gov/abs/1411.3324}{{\normalfont
  [arXiv:astro-ph.GA/1411.3324]}}.
\newblock
  doi:{\changeurlcolor{black}\href{https://doi.org/10.1088/2041-8205/799/1/L13}{\detokenize{10.1088/2041-8205/799/1/L13}}}.

\bibitem[{York} \em{et~al.}(2000){York}, {Adelman}, {Anderson}, {Anderson},
  {Annis}, {Bahcall}, {Bakken}, {Barkhouser}, {Bastian}, {Berman}, {Boroski},
  {Bracker}, {Briegel}, {Briggs}, {Brinkmann}, {Brunner}, {Burles}, {Carey},
  {Carr}, {Castander}, {Chen}, {Colestock}, {Connolly}, {Crocker}, {Csabai},
  {Czarapata}, {Davis}, {Doi}, {Dombeck}, {Eisenstein}, {Ellman}, {Elms},
  {Evans}, {Fan}, {Federwitz}, {Fiscelli}, {Friedman}, {Frieman}, {Fukugita},
  {Gillespie}, {Gunn}, {Gurbani}, {de Haas}, {Haldeman}, {Harris}, {Hayes},
  {Heckman}, {Hennessy}, {Hindsley}, {Holm}, {Holmgren}, {Huang}, {Hull},
  {Husby}, {Ichikawa}, {Ichikawa}, {Ivezi{\'c}}, {Kent}, {Kim}, {Kinney},
  {Klaene}, {Kleinman}, {Kleinman}, {Knapp}, {Korienek}, {Kron}, {Kunszt},
  {Lamb}, {Lee}, {Leger}, {Limmongkol}, {Lindenmeyer}, {Long}, {Loomis},
  {Loveday}, {Lucinio}, {Lupton}, {MacKinnon}, {Mannery}, {Mantsch}, {Margon},
  {McGehee}, {McKay}, {Meiksin}, {Merelli}, {Monet}, {Munn}, {Narayanan},
  {Nash}, {Neilsen}, {Neswold}, {Newberg}, {Nichol}, {Nicinski}, {Nonino},
  {Okada}, {Okamura}, {Ostriker}, {Owen}, {Pauls}, {Peoples}, {Peterson},
  {Petravick}, {Pier}, {Pope}, {Pordes}, {Prosapio}, {Rechenmacher}, {Quinn},
  {Richards}, {Richmond}, {Rivetta}, {Rockosi}, {Ruthmansdorfer}, {Sandford},
  {Schlegel}, {Schneider}, {Sekiguchi}, {Sergey}, {Shimasaku}, {Siegmund},
  {Smee}, {Smith}, {Snedden}, {Stone}, {Stoughton}, {Strauss}, {Stubbs},
  {SubbaRao}, {Szalay}, {Szapudi}, {Szokoly}, {Thakar}, {Tremonti}, {Tucker},
  {Uomoto}, {Vanden Berk}, {Vogeley}, {Waddell}, {Wang}, {Watanabe},
  {Weinberg}, {Yanny}, {Yasuda}, and {SDSS Collaboration}]{2000AJ....120.1579Y}
{York}, D.G.; {Adelman}, J.; {Anderson}, John~E., J.; {Anderson}, S.F.;
  {Annis}, J.; {Bahcall}, N.A.; {Bakken}, J.A.; {Barkhouser}, R.; {Bastian},
  S.; {Berman}, E.; {Boroski}, W.N.; {Bracker}, S.; {Briegel}, C.; {Briggs},
  J.W.; {Brinkmann}, J.; {Brunner}, R.; {Burles}, S.; {Carey}, L.; {Carr},
  M.A.; {Castander}, F.J.; {Chen}, B.; {Colestock}, P.L.; {Connolly}, A.J.;
  {Crocker}, J.H.; {Csabai}, I.; {Czarapata}, P.C.; {Davis}, J.E.; {Doi}, M.;
  {Dombeck}, T.; {Eisenstein}, D.; {Ellman}, N.; {Elms}, B.R.; {Evans}, M.L.;
  {Fan}, X.; {Federwitz}, G.R.; {Fiscelli}, L.; {Friedman}, S.; {Frieman},
  J.A.; {Fukugita}, M.; {Gillespie}, B.; {Gunn}, J.E.; {Gurbani}, V.K.; {de
  Haas}, E.; {Haldeman}, M.; {Harris}, F.H.; {Hayes}, J.; {Heckman}, T.M.;
  {Hennessy}, G.S.; {Hindsley}, R.B.; {Holm}, S.; {Holmgren}, D.J.; {Huang},
  C.h.; {Hull}, C.; {Husby}, D.; {Ichikawa}, S.I.; {Ichikawa}, T.;
  {Ivezi{\'c}}, {\v{Z}}.; {Kent}, S.; {Kim}, R.S.J.; {Kinney}, E.; {Klaene},
  M.; {Kleinman}, A.N.; {Kleinman}, S.; {Knapp}, G.R.; {Korienek}, J.; {Kron},
  R.G.; {Kunszt}, P.Z.; {Lamb}, D.Q.; {Lee}, B.; {Leger}, R.F.; {Limmongkol},
  S.; {Lindenmeyer}, C.; {Long}, D.C.; {Loomis}, C.; {Loveday}, J.; {Lucinio},
  R.; {Lupton}, R.H.; {MacKinnon}, B.; {Mannery}, E.J.; {Mantsch}, P.M.;
  {Margon}, B.; {McGehee}, P.; {McKay}, T.A.; {Meiksin}, A.; {Merelli}, A.;
  {Monet}, D.G.; {Munn}, J.A.; {Narayanan}, V.K.; {Nash}, T.; {Neilsen}, E.;
  {Neswold}, R.; {Newberg}, H.J.; {Nichol}, R.C.; {Nicinski}, T.; {Nonino}, M.;
  {Okada}, N.; {Okamura}, S.; {Ostriker}, J.P.; {Owen}, R.; {Pauls}, A.G.;
  {Peoples}, J.; {Peterson}, R.L.; {Petravick}, D.; {Pier}, J.R.; {Pope}, A.;
  {Pordes}, R.; {Prosapio}, A.; {Rechenmacher}, R.; {Quinn}, T.R.; {Richards},
  G.T.; {Richmond}, M.W.; {Rivetta}, C.H.; {Rockosi}, C.M.; {Ruthmansdorfer},
  K.; {Sandford}, D.; {Schlegel}, D.J.; {Schneider}, D.P.; {Sekiguchi}, M.;
  {Sergey}, G.; {Shimasaku}, K.; {Siegmund}, W.A.; {Smee}, S.; {Smith}, J.A.;
  {Snedden}, S.; {Stone}, R.; {Stoughton}, C.; {Strauss}, M.A.; {Stubbs}, C.;
  {SubbaRao}, M.; {Szalay}, A.S.; {Szapudi}, I.; {Szokoly}, G.P.; {Thakar},
  A.R.; {Tremonti}, C.; {Tucker}, D.L.; {Uomoto}, A.; {Vanden Berk}, D.;
  {Vogeley}, M.S.; {Waddell}, P.; {Wang}, S.i.; {Watanabe}, M.; {Weinberg},
  D.H.; {Yanny}, B.; {Yasuda}, N.; {SDSS Collaboration}.
\newblock {The Sloan Digital Sky Survey: Technical Summary}.
\newblock {\em \aj} {\bf 2000}, {\em 120},~1579--1587,
  \href{http://xxx.lanl.gov/abs/astro-ph/0006396}{{\normalfont
  [arXiv:astro-ph/astro-ph/0006396]}}.
\newblock
  doi:{\changeurlcolor{black}\href{https://doi.org/10.1086/301513}{\detokenize{10.1086/301513}}}.

\bibitem[{Pawlowski}(2016)]{2016MNRAS.456..448P}
{Pawlowski}, M.S.
\newblock {The alignment of SDSS satellites with the VPOS: effects of the
  survey footprint shape}.
\newblock {\em \mnras} {\bf 2016}, {\em 456},~448--458,
  \href{http://xxx.lanl.gov/abs/1511.05557}{{\normalfont
  [arXiv:astro-ph.GA/1511.05557]}}.
\newblock
  doi:{\changeurlcolor{black}\href{https://doi.org/10.1093/mnras/stv2673}{\detokenize{10.1093/mnras/stv2673}}}.

\bibitem[{Bechtol} \em{et~al.}(2015){Bechtol}, {Drlica-Wagner}, {Balbinot},
  {Pieres}, {Simon}, {Yanny}, {Santiago}, {Wechsler}, {Frieman}, {Walker},
  {Williams}, {Rozo}, {Rykoff}, {Queiroz}, {Luque}, {Benoit-L{\'e}vy},
  {Tucker}, {Sevilla}, {Gruendl}, {da Costa}, {Fausti Neto}, {Maia}, {Abbott},
  {Allam}, {Armstrong}, {Bauer}, {Bernstein}, {Bernstein}, {Bertin}, {Brooks},
  {Buckley-Geer}, {Burke}, {Carnero Rosell}, {Castander}, {Covarrubias},
  {D'Andrea}, {DePoy}, {Desai}, {Diehl}, {Eifler}, {Estrada}, {Evrard},
  {Fernandez}, {Finley}, {Flaugher}, {Gaztanaga}, {Gerdes}, {Girardi},
  {Gladders}, {Gruen}, {Gutierrez}, {Hao}, {Honscheid}, {Jain}, {James},
  {Kent}, {Kron}, {Kuehn}, {Kuropatkin}, {Lahav}, {Li}, {Lin}, {Makler},
  {March}, {Marshall}, {Martini}, {Merritt}, {Miller}, {Miquel}, {Mohr},
  {Neilsen}, {Nichol}, {Nord}, {Ogando}, {Peoples}, {Petravick}, {Plazas},
  {Romer}, {Roodman}, {Sako}, {Sanchez}, {Scarpine}, {Schubnell}, {Smith},
  {Soares-Santos}, {Sobreira}, {Suchyta}, {Swanson}, {Tarle}, {Thaler},
  {Thomas}, {Wester}, {Zuntz}, and {DES Collaboration}]{2015ApJ...807...50B}
{Bechtol}, K.; {Drlica-Wagner}, A.; {Balbinot}, E.; {Pieres}, A.; {Simon},
  J.D.; {Yanny}, B.; {Santiago}, B.; {Wechsler}, R.H.; {Frieman}, J.; {Walker},
  A.R.; {Williams}, P.; {Rozo}, E.; {Rykoff}, E.S.; {Queiroz}, A.; {Luque}, E.;
  {Benoit-L{\'e}vy}, A.; {Tucker}, D.; {Sevilla}, I.; {Gruendl}, R.A.; {da
  Costa}, L.N.; {Fausti Neto}, A.; {Maia}, M.A.G.; {Abbott}, T.; {Allam}, S.;
  {Armstrong}, R.; {Bauer}, A.H.; {Bernstein}, G.M.; {Bernstein}, R.A.;
  {Bertin}, E.; {Brooks}, D.; {Buckley-Geer}, E.; {Burke}, D.L.; {Carnero
  Rosell}, A.; {Castander}, F.J.; {Covarrubias}, R.; {D'Andrea}, C.B.; {DePoy},
  D.L.; {Desai}, S.; {Diehl}, H.T.; {Eifler}, T.F.; {Estrada}, J.; {Evrard},
  A.E.; {Fernandez}, E.; {Finley}, D.A.; {Flaugher}, B.; {Gaztanaga}, E.;
  {Gerdes}, D.; {Girardi}, L.; {Gladders}, M.; {Gruen}, D.; {Gutierrez}, G.;
  {Hao}, J.; {Honscheid}, K.; {Jain}, B.; {James}, D.; {Kent}, S.; {Kron}, R.;
  {Kuehn}, K.; {Kuropatkin}, N.; {Lahav}, O.; {Li}, T.S.; {Lin}, H.; {Makler},
  M.; {March}, M.; {Marshall}, J.; {Martini}, P.; {Merritt}, K.W.; {Miller},
  C.; {Miquel}, R.; {Mohr}, J.; {Neilsen}, E.; {Nichol}, R.; {Nord}, B.;
  {Ogando}, R.; {Peoples}, J.; {Petravick}, D.; {Plazas}, A.A.; {Romer}, A.K.;
  {Roodman}, A.; {Sako}, M.; {Sanchez}, E.; {Scarpine}, V.; {Schubnell}, M.;
  {Smith}, R.C.; {Soares-Santos}, M.; {Sobreira}, F.; {Suchyta}, E.; {Swanson},
  M.E.C.; {Tarle}, G.; {Thaler}, J.; {Thomas}, D.; {Wester}, W.; {Zuntz}, J.;
  {DES Collaboration}.
\newblock {Eight New Milky Way Companions Discovered in First-year Dark Energy
  Survey Data}.
\newblock {\em \apj} {\bf 2015}, {\em 807},~50,
  \href{http://xxx.lanl.gov/abs/1503.02584}{{\normalfont
  [arXiv:astro-ph.GA/1503.02584]}}.
\newblock
  doi:{\changeurlcolor{black}\href{https://doi.org/10.1088/0004-637X/807/1/50}{\detokenize{10.1088/0004-637X/807/1/50}}}.

\bibitem[{Pawlowski} \em{et~al.}(2015){Pawlowski}, {McGaugh}, and
  {Jerjen}]{2015MNRAS.453.1047P}
{Pawlowski}, M.S.; {McGaugh}, S.S.; {Jerjen}, H.
\newblock {The new Milky Way satellites: alignment with the VPOS and
  predictions for proper motions and velocity dispersions}.
\newblock {\em \mnras} {\bf 2015}, {\em 453},~1047--1061,
  \href{http://xxx.lanl.gov/abs/1505.07465}{{\normalfont
  [arXiv:astro-ph.GA/1505.07465]}}.
\newblock
  doi:{\changeurlcolor{black}\href{https://doi.org/10.1093/mnras/stv1588}{\detokenize{10.1093/mnras/stv1588}}}.

\bibitem[{Simon}(2019)]{2019ARA&A..57..375S}
{Simon}, J.D.
\newblock {The Faintest Dwarf Galaxies}.
\newblock {\em \araa} {\bf 2019}, {\em 57},~375--415,
  \href{http://xxx.lanl.gov/abs/1901.05465}{{\normalfont
  [arXiv:astro-ph.GA/1901.05465]}}.
\newblock
  doi:{\changeurlcolor{black}\href{https://doi.org/10.1146/annurev-astro-091918-104453}{\detokenize{10.1146/annurev-astro-091918-104453}}}.

\bibitem[{Martin} \em{et~al.}(2013){Martin}, {Ibata}, {McConnachie}, {Mackey},
  {Ferguson}, {Irwin}, {Lewis}, and {Fardal}]{2013ApJ...776...80M}
{Martin}, N.F.; {Ibata}, R.A.; {McConnachie}, A.W.; {Mackey}, A.D.; {Ferguson},
  A.M.N.; {Irwin}, M.J.; {Lewis}, G.F.; {Fardal}, M.A.
\newblock {The PAndAS View of the Andromeda Satellite System. I. A Bayesian
  Search for Dwarf Galaxies Using Spatial and Color-Magnitude Information}.
\newblock {\em \apj} {\bf 2013}, {\em 776},~80,
  \href{http://xxx.lanl.gov/abs/1307.7626}{{\normalfont
  [arXiv:astro-ph.GA/1307.7626]}}.
\newblock
  doi:{\changeurlcolor{black}\href{https://doi.org/10.1088/0004-637X/776/2/80}{\detokenize{10.1088/0004-637X/776/2/80}}}.

\bibitem[{Crnojevi{\'c}} \em{et~al.}(2016){Crnojevi{\'c}}, {Sand}, {Spekkens},
  {Caldwell}, {Guhathakurta}, {McLeod}, {Seth}, {Simon}, {Strader}, and
  {Toloba}]{2016ApJ...823...19C}
{Crnojevi{\'c}}, D.; {Sand}, D.J.; {Spekkens}, K.; {Caldwell}, N.;
  {Guhathakurta}, P.; {McLeod}, B.; {Seth}, A.; {Simon}, J.D.; {Strader}, J.;
  {Toloba}, E.
\newblock {The Extended Halo of Centaurus A: Uncovering Satellites, Streams,
  and Substructures}.
\newblock {\em \apj} {\bf 2016}, {\em 823},~19,
  \href{http://xxx.lanl.gov/abs/1512.05366}{{\normalfont
  [arXiv:astro-ph.GA/1512.05366]}}.
\newblock
  doi:{\changeurlcolor{black}\href{https://doi.org/10.3847/0004-637X/823/1/19}{\detokenize{10.3847/0004-637X/823/1/19}}}.

\bibitem[{M{\"u}ller} \em{et~al.}(2017){M{\"u}ller}, {Jerjen}, and
  {Binggeli}]{2017A&A...597A...7M}
{M{\"u}ller}, O.; {Jerjen}, H.; {Binggeli}, B.
\newblock {New low surface brightness dwarf galaxies in the Centaurus group}.
\newblock {\em \aap} {\bf 2017}, {\em 597},~A7,
  \href{http://xxx.lanl.gov/abs/1605.04130}{{\normalfont
  [arXiv:astro-ph.GA/1605.04130]}}.
\newblock
  doi:{\changeurlcolor{black}\href{https://doi.org/10.1051/0004-6361/201628921}{\detokenize{10.1051/0004-6361/201628921}}}.

\bibitem[{Chiboucas} \em{et~al.}(2013){Chiboucas}, {Jacobs}, {Tully}, and
  {Karachentsev}]{2013AJ....146..126C}
{Chiboucas}, K.; {Jacobs}, B.A.; {Tully}, R.B.; {Karachentsev}, I.D.
\newblock {Confirmation of Faint Dwarf Galaxies in the M81 Group}.
\newblock {\em \aj} {\bf 2013}, {\em 146},~126,
  \href{http://xxx.lanl.gov/abs/1309.4130}{{\normalfont
  [arXiv:astro-ph.CO/1309.4130]}}.
\newblock
  doi:{\changeurlcolor{black}\href{https://doi.org/10.1088/0004-6256/146/5/126}{\detokenize{10.1088/0004-6256/146/5/126}}}.

\bibitem[{Merritt} \em{et~al.}(2014){Merritt}, {van Dokkum}, and
  {Abraham}]{2014ApJ...787L..37M}
{Merritt}, A.; {van Dokkum}, P.; {Abraham}, R.
\newblock {The Discovery of Seven Extremely Low Surface Brightness Galaxies in
  the Field of the Nearby Spiral Galaxy M101}.
\newblock {\em \apjl} {\bf 2014}, {\em 787},~L37,
  \href{http://xxx.lanl.gov/abs/1406.2315}{{\normalfont
  [arXiv:astro-ph.GA/1406.2315]}}.
\newblock
  doi:{\changeurlcolor{black}\href{https://doi.org/10.1088/2041-8205/787/2/L37}{\detokenize{10.1088/2041-8205/787/2/L37}}}.

\bibitem[{M{\"u}ller} \em{et~al.}(2017){M{\"u}ller}, {Scalera}, {Binggeli}, and
  {Jerjen}]{2017A&A...602A.119M}
{M{\"u}ller}, O.; {Scalera}, R.; {Binggeli}, B.; {Jerjen}, H.
\newblock {The M 101 group complex: new dwarf galaxy candidates and spatial
  structure}.
\newblock {\em \aap} {\bf 2017}, {\em 602},~A119,
  \href{http://xxx.lanl.gov/abs/1701.03681}{{\normalfont
  [arXiv:astro-ph.GA/1701.03681]}}.
\newblock
  doi:{\changeurlcolor{black}\href{https://doi.org/10.1051/0004-6361/201730434}{\detokenize{10.1051/0004-6361/201730434}}}.

\bibitem[{Carlsten} \em{et~al.}(2021){Carlsten}, {Greene}, {Greco}, {Beaton},
  and {Kado-Fong}]{2021arXiv210503435C}
{Carlsten}, S.G.; {Greene}, J.E.; {Greco}, J.P.; {Beaton}, R.L.; {Kado-Fong},
  E.
\newblock {ELVES I: Structures of Dwarf Satellites of MW-like Galaxies;
  Morphology, Scaling Relations, and Intrinsic Shapes}.
\newblock {\em arXiv e-prints} {\bf 2021}, p. arXiv:2105.03435,
  \href{http://xxx.lanl.gov/abs/2105.03435}{{\normalfont
  [arXiv:astro-ph.GA/2105.03435]}}.

\bibitem[{Geha} \em{et~al.}(2017){Geha}, {Wechsler}, {Mao}, {Tollerud},
  {Weiner}, {Bernstein}, {Hoyle}, {Marchi}, {Marshall}, {Mu{\~n}oz}, and
  {Lu}]{2017ApJ...847....4G}
{Geha}, M.; {Wechsler}, R.H.; {Mao}, Y.Y.; {Tollerud}, E.J.; {Weiner}, B.;
  {Bernstein}, R.; {Hoyle}, B.; {Marchi}, S.; {Marshall}, P.J.; {Mu{\~n}oz},
  R.; {Lu}, Y.
\newblock {The SAGA Survey. I. Satellite Galaxy Populations around Eight Milky
  Way Analogs}.
\newblock {\em \apj} {\bf 2017}, {\em 847},~4,
  \href{http://xxx.lanl.gov/abs/1705.06743}{{\normalfont
  [arXiv:astro-ph.GA/1705.06743]}}.
\newblock
  doi:{\changeurlcolor{black}\href{https://doi.org/10.3847/1538-4357/aa8626}{\detokenize{10.3847/1538-4357/aa8626}}}.

\bibitem[{Mao} \em{et~al.}(2021){Mao}, {Geha}, {Wechsler}, {Weiner},
  {Tollerud}, {Nadler}, and {Kallivayalil}]{2021ApJ...907...85M}
{Mao}, Y.Y.; {Geha}, M.; {Wechsler}, R.H.; {Weiner}, B.; {Tollerud}, E.J.;
  {Nadler}, E.O.; {Kallivayalil}, N.
\newblock {The SAGA Survey. II. Building a Statistical Sample of Satellite
  Systems around Milky Way-like Galaxies}.
\newblock {\em \apj} {\bf 2021}, {\em 907},~85,
  \href{http://xxx.lanl.gov/abs/2008.12783}{{\normalfont
  [arXiv:astro-ph.GA/2008.12783]}}.
\newblock
  doi:{\changeurlcolor{black}\href{https://doi.org/10.3847/1538-4357/abce58}{\detokenize{10.3847/1538-4357/abce58}}}.

\bibitem[{Habas} \em{et~al.}(2020){Habas}, {Marleau}, {Duc}, {Durrell},
  {Paudel}, {Poulain}, {S{\'a}nchez-Janssen}, {Sreejith}, {Ramasawmy},
  {Stemock}, {Leach}, {Cuillandre}, {Gwyn}, {Agnello}, {B{\'\i}lek}, {Fensch},
  {M{\"u}ller}, {Peng}, and {van der Burg}]{2020MNRAS.491.1901H}
{Habas}, R.; {Marleau}, F.R.; {Duc}, P.A.; {Durrell}, P.R.; {Paudel}, S.;
  {Poulain}, M.; {S{\'a}nchez-Janssen}, R.; {Sreejith}, S.; {Ramasawmy}, J.;
  {Stemock}, B.; {Leach}, C.; {Cuillandre}, J.C.; {Gwyn}, S.; {Agnello}, A.;
  {B{\'\i}lek}, M.; {Fensch}, J.; {M{\"u}ller}, O.; {Peng}, E.W.; {van der
  Burg}, R.F.J.
\newblock {Newly discovered dwarf galaxies in the MATLAS low-density fields}.
\newblock {\em \mnras} {\bf 2020}, {\em 491},~1901--1919,
  \href{http://xxx.lanl.gov/abs/1910.13462}{{\normalfont
  [arXiv:astro-ph.GA/1910.13462]}}.
\newblock
  doi:{\changeurlcolor{black}\href{https://doi.org/10.1093/mnras/stz3045}{\detokenize{10.1093/mnras/stz3045}}}.

\bibitem[{Javanmardi} \em{et~al.}(2016){Javanmardi}, {Martinez-Delgado},
  {Kroupa}, {Henkel}, {Crawford}, {Teuwen}, {Gabany}, {Hanson}, {Chonis}, and
  {Neyer}]{2016A&A...588A..89J}
{Javanmardi}, B.; {Martinez-Delgado}, D.; {Kroupa}, P.; {Henkel}, C.;
  {Crawford}, K.; {Teuwen}, K.; {Gabany}, R.J.; {Hanson}, M.; {Chonis}, T.S.;
  {Neyer}, F.
\newblock {DGSAT: Dwarf Galaxy Survey with Amateur Telescopes. I. Discovery of
  low surface brightness systems around nearby spiral galaxies}.
\newblock {\em \aap} {\bf 2016}, {\em 588},~A89,
  \href{http://xxx.lanl.gov/abs/1511.04446}{{\normalfont
  [arXiv:astro-ph.GA/1511.04446]}}.
\newblock
  doi:{\changeurlcolor{black}\href{https://doi.org/10.1051/0004-6361/201527745}{\detokenize{10.1051/0004-6361/201527745}}}.

\bibitem[{Henkel} \em{et~al.}(2017){Henkel}, {Javanmardi},
  {Mart{\'\i}nez-Delgado}, {Kroupa}, and {Teuwen}]{2017A&A...603A..18H}
{Henkel}, C.; {Javanmardi}, B.; {Mart{\'\i}nez-Delgado}, D.; {Kroupa}, P.;
  {Teuwen}, K.
\newblock {DGSAT: Dwarf Galaxy Survey with Amateur Telescopes. II. A catalogue
  of isolated nearby edge-on disk galaxies and the discovery of new low surface
  brightness systems}.
\newblock {\em \aap} {\bf 2017}, {\em 603},~A18,
  \href{http://xxx.lanl.gov/abs/1703.05356}{{\normalfont
  [arXiv:astro-ph.GA/1703.05356]}}.
\newblock
  doi:{\changeurlcolor{black}\href{https://doi.org/10.1051/0004-6361/201730539}{\detokenize{10.1051/0004-6361/201730539}}}.

\bibitem[{Pawlowski} \em{et~al.}(2014){Pawlowski}, {Famaey}, {Jerjen},
  {Merritt}, {Kroupa}, {Dabringhausen}, {L{\"u}ghausen}, {Forbes}, {Hensler},
  {Hammer}, {Puech}, {Fouquet}, {Flores}, and {Yang}]{2014MNRAS.442.2362P}
{Pawlowski}, M.S.; {Famaey}, B.; {Jerjen}, H.; {Merritt}, D.; {Kroupa}, P.;
  {Dabringhausen}, J.; {L{\"u}ghausen}, F.; {Forbes}, D.A.; {Hensler}, G.;
  {Hammer}, F.; {Puech}, M.; {Fouquet}, S.; {Flores}, H.; {Yang}, Y.
\newblock {Co-orbiting satellite galaxy structures are still in conflict with
  the distribution of primordial dwarf galaxies}.
\newblock {\em \mnras} {\bf 2014}, {\em 442},~2362--2380,
  \href{http://xxx.lanl.gov/abs/1406.1799}{{\normalfont
  [arXiv:astro-ph.GA/1406.1799]}}.
\newblock
  doi:{\changeurlcolor{black}\href{https://doi.org/10.1093/mnras/stu1005}{\detokenize{10.1093/mnras/stu1005}}}.

\bibitem[{Piatek} \em{et~al.}(2006){Piatek}, {Pryor}, {Bristow}, {Olszewski},
  {Harris}, {Mateo}, {Minniti}, and {Tinney}]{2006AJ....131.1445P}
{Piatek}, S.; {Pryor}, C.; {Bristow}, P.; {Olszewski}, E.W.; {Harris}, H.C.;
  {Mateo}, M.; {Minniti}, D.; {Tinney}, C.G.
\newblock {Proper Motions of Dwarf Spheroidal Galaxies from Hubble Space
  Telescope Imaging. IV. Measurement for Sculptor}.
\newblock {\em \aj} {\bf 2006}, {\em 131},~1445--1460,
  \href{http://xxx.lanl.gov/abs/astro-ph/0601547}{{\normalfont
  [arXiv:astro-ph/astro-ph/0601547]}}.
\newblock
  doi:{\changeurlcolor{black}\href{https://doi.org/10.1086/499526}{\detokenize{10.1086/499526}}}.

\bibitem[{Kallivayalil} \em{et~al.}(2013){Kallivayalil}, {van der Marel},
  {Besla}, {Anderson}, and {Alcock}]{2013ApJ...764..161K}
{Kallivayalil}, N.; {van der Marel}, R.P.; {Besla}, G.; {Anderson}, J.;
  {Alcock}, C.
\newblock {Third-epoch Magellanic Cloud Proper Motions. I. Hubble Space
  Telescope/WFC3 Data and Orbit Implications}.
\newblock {\em \apj} {\bf 2013}, {\em 764},~161,
  \href{http://xxx.lanl.gov/abs/1301.0832}{{\normalfont
  [arXiv:astro-ph.CO/1301.0832]}}.
\newblock
  doi:{\changeurlcolor{black}\href{https://doi.org/10.1088/0004-637X/764/2/161}{\detokenize{10.1088/0004-637X/764/2/161}}}.

\bibitem[{Sohn} \em{et~al.}(2013){Sohn}, {Besla}, {van der Marel},
  {Boylan-Kolchin}, {Majewski}, and {Bullock}]{2013ApJ...768..139S}
{Sohn}, S.T.; {Besla}, G.; {van der Marel}, R.P.; {Boylan-Kolchin}, M.;
  {Majewski}, S.R.; {Bullock}, J.S.
\newblock {The Space Motion of Leo I: Hubble Space Telescope Proper Motion and
  Implied Orbit}.
\newblock {\em \apj} {\bf 2013}, {\em 768},~139,
  \href{http://xxx.lanl.gov/abs/1210.6039}{{\normalfont
  [arXiv:astro-ph.GA/1210.6039]}}.
\newblock
  doi:{\changeurlcolor{black}\href{https://doi.org/10.1088/0004-637X/768/2/139}{\detokenize{10.1088/0004-637X/768/2/139}}}.

\bibitem[{Pryor} \em{et~al.}(2015){Pryor}, {Piatek}, and
  {Olszewski}]{2015AJ....149...42P}
{Pryor}, C.; {Piatek}, S.; {Olszewski}, E.W.
\newblock {Proper Motion of the Draco Dwarf Galaxy Based On Hubble Space
  Telescope Imaging}.
\newblock {\em \aj} {\bf 2015}, {\em 149},~42,
  \href{http://xxx.lanl.gov/abs/1407.3509}{{\normalfont
  [arXiv:astro-ph.GA/1407.3509]}}.
\newblock
  doi:{\changeurlcolor{black}\href{https://doi.org/10.1088/0004-6256/149/2/42}{\detokenize{10.1088/0004-6256/149/2/42}}}.

\bibitem[{Sohn} \em{et~al.}(2017){Sohn}, {Patel}, {Besla}, {van der Marel},
  {Bullock}, {Strigari}, {van de Ven}, {Walker}, and
  {Bellini}]{2017ApJ...849...93S}
{Sohn}, S.T.; {Patel}, E.; {Besla}, G.; {van der Marel}, R.P.; {Bullock}, J.S.;
  {Strigari}, L.E.; {van de Ven}, G.; {Walker}, M.G.; {Bellini}, A.
\newblock {Space Motions of the Dwarf Spheroidal Galaxies Draco and Sculptor
  Based on HST Proper Motions with a {\ensuremath{\sim}}10 yr Time Baseline}.
\newblock {\em \apj} {\bf 2017}, {\em 849},~93,
  \href{http://xxx.lanl.gov/abs/1707.02593}{{\normalfont
  [arXiv:astro-ph.GA/1707.02593]}}.
\newblock
  doi:{\changeurlcolor{black}\href{https://doi.org/10.3847/1538-4357/aa917b}{\detokenize{10.3847/1538-4357/aa917b}}}.

\bibitem[{Gaia Collaboration} \em{et~al.}(2016){Gaia Collaboration}, {Prusti},
  {de Bruijne}, {Brown}, {Vallenari}, {Babusiaux}, {Bailer-Jones}, {Bastian},
  {Biermann}, {Evans}, and et~al.]{2016A&A...595A...1G}
{Gaia Collaboration}.; {Prusti}, T.; {de Bruijne}, J.H.J.; {Brown}, A.G.A.;
  {Vallenari}, A.; {Babusiaux}, C.; {Bailer-Jones}, C.A.L.; {Bastian}, U.;
  {Biermann}, M.; {Evans}, D.W.; et~al..
\newblock {The Gaia mission}.
\newblock {\em \aap} {\bf 2016}, {\em 595},~A1,
  \href{http://xxx.lanl.gov/abs/1609.04153}{{\normalfont
  [arXiv:astro-ph.IM/1609.04153]}}.
\newblock
  doi:{\changeurlcolor{black}\href{https://doi.org/10.1051/0004-6361/201629272}{\detokenize{10.1051/0004-6361/201629272}}}.

\bibitem[{Simon}(2018)]{2018ApJ...863...89S}
{Simon}, J.D.
\newblock {Gaia Proper Motions and Orbits of the Ultra-faint Milky Way
  Satellites}.
\newblock {\em \apj} {\bf 2018}, {\em 863},~89,
  \href{http://xxx.lanl.gov/abs/1804.10230}{{\normalfont
  [arXiv:astro-ph.GA/1804.10230]}}.
\newblock
  doi:{\changeurlcolor{black}\href{https://doi.org/10.3847/1538-4357/aacdfb}{\detokenize{10.3847/1538-4357/aacdfb}}}.

\bibitem[{Fritz} \em{et~al.}(2018){Fritz}, {Battaglia}, {Pawlowski},
  {Kallivayalil}, {van der Marel}, {Sohn}, {Brook}, and
  {Besla}]{2018A&A...619A.103F}
{Fritz}, T.K.; {Battaglia}, G.; {Pawlowski}, M.S.; {Kallivayalil}, N.; {van der
  Marel}, R.; {Sohn}, S.T.; {Brook}, C.; {Besla}, G.
\newblock {Gaia DR2 proper motions of dwarf galaxies within 420 kpc. Orbits,
  Milky Way mass, tidal influences, planar alignments, and group infall}.
\newblock {\em \aap} {\bf 2018}, {\em 619},~A103,
  \href{http://xxx.lanl.gov/abs/1805.00908}{{\normalfont
  [arXiv:astro-ph.GA/1805.00908]}}.
\newblock
  doi:{\changeurlcolor{black}\href{https://doi.org/10.1051/0004-6361/201833343}{\detokenize{10.1051/0004-6361/201833343}}}.

\bibitem[{Kallivayalil} \em{et~al.}(2018){Kallivayalil}, {Sales}, {Zivick},
  {Fritz}, {Del Pino}, {Sohn}, {Besla}, {van der Marel}, {Navarro}, and
  {Sacchi}]{2018ApJ...867...19K}
{Kallivayalil}, N.; {Sales}, L.V.; {Zivick}, P.; {Fritz}, T.K.; {Del Pino}, A.;
  {Sohn}, S.T.; {Besla}, G.; {van der Marel}, R.P.; {Navarro}, J.F.; {Sacchi},
  E.
\newblock {The Missing Satellites of the Magellanic Clouds? Gaia Proper Motions
  of the Recently Discovered Ultra-faint Galaxies}.
\newblock {\em \apj} {\bf 2018}, {\em 867},~19,
  \href{http://xxx.lanl.gov/abs/1805.01448}{{\normalfont
  [arXiv:astro-ph.GA/1805.01448]}}.
\newblock
  doi:{\changeurlcolor{black}\href{https://doi.org/10.3847/1538-4357/aadfee}{\detokenize{10.3847/1538-4357/aadfee}}}.

\bibitem[{Massari} and {Helmi}(2018)]{2018A&A...620A.155M}
{Massari}, D.; {Helmi}, A.
\newblock {With and without spectroscopy: Gaia DR2 proper motions of seven
  ultra-faint dwarf galaxies}.
\newblock {\em \aap} {\bf 2018}, {\em 620},~A155,
  \href{http://xxx.lanl.gov/abs/1805.01839}{{\normalfont
  [arXiv:astro-ph.GA/1805.01839]}}.
\newblock
  doi:{\changeurlcolor{black}\href{https://doi.org/10.1051/0004-6361/201833367}{\detokenize{10.1051/0004-6361/201833367}}}.

\bibitem[{Pace} and {Li}(2019)]{2019ApJ...875...77P}
{Pace}, A.B.; {Li}, T.S.
\newblock {Proper Motions of Milky Way Ultra-faint Satellites with Gaia DR2 {x}
  DES DR1}.
\newblock {\em \apj} {\bf 2019}, {\em 875},~77,
  \href{http://xxx.lanl.gov/abs/1806.02345}{{\normalfont
  [arXiv:astro-ph.GA/1806.02345]}}.
\newblock
  doi:{\changeurlcolor{black}\href{https://doi.org/10.3847/1538-4357/ab0aee}{\detokenize{10.3847/1538-4357/ab0aee}}}.

\bibitem[{Fritz} \em{et~al.}(2019){Fritz}, {Carrera}, {Battaglia}, and
  {Taibi}]{2019A&A...623A.129F}
{Fritz}, T.K.; {Carrera}, R.; {Battaglia}, G.; {Taibi}, S.
\newblock {Gaia DR 2 and VLT/FLAMES search for new satellites of the LMC}.
\newblock {\em \aap} {\bf 2019}, {\em 623},~A129,
  \href{http://xxx.lanl.gov/abs/1805.07350}{{\normalfont
  [arXiv:astro-ph.GA/1805.07350]}}.
\newblock
  doi:{\changeurlcolor{black}\href{https://doi.org/10.1051/0004-6361/201833458}{\detokenize{10.1051/0004-6361/201833458}}}.

\bibitem[{Li} \em{et~al.}(2021){Li}, {Hammer}, {Babusiaux}, {Pawlowski},
  {Yang}, {Arenou}, {Du}, and {Wang}]{2021ApJ...916....8L}
{Li}, H.; {Hammer}, F.; {Babusiaux}, C.; {Pawlowski}, M.S.; {Yang}, Y.;
  {Arenou}, F.; {Du}, C.; {Wang}, J.
\newblock {Gaia EDR3 Proper Motions of Milky Way Dwarfs. I. 3D Motions and
  Orbits}.
\newblock {\em \apj} {\bf 2021}, {\em 916},~8.
\newblock
  doi:{\changeurlcolor{black}\href{https://doi.org/10.3847/1538-4357/ac0436}{\detokenize{10.3847/1538-4357/ac0436}}}.

\bibitem[{McConnachie} and {Venn}(2020)]{2020RNAAS...4..229M}
{McConnachie}, A.W.; {Venn}, K.A.
\newblock {Updated Proper Motions for Local Group Dwarf Galaxies Using Gaia
  Early Data Release 3}.
\newblock {\em Research Notes of the American Astronomical Society} {\bf 2020},
  {\em 4},~229,  \href{http://xxx.lanl.gov/abs/2012.03904}{{\normalfont
  [arXiv:astro-ph.GA/2012.03904]}}.
\newblock
  doi:{\changeurlcolor{black}\href{https://doi.org/10.3847/2515-5172/abd18b}{\detokenize{10.3847/2515-5172/abd18b}}}.

\bibitem[{Battaglia} \em{et~al.}(2021){Battaglia}, {Taibi}, {Thomas}, and
  {Fritz}]{2021arXiv210608819B}
{Battaglia}, G.; {Taibi}, S.; {Thomas}, G.F.; {Fritz}, T.K.
\newblock {Gaia early DR3 systemic motions of Local Group dwarf galaxies and
  orbital properties with a massive Large Magellanic Cloud}.
\newblock {\em arXiv e-prints} {\bf 2021}, p. arXiv:2106.08819,
  \href{http://xxx.lanl.gov/abs/2106.08819}{{\normalfont
  [arXiv:astro-ph.GA/2106.08819]}}.

\bibitem[{Pawlowski} and {Kroupa}(2020)]{2020MNRAS.491.3042P}
{Pawlowski}, M.S.; {Kroupa}, P.
\newblock {The Milky Way's disc of classical satellite galaxies in light of
  Gaia DR2}.
\newblock {\em \mnras} {\bf 2020}, {\em 491},~3042--3059,
  \href{http://xxx.lanl.gov/abs/1911.05081}{{\normalfont
  [arXiv:astro-ph.GA/1911.05081]}}.
\newblock
  doi:{\changeurlcolor{black}\href{https://doi.org/10.1093/mnras/stz3163}{\detokenize{10.1093/mnras/stz3163}}}.

\bibitem[{Patel} \em{et~al.}(2020){Patel}, {Kallivayalil}, {Garavito-Camargo},
  {Besla}, {Weisz}, {van der Marel}, {Boylan-Kolchin}, {Pawlowski}, and
  {G{\'o}mez}]{2020ApJ...893..121P}
{Patel}, E.; {Kallivayalil}, N.; {Garavito-Camargo}, N.; {Besla}, G.; {Weisz},
  D.R.; {van der Marel}, R.P.; {Boylan-Kolchin}, M.; {Pawlowski}, M.S.;
  {G{\'o}mez}, F.A.
\newblock {The Orbital Histories of Magellanic Satellites Using Gaia DR2 Proper
  Motions}.
\newblock {\em \apj} {\bf 2020}, {\em 893},~121,
  \href{http://xxx.lanl.gov/abs/2001.01746}{{\normalfont
  [arXiv:astro-ph.GA/2001.01746]}}.
\newblock
  doi:{\changeurlcolor{black}\href{https://doi.org/10.3847/1538-4357/ab7b75}{\detokenize{10.3847/1538-4357/ab7b75}}}.

\bibitem[{Kunkel} and {Demers}(1976)]{1976RGOB..182..241K}
{Kunkel}, W.E.; {Demers}, S.
\newblock {The Magellanic Plane}.
\newblock  The Galaxy and the Local Group,  1976, Vol. 182, p. 241.

\bibitem[{Lynden-Bell}(1976)]{1976MNRAS.174..695L}
{Lynden-Bell}, D.
\newblock {Dwarf galaxies and globular clusters in high velocity hydrogen
  streams.}
\newblock {\em \mnras} {\bf 1976}, {\em 174},~695--710.
\newblock
  doi:{\changeurlcolor{black}\href{https://doi.org/10.1093/mnras/174.3.695}{\detokenize{10.1093/mnras/174.3.695}}}.

\bibitem[{Pawlowski} \em{et~al.}(2012){Pawlowski}, {Pflamm-Altenburg}, and
  {Kroupa}]{2012MNRAS.423.1109P}
{Pawlowski}, M.S.; {Pflamm-Altenburg}, J.; {Kroupa}, P.
\newblock {The VPOS: a vast polar structure of satellite galaxies, globular
  clusters and streams around the Milky Way}.
\newblock {\em \mnras} {\bf 2012}, {\em 423},~1109--1126,
  \href{http://xxx.lanl.gov/abs/1204.5176}{{\normalfont
  [arXiv:astro-ph.GA/1204.5176]}}.
\newblock
  doi:{\changeurlcolor{black}\href{https://doi.org/10.1111/j.1365-2966.2012.20937.x}{\detokenize{10.1111/j.1365-2966.2012.20937.x}}}.

\bibitem[{Riley} and {Strigari}(2020)]{2020MNRAS.494..983R}
{Riley}, A.H.; {Strigari}, L.E.
\newblock {The Milky Way's stellar streams and globular clusters do not align
  in a Vast Polar Structure}.
\newblock {\em \mnras} {\bf 2020}, {\em 494},~983--1001,
  \href{http://xxx.lanl.gov/abs/2001.11564}{{\normalfont
  [arXiv:astro-ph.GA/2001.11564]}}.
\newblock
  doi:{\changeurlcolor{black}\href{https://doi.org/10.1093/mnras/staa710}{\detokenize{10.1093/mnras/staa710}}}.

\bibitem[{Metz} \em{et~al.}(2009){Metz}, {Kroupa}, and
  {Jerjen}]{2009MNRAS.394.2223M}
{Metz}, M.; {Kroupa}, P.; {Jerjen}, H.
\newblock {Discs of satellites: the new dwarf spheroidals}.
\newblock {\em \mnras} {\bf 2009}, {\em 394},~2223--2228,
  \href{http://xxx.lanl.gov/abs/0901.1658}{{\normalfont
  [arXiv:astro-ph.GA/0901.1658]}}.
\newblock
  doi:{\changeurlcolor{black}\href{https://doi.org/10.1111/j.1365-2966.2009.14489.x}{\detokenize{10.1111/j.1365-2966.2009.14489.x}}}.

\bibitem[{Pawlowski} and {Kroupa}(2013)]{2013MNRAS.435.2116P}
{Pawlowski}, M.S.; {Kroupa}, P.
\newblock {The rotationally stabilized VPOS and predicted proper motions of the
  Milky Way satellite galaxies}.
\newblock {\em \mnras} {\bf 2013}, {\em 435},~2116--2131,
  \href{http://xxx.lanl.gov/abs/1309.1159}{{\normalfont
  [arXiv:astro-ph.CO/1309.1159]}}.
\newblock
  doi:{\changeurlcolor{black}\href{https://doi.org/10.1093/mnras/stt1429}{\detokenize{10.1093/mnras/stt1429}}}.

\bibitem[{Sohn} \em{et~al.}(2020){Sohn}, {Patel}, {Fardal}, {Besla}, {van der
  Marel}, {Geha}, and {Guhathakurta}]{2020ApJ...901...43S}
{Sohn}, S.T.; {Patel}, E.; {Fardal}, M.A.; {Besla}, G.; {van der Marel}, R.P.;
  {Geha}, M.; {Guhathakurta}, P.
\newblock {HST Proper Motions of NGC 147 and NGC 185: Orbital Histories and
  Tests of a Dynamically Coherent Andromeda Satellite Plane}.
\newblock {\em \apj} {\bf 2020}, {\em 901},~43,
  \href{http://xxx.lanl.gov/abs/2008.06055}{{\normalfont
  [arXiv:astro-ph.GA/2008.06055]}}.
\newblock
  doi:{\changeurlcolor{black}\href{https://doi.org/10.3847/1538-4357/abaf49}{\detokenize{10.3847/1538-4357/abaf49}}}.

\bibitem[{Ibata} \em{et~al.}(2013){Ibata}, {Lewis}, {Conn}, {Irwin},
  {McConnachie}, {Chapman}, {Collins}, {Fardal}, {Ferguson}, {Ibata}, {Mackey},
  {Martin}, {Navarro}, {Rich}, {Valls-Gabaud}, and
  {Widrow}]{2013Natur.493...62I}
{Ibata}, R.A.; {Lewis}, G.F.; {Conn}, A.R.; {Irwin}, M.J.; {McConnachie}, A.W.;
  {Chapman}, S.C.; {Collins}, M.L.; {Fardal}, M.; {Ferguson}, A.M.N.; {Ibata},
  N.G.; {Mackey}, A.D.; {Martin}, N.F.; {Navarro}, J.; {Rich}, R.M.;
  {Valls-Gabaud}, D.; {Widrow}, L.M.
\newblock {A vast, thin plane of corotating dwarf galaxies orbiting the
  Andromeda galaxy}.
\newblock {\em \nat} {\bf 2013}, {\em 493},~62--65,
  \href{http://xxx.lanl.gov/abs/1301.0446}{{\normalfont
  [arXiv:astro-ph.CO/1301.0446]}}.
\newblock
  doi:{\changeurlcolor{black}\href{https://doi.org/10.1038/nature11717}{\detokenize{10.1038/nature11717}}}.

\bibitem[{Conn} \em{et~al.}(2013){Conn}, {Lewis}, {Ibata}, {Parker}, {Zucker},
  {McConnachie}, {Martin}, {Valls-Gabaud}, {Tanvir}, {Irwin}, {Ferguson}, and
  {Chapman}]{2013ApJ...766..120C}
{Conn}, A.R.; {Lewis}, G.F.; {Ibata}, R.A.; {Parker}, Q.A.; {Zucker}, D.B.;
  {McConnachie}, A.W.; {Martin}, N.F.; {Valls-Gabaud}, D.; {Tanvir}, N.;
  {Irwin}, M.J.; {Ferguson}, A.M.N.; {Chapman}, S.C.
\newblock {The Three-dimensional Structure of the M31 Satellite System; Strong
  Evidence for an Inhomogeneous Distribution of Satellites}.
\newblock {\em \apj} {\bf 2013}, {\em 766},~120,
  \href{http://xxx.lanl.gov/abs/1301.7131}{{\normalfont
  [arXiv:astro-ph.CO/1301.7131]}}.
\newblock
  doi:{\changeurlcolor{black}\href{https://doi.org/10.1088/0004-637X/766/2/120}{\detokenize{10.1088/0004-637X/766/2/120}}}.

\bibitem[{Gillet} \em{et~al.}(2015){Gillet}, {Ocvirk}, {Aubert}, {Knebe},
  {Libeskind}, {Yepes}, {Gottl{\"o}ber}, and {Hoffman}]{2015ApJ...800...34G}
{Gillet}, N.; {Ocvirk}, P.; {Aubert}, D.; {Knebe}, A.; {Libeskind}, N.;
  {Yepes}, G.; {Gottl{\"o}ber}, S.; {Hoffman}, Y.
\newblock {Vast Planes of Satellites in a High-resolution Simulation of the
  Local Group: Comparison to Andromeda}.
\newblock {\em \apj} {\bf 2015}, {\em 800},~34,
  \href{http://xxx.lanl.gov/abs/1412.3110}{{\normalfont
  [arXiv:astro-ph.GA/1412.3110]}}.
\newblock
  doi:{\changeurlcolor{black}\href{https://doi.org/10.1088/0004-637X/800/1/34}{\detokenize{10.1088/0004-637X/800/1/34}}}.

\bibitem[{Buck} \em{et~al.}(2016){Buck}, {Dutton}, and
  {Macci{\`o}}]{2016MNRAS.460.4348B}
{Buck}, T.; {Dutton}, A.A.; {Macci{\`o}}, A.V.
\newblock {Simulated {\ensuremath{\Lambda}}CDM analogues of the thin plane of
  satellites around the Andromeda galaxy are not kinematically coherent
  structures}.
\newblock {\em \mnras} {\bf 2016}, {\em 460},~4348--4365,
  \href{http://xxx.lanl.gov/abs/1510.06028}{{\normalfont
  [arXiv:astro-ph.GA/1510.06028]}}.
\newblock
  doi:{\changeurlcolor{black}\href{https://doi.org/10.1093/mnras/stw1232}{\detokenize{10.1093/mnras/stw1232}}}.

\bibitem[{Fernando} \em{et~al.}(2017){Fernando}, {Arias}, {Guglielmo}, {Lewis},
  {Ibata}, and {Power}]{2017MNRAS.465..641F}
{Fernando}, N.; {Arias}, V.; {Guglielmo}, M.; {Lewis}, G.F.; {Ibata}, R.A.;
  {Power}, C.
\newblock {On the stability of satellite planes - I. Effects of mass, velocity,
  halo shape and alignment}.
\newblock {\em \mnras} {\bf 2017}, {\em 465},~641--652,
  \href{http://xxx.lanl.gov/abs/1610.05393}{{\normalfont
  [arXiv:astro-ph.GA/1610.05393]}}.
\newblock
  doi:{\changeurlcolor{black}\href{https://doi.org/10.1093/mnras/stw2694}{\detokenize{10.1093/mnras/stw2694}}}.

\bibitem[{Fernando} \em{et~al.}(2018){Fernando}, {Arias}, {Lewis}, {Ibata}, and
  {Power}]{2018MNRAS.473.2212F}
{Fernando}, N.; {Arias}, V.; {Lewis}, G.F.; {Ibata}, R.A.; {Power}, C.
\newblock {Stability of satellite planes in M31 II: effects of the dark subhalo
  population}.
\newblock {\em \mnras} {\bf 2018}, {\em 473},~2212--2221,
  \href{http://xxx.lanl.gov/abs/1709.08010}{{\normalfont
  [arXiv:astro-ph.GA/1709.08010]}}.
\newblock
  doi:{\changeurlcolor{black}\href{https://doi.org/10.1093/mnras/stx2483}{\detokenize{10.1093/mnras/stx2483}}}.

\bibitem[{Pawlowski} \em{et~al.}(2013){Pawlowski}, {Kroupa}, and
  {Jerjen}]{2013MNRAS.435.1928P}
{Pawlowski}, M.S.; {Kroupa}, P.; {Jerjen}, H.
\newblock {Dwarf galaxy planes: the discovery of symmetric structures in the
  Local Group}.
\newblock {\em \mnras} {\bf 2013}, {\em 435},~1928--1957,
  \href{http://xxx.lanl.gov/abs/1307.6210}{{\normalfont
  [arXiv:astro-ph.CO/1307.6210]}}.
\newblock
  doi:{\changeurlcolor{black}\href{https://doi.org/10.1093/mnras/stt1384}{\detokenize{10.1093/mnras/stt1384}}}.

\bibitem[{M{\"u}ller} \em{et~al.}(2021){M{\"u}ller}, {Pawlowski}, {Lelli},
  {Fahrion}, {Rejkuba}, {Hilker}, {Kanehisa}, {Libeskind}, and
  {Jerjen}]{2021AnA...645L...5M}
{M{\"u}ller}, O.; {Pawlowski}, M.S.; {Lelli}, F.; {Fahrion}, K.; {Rejkuba}, M.;
  {Hilker}, M.; {Kanehisa}, J.; {Libeskind}, N.; {Jerjen}, H.
\newblock {The coherent motion of Cen A dwarf satellite galaxies remains a
  challenge for {\ensuremath{\Lambda}}CDM cosmology}.
\newblock {\em \aap} {\bf 2021}, {\em 645},~L5,
  \href{http://xxx.lanl.gov/abs/2012.08138}{{\normalfont
  [arXiv:astro-ph.GA/2012.08138]}}.
\newblock
  doi:{\changeurlcolor{black}\href{https://doi.org/10.1051/0004-6361/202039973}{\detokenize{10.1051/0004-6361/202039973}}}.

\bibitem[{M{\"u}ller} \em{et~al.}(2018){M{\"u}ller}, {Rejkuba}, and
  {Jerjen}]{2018AnA...615A..96M}
{M{\"u}ller}, O.; {Rejkuba}, M.; {Jerjen}, H.
\newblock {Distances from the tip of the red giant branch to the dwarf galaxies
  dw1335-29 and dw1340-30 in the Centaurus group}.
\newblock {\em \aap} {\bf 2018}, {\em 615},~A96,
  \href{http://xxx.lanl.gov/abs/1803.02406}{{\normalfont
  [arXiv:astro-ph.GA/1803.02406]}}.
\newblock
  doi:{\changeurlcolor{black}\href{https://doi.org/10.1051/0004-6361/201732455}{\detokenize{10.1051/0004-6361/201732455}}}.

\bibitem[{M{\"u}ller} \em{et~al.}(2017){M{\"u}ller}, {Scalera}, {Binggeli}, and
  {Jerjen}]{2017AnA...602A.119M}
{M{\"u}ller}, O.; {Scalera}, R.; {Binggeli}, B.; {Jerjen}, H.
\newblock {The M 101 group complex: new dwarf galaxy candidates and spatial
  structure}.
\newblock {\em \aap} {\bf 2017}, {\em 602},~A119,
  \href{http://xxx.lanl.gov/abs/1701.03681}{{\normalfont
  [arXiv:astro-ph.GA/1701.03681]}}.
\newblock
  doi:{\changeurlcolor{black}\href{https://doi.org/10.1051/0004-6361/201730434}{\detokenize{10.1051/0004-6361/201730434}}}.

\bibitem[{Mart{\'\i}nez-Delgado} \em{et~al.}(2021){Mart{\'\i}nez-Delgado},
  {Makarov}, {Javanmardi}, {Pawlowski}, {Makarova}, {Donatiello}, {Lang},
  {Rom{\'a}n}, {Vivas}, and {Carballo-Bello}]{2021arXiv210608868M}
{Mart{\'\i}nez-Delgado}, D.; {Makarov}, D.; {Javanmardi}, B.; {Pawlowski},
  M.S.; {Makarova}, L.; {Donatiello}, G.; {Lang}, D.; {Rom{\'a}n}, J.; {Vivas},
  K.; {Carballo-Bello}, J.A.
\newblock {Tracing satellite planes in the Sculptor group. I. Discovery of
  three faint dwarf galaxies around NGC 253}.
\newblock {\em \aap} {\bf 2021}, {\em 652},~A48,
  \href{http://xxx.lanl.gov/abs/2106.08868}{{\normalfont
  [arXiv:astro-ph.GA/2106.08868]}}.
\newblock
  doi:{\changeurlcolor{black}\href{https://doi.org/10.1051/0004-6361/202141242}{\detokenize{10.1051/0004-6361/202141242}}}.

\bibitem[{Paudel} \em{et~al.}(2021){Paudel}, {Yoon}, and
  {Smith}]{2021ApJ...917L..18P}
{Paudel}, S.; {Yoon}, S.J.; {Smith}, R.
\newblock {A Corotating Group of Dwarf Galaxies around NGC 2750 as a Centaurus
  A Analog}.
\newblock {\em \apjl} {\bf 2021}, {\em 917},~L18.
\newblock
  doi:{\changeurlcolor{black}\href{https://doi.org/10.3847/2041-8213/ac1866}{\detokenize{10.3847/2041-8213/ac1866}}}.

\bibitem[{Mutlu-Pakdil} \em{et~al.}(2021){Mutlu-Pakdil}, {Sand},
  {Crnojevi{\'c}}, {Jones}, {Caldwell}, {Guhathakurta}, {Seth}, {Simon},
  {Spekkens}, {Strader}, and {Toloba}]{arXiv:2108.09312}
{Mutlu-Pakdil}, B.; {Sand}, D.J.; {Crnojevi{\'c}}, D.; {Jones}, M.G.;
  {Caldwell}, N.; {Guhathakurta}, P.; {Seth}, A.C.; {Simon}, J.D.; {Spekkens},
  K.; {Strader}, J.; {Toloba}, E.
\newblock {Hubble Space Telescope Observations of NGC 253 Dwarf Satellites:
  Discovery of Three Ultra-faint Dwarf Galaxies}.
\newblock {\em arXiv e-prints} {\bf 2021}, p. arXiv:2108.09312,
  \href{http://xxx.lanl.gov/abs/2108.09312}{{\normalfont
  [arXiv:astro-ph.GA/2108.09312]}}.

\bibitem[{Tully} \em{et~al.}(2015){Tully}, {Libeskind}, {Karachentsev},
  {Karachentseva}, {Rizzi}, and {Shaya}]{2015ApJ...802L..25T}
{Tully}, R.B.; {Libeskind}, N.I.; {Karachentsev}, I.D.; {Karachentseva}, V.E.;
  {Rizzi}, L.; {Shaya}, E.J.
\newblock {Two Planes of Satellites in the Centaurus A Group}.
\newblock {\em \apjl} {\bf 2015}, {\em 802},~L25,
  \href{http://xxx.lanl.gov/abs/1503.05599}{{\normalfont
  [arXiv:astro-ph.GA/1503.05599]}}.
\newblock
  doi:{\changeurlcolor{black}\href{https://doi.org/10.1088/2041-8205/802/2/L25}{\detokenize{10.1088/2041-8205/802/2/L25}}}.

\bibitem[{M{\"u}ller} \em{et~al.}(2016){M{\"u}ller}, {Jerjen}, {Pawlowski}, and
  {Binggeli}]{2016A&A...595A.119M}
{M{\"u}ller}, O.; {Jerjen}, H.; {Pawlowski}, M.S.; {Binggeli}, B.
\newblock {Testing the two planes of satellites in the Centaurus group}.
\newblock {\em \aap} {\bf 2016}, {\em 595},~A119,
  \href{http://xxx.lanl.gov/abs/1607.04024}{{\normalfont
  [arXiv:astro-ph.CO/1607.04024]}}.
\newblock
  doi:{\changeurlcolor{black}\href{https://doi.org/10.1051/0004-6361/201629298}{\detokenize{10.1051/0004-6361/201629298}}}.

\bibitem[{M{\"u}ller} \em{et~al.}(2018){M{\"u}ller}, {Pawlowski}, {Jerjen}, and
  {Lelli}]{2018Sci...359..534M}
{M{\"u}ller}, O.; {Pawlowski}, M.S.; {Jerjen}, H.; {Lelli}, F.
\newblock {A whirling plane of satellite galaxies around Centaurus A challenges
  cold dark matter cosmology}.
\newblock {\em Science} {\bf 2018}, {\em 359},~534--537,
  \href{http://xxx.lanl.gov/abs/1802.00081}{{\normalfont
  [arXiv:astro-ph.GA/1802.00081]}}.
\newblock
  doi:{\changeurlcolor{black}\href{https://doi.org/10.1126/science.aao1858}{\detokenize{10.1126/science.aao1858}}}.

\bibitem[{M{\"u}ller} \em{et~al.}(2021{\natexlab{a}}){M{\"u}ller}, {Fahrion},
  {Rejkuba}, {Hilker}, {Lelli}, {Lutz}, {Pawlowski}, {Coccato}, {Anand}, and
  {Jerjen}]{2021A&A...645A..92M}
{M{\"u}ller}, O.; {Fahrion}, K.; {Rejkuba}, M.; {Hilker}, M.; {Lelli}, F.;
  {Lutz}, K.; {Pawlowski}, M.S.; {Coccato}, L.; {Anand}, G.S.; {Jerjen}, H.
\newblock {The properties of dwarf spheroidal galaxies in the Cen A group.
  Stellar populations, internal dynamics, and a heart-shaped
  H{\ensuremath{\alpha}} ring}.
\newblock {\em \aap} {\bf 2021}, {\em 645},~A92,
  \href{http://xxx.lanl.gov/abs/2011.04990}{{\normalfont
  [arXiv:astro-ph.GA/2011.04990]}}.
\newblock
  doi:{\changeurlcolor{black}\href{https://doi.org/10.1051/0004-6361/202039359}{\detokenize{10.1051/0004-6361/202039359}}}.

\bibitem[{M{\"u}ller} \em{et~al.}(2021{\natexlab{b}}){M{\"u}ller}, {Pawlowski},
  {Lelli}, {Fahrion}, {Rejkuba}, {Hilker}, {Kanehisa}, {Libeskind}, and
  {Jerjen}]{2021A&A...645L...5M}
{M{\"u}ller}, O.; {Pawlowski}, M.S.; {Lelli}, F.; {Fahrion}, K.; {Rejkuba}, M.;
  {Hilker}, M.; {Kanehisa}, J.; {Libeskind}, N.; {Jerjen}, H.
\newblock {The coherent motion of Cen A dwarf satellite galaxies remains a
  challenge for {\ensuremath{\Lambda}}CDM cosmology}.
\newblock {\em \aap} {\bf 2021}, {\em 645},~L5.
\newblock
  doi:{\changeurlcolor{black}\href{https://doi.org/10.1051/0004-6361/202039973}{\detokenize{10.1051/0004-6361/202039973}}}.

\bibitem[{Wang} \em{et~al.}(2020){Wang}, {Hammer}, {Rejkuba}, {Crnojevi{\'c}},
  and {Yang}]{2020MNRAS.498.2766W}
{Wang}, J.; {Hammer}, F.; {Rejkuba}, M.; {Crnojevi{\'c}}, D.; {Yang}, Y.
\newblock {A recent major merger tale for the closest giant elliptical galaxy
  Centaurus A}.
\newblock {\em \mnras} {\bf 2020}, {\em 498},~2766--2777,
  \href{http://xxx.lanl.gov/abs/2008.13418}{{\normalfont
  [arXiv:astro-ph.GA/2008.13418]}}.
\newblock
  doi:{\changeurlcolor{black}\href{https://doi.org/10.1093/mnras/staa2508}{\detokenize{10.1093/mnras/staa2508}}}.

\bibitem[{Ibata} \em{et~al.}(2014){Ibata}, {Ibata}, {Famaey}, and
  {Lewis}]{2014Natur.511..563I}
{Ibata}, N.G.; {Ibata}, R.A.; {Famaey}, B.; {Lewis}, G.F.
\newblock {Velocity anti-correlation of diametrically opposed galaxy satellites
  in the low-redshift Universe}.
\newblock {\em \nat} {\bf 2014}, {\em 511},~563--566,
  \href{http://xxx.lanl.gov/abs/1407.8178}{{\normalfont
  [arXiv:astro-ph.GA/1407.8178]}}.
\newblock
  doi:{\changeurlcolor{black}\href{https://doi.org/10.1038/nature13481}{\detokenize{10.1038/nature13481}}}.

\bibitem[{Phillips} \em{et~al.}(2015){Phillips}, {Cooper}, {Bullock}, and
  {Boylan-Kolchin}]{2015MNRAS.453.3839P}
{Phillips}, J.I.; {Cooper}, M.C.; {Bullock}, J.S.; {Boylan-Kolchin}, M.
\newblock {Are rotating planes of satellite galaxies ubiquitous?}
\newblock {\em \mnras} {\bf 2015}, {\em 453},~3839--3847,
  \href{http://xxx.lanl.gov/abs/1505.05876}{{\normalfont
  [arXiv:astro-ph.GA/1505.05876]}}.
\newblock
  doi:{\changeurlcolor{black}\href{https://doi.org/10.1093/mnras/stv1770}{\detokenize{10.1093/mnras/stv1770}}}.

\bibitem[{Cautun} \em{et~al.}(2015){Cautun}, {Wang}, {Frenk}, and
  {Sawala}]{2015MNRAS.449.2576C}
{Cautun}, M.; {Wang}, W.; {Frenk}, C.S.; {Sawala}, T.
\newblock {A new spin on discs of satellite galaxies}.
\newblock {\em \mnras} {\bf 2015}, {\em 449},~2576--2587,
  \href{http://xxx.lanl.gov/abs/1410.7778}{{\normalfont
  [arXiv:astro-ph.GA/1410.7778]}}.
\newblock
  doi:{\changeurlcolor{black}\href{https://doi.org/10.1093/mnras/stv490}{\detokenize{10.1093/mnras/stv490}}}.

\bibitem[{Ibata} \em{et~al.}(2015){Ibata}, {Famaey}, {Lewis}, {Ibata}, and
  {Martin}]{2015ApJ...805...67I}
{Ibata}, R.A.; {Famaey}, B.; {Lewis}, G.F.; {Ibata}, N.G.; {Martin}, N.
\newblock {Eppur si Muove: Positional and Kinematic Correlations of Satellite
  Pairs in the Low Z Universe}.
\newblock {\em \apj} {\bf 2015}, {\em 805},~67,
  \href{http://xxx.lanl.gov/abs/1411.3718}{{\normalfont
  [arXiv:astro-ph.GA/1411.3718]}}.
\newblock
  doi:{\changeurlcolor{black}\href{https://doi.org/10.1088/0004-637X/805/1/67}{\detokenize{10.1088/0004-637X/805/1/67}}}.

\bibitem[{Heesters} \em{et~al.}(2021){Heesters}, {Habas}, {Marleau},
  {M{\"u}ller}, {Duc}, {Poulain}, {Durrell}, {S{\'a}nchez-Janssen}, and
  {Paudel}]{2021arXiv210810189H}
{Heesters}, N.; {Habas}, R.; {Marleau}, F.R.; {M{\"u}ller}, O.; {Duc}, P.A.;
  {Poulain}, M.; {Durrell}, P.; {S{\'a}nchez-Janssen}, R.; {Paudel}, S.
\newblock {Flattened structures of dwarf satellites around massive host
  galaxies in the MATLAS low-to-moderate density fields}.
\newblock {\em arXiv e-prints} {\bf 2021}, p. arXiv:2108.10189,
  \href{http://xxx.lanl.gov/abs/2108.10189}{{\normalfont
  [arXiv:astro-ph.GA/2108.10189]}}.

\bibitem[{Kroupa} \em{et~al.}(2005){Kroupa}, {Theis}, and
  {Boily}]{2005A&A...431..517K}
{Kroupa}, P.; {Theis}, C.; {Boily}, C.M.
\newblock {The great disk of Milky-Way satellites and cosmological
  sub-structures}.
\newblock {\em \aap} {\bf 2005}, {\em 431},~517--521,
  \href{http://xxx.lanl.gov/abs/astro-ph/0410421}{{\normalfont
  [arXiv:astro-ph/astro-ph/0410421]}}.
\newblock
  doi:{\changeurlcolor{black}\href{https://doi.org/10.1051/0004-6361:20041122}{\detokenize{10.1051/0004-6361:20041122}}}.

\bibitem[{Zentner} \em{et~al.}(2005){Zentner}, {Kravtsov}, {Gnedin}, and
  {Klypin}]{2005ApJ...629..219Z}
{Zentner}, A.R.; {Kravtsov}, A.V.; {Gnedin}, O.Y.; {Klypin}, A.A.
\newblock {The Anisotropic Distribution of Galactic Satellites}.
\newblock {\em \apj} {\bf 2005}, {\em 629},~219--232,
  \href{http://xxx.lanl.gov/abs/astro-ph/0502496}{{\normalfont
  [arXiv:astro-ph/astro-ph/0502496]}}.
\newblock
  doi:{\changeurlcolor{black}\href{https://doi.org/10.1086/431355}{\detokenize{10.1086/431355}}}.

\bibitem[{Libeskind} \em{et~al.}(2011){Libeskind}, {Knebe}, {Hoffman},
  {Gottl{\"o}ber}, {Yepes}, and {Steinmetz}]{2011MNRAS.411.1525L}
{Libeskind}, N.I.; {Knebe}, A.; {Hoffman}, Y.; {Gottl{\"o}ber}, S.; {Yepes},
  G.; {Steinmetz}, M.
\newblock {The preferred direction of infalling satellite galaxies in the Local
  Group}.
\newblock {\em \mnras} {\bf 2011}, {\em 411},~1525--1535,
  \href{http://xxx.lanl.gov/abs/1010.1531}{{\normalfont
  [arXiv:astro-ph.CO/1010.1531]}}.
\newblock
  doi:{\changeurlcolor{black}\href{https://doi.org/10.1111/j.1365-2966.2010.17786.x}{\detokenize{10.1111/j.1365-2966.2010.17786.x}}}.

\bibitem[{Garaldi} \em{et~al.}(2018){Garaldi}, {Romano-D{\'\i}az},
  {Borzyszkowski}, and {Porciani}]{2018MNRAS.473.2234G}
{Garaldi}, E.; {Romano-D{\'\i}az}, E.; {Borzyszkowski}, M.; {Porciani}, C.
\newblock {ZOMG - III. The effect of halo assembly on the satellite
  population}.
\newblock {\em \mnras} {\bf 2018}, {\em 473},~2234--2250,
  \href{http://xxx.lanl.gov/abs/1707.01108}{{\normalfont
  [arXiv:astro-ph.GA/1707.01108]}}.
\newblock
  doi:{\changeurlcolor{black}\href{https://doi.org/10.1093/mnras/stx2489}{\detokenize{10.1093/mnras/stx2489}}}.

\bibitem[{Pawlowski} \em{et~al.}(2012){Pawlowski}, {Kroupa}, {Angus}, {de
  Boer}, {Famaey}, and {Hensler}]{2012MNRAS.424...80P}
{Pawlowski}, M.S.; {Kroupa}, P.; {Angus}, G.; {de Boer}, K.S.; {Famaey}, B.;
  {Hensler}, G.
\newblock {Filamentary accretion cannot explain the orbital poles of the Milky
  Way satellites}.
\newblock {\em \mnras} {\bf 2012}, {\em 424},~80--92,
  \href{http://xxx.lanl.gov/abs/1204.6039}{{\normalfont
  [arXiv:astro-ph.CO/1204.6039]}}.
\newblock
  doi:{\changeurlcolor{black}\href{https://doi.org/10.1111/j.1365-2966.2012.21169.x}{\detokenize{10.1111/j.1365-2966.2012.21169.x}}}.

\bibitem[{Nelson} \em{et~al.}(2019){Nelson}, {Springel}, {Pillepich},
  {Rodriguez-Gomez}, {Torrey}, {Genel}, {Vogelsberger}, {Pakmor}, {Marinacci},
  {Weinberger}, {Kelley}, {Lovell}, {Diemer}, and
  {Hernquist}]{2019ComAC...6....2N}
{Nelson}, D.; {Springel}, V.; {Pillepich}, A.; {Rodriguez-Gomez}, V.; {Torrey},
  P.; {Genel}, S.; {Vogelsberger}, M.; {Pakmor}, R.; {Marinacci}, F.;
  {Weinberger}, R.; {Kelley}, L.; {Lovell}, M.; {Diemer}, B.; {Hernquist}, L.
\newblock {The IllustrisTNG simulations: public data release}.
\newblock {\em Computational Astrophysics and Cosmology} {\bf 2019}, {\em
  6},~2,  \href{http://xxx.lanl.gov/abs/1812.05609}{{\normalfont
  [arXiv:astro-ph.GA/1812.05609]}}.
\newblock
  doi:{\changeurlcolor{black}\href{https://doi.org/10.1186/s40668-019-0028-x}{\detokenize{10.1186/s40668-019-0028-x}}}.

\bibitem[{Pawlowski} and {Sohn}(2021)]{PawlowskiSohn2021}
{Pawlowski}, M.S.; {Sohn}, S.T.
\newblock {On the Co-Orbitation of Satellite Galaxies Along the Great Plane of
  Andromeda: NGC 147, NGC 185, and Expectations from Cosmological Simulations}.
\newblock {\em \apj} {\bf 2021}, {\em submitted},~1--25.

\bibitem[{Ibata} \em{et~al.}(2014){Ibata}, {Ibata}, {Lewis}, {Martin}, {Conn},
  {Elahi}, {Arias}, and {Fernando}]{2014ApJ...784L...6I}
{Ibata}, R.A.; {Ibata}, N.G.; {Lewis}, G.F.; {Martin}, N.F.; {Conn}, A.;
  {Elahi}, P.; {Arias}, V.; {Fernando}, N.
\newblock {A Thousand Shadows of Andromeda: Rotating Planes of Satellites in
  the Millennium-II Cosmological Simulation}.
\newblock {\em \apjl} {\bf 2014}, {\em 784},~L6,
  \href{http://xxx.lanl.gov/abs/1403.2389}{{\normalfont
  [arXiv:astro-ph.GA/1403.2389]}}.
\newblock
  doi:{\changeurlcolor{black}\href{https://doi.org/10.1088/2041-8205/784/1/L6}{\detokenize{10.1088/2041-8205/784/1/L6}}}.

\bibitem[{Pawlowski} and {McGaugh}(2014)]{2014ApJ...789L..24P}
{Pawlowski}, M.S.; {McGaugh}, S.S.
\newblock {Co-orbiting Planes of Sub-halos are Similarly Unlikely around Paired
  and Isolated Hosts}.
\newblock {\em \apjl} {\bf 2014}, {\em 789},~L24,
  \href{http://xxx.lanl.gov/abs/1406.6062}{{\normalfont
  [arXiv:astro-ph.GA/1406.6062]}}.
\newblock
  doi:{\changeurlcolor{black}\href{https://doi.org/10.1088/2041-8205/789/1/L24}{\detokenize{10.1088/2041-8205/789/1/L24}}}.

\bibitem[{Pawlowski} \em{et~al.}(2019){Pawlowski}, {Bullock}, {Kelley}, and
  {Famaey}]{2019ApJ...875..105P}
{Pawlowski}, M.S.; {Bullock}, J.S.; {Kelley}, T.; {Famaey}, B.
\newblock {Do Halos that Form Early, Have High Concentration, Are Part of a
  Pair, or Contain a Central Galaxy Potential Host More Pronounced Planes of
  Satellite Galaxies?}
\newblock {\em \apj} {\bf 2019}, {\em 875},~105,
  \href{http://xxx.lanl.gov/abs/1903.10513}{{\normalfont
  [arXiv:astro-ph.GA/1903.10513]}}.
\newblock
  doi:{\changeurlcolor{black}\href{https://doi.org/10.3847/1538-4357/ab10e0}{\detokenize{10.3847/1538-4357/ab10e0}}}.

\bibitem[{Samuel} \em{et~al.}(2021){Samuel}, {Wetzel}, {Chapman}, {Tollerud},
  {Hopkins}, {Boylan-Kolchin}, {Bailin}, and
  {Faucher-Gigu{\`e}re}]{2021MNRAS.504.1379S}
{Samuel}, J.; {Wetzel}, A.; {Chapman}, S.; {Tollerud}, E.; {Hopkins}, P.F.;
  {Boylan-Kolchin}, M.; {Bailin}, J.; {Faucher-Gigu{\`e}re}, C.A.
\newblock {Planes of satellites around Milky Way/M31-mass galaxies in the FIRE
  simulations and comparisons with the Local Group}.
\newblock {\em \mnras} {\bf 2021}, {\em 504},~1379--1397,
  \href{http://xxx.lanl.gov/abs/2010.08571}{{\normalfont
  [arXiv:astro-ph.GA/2010.08571]}}.
\newblock
  doi:{\changeurlcolor{black}\href{https://doi.org/10.1093/mnras/stab955}{\detokenize{10.1093/mnras/stab955}}}.

\bibitem[{Shao} \em{et~al.}(2019){Shao}, {Cautun}, and
  {Frenk}]{2019MNRAS.488.1166S}
{Shao}, S.; {Cautun}, M.; {Frenk}, C.S.
\newblock {Evolution of galactic planes of satellites in the EAGLE simulation}.
\newblock {\em \mnras} {\bf 2019}, {\em 488},~1166--1179,
  \href{http://xxx.lanl.gov/abs/1904.02719}{{\normalfont
  [arXiv:astro-ph.GA/1904.02719]}}.
\newblock
  doi:{\changeurlcolor{black}\href{https://doi.org/10.1093/mnras/stz1741}{\detokenize{10.1093/mnras/stz1741}}}.

\bibitem[{Lovell} \em{et~al.}(2011){Lovell}, {Eke}, {Frenk}, and
  {Jenkins}]{2011MNRAS.413.3013L}
{Lovell}, M.R.; {Eke}, V.R.; {Frenk}, C.S.; {Jenkins}, A.
\newblock {The link between galactic satellite orbits and subhalo accretion}.
\newblock {\em \mnras} {\bf 2011}, {\em 413},~3013--3021,
  \href{http://xxx.lanl.gov/abs/1008.0484}{{\normalfont
  [arXiv:astro-ph.CO/1008.0484]}}.
\newblock
  doi:{\changeurlcolor{black}\href{https://doi.org/10.1111/j.1365-2966.2011.18377.x}{\detokenize{10.1111/j.1365-2966.2011.18377.x}}}.

\bibitem[{Wang} \em{et~al.}(2013){Wang}, {Frenk}, and
  {Cooper}]{2013MNRAS.429.1502W}
{Wang}, J.; {Frenk}, C.S.; {Cooper}, A.P.
\newblock {The spatial distribution of galactic satellites in the
  {\ensuremath{\Lambda}} cold dark matter cosmology}.
\newblock {\em \mnras} {\bf 2013}, {\em 429},~1502--1513,
  \href{http://xxx.lanl.gov/abs/1206.1340}{{\normalfont
  [arXiv:astro-ph.GA/1206.1340]}}.
\newblock
  doi:{\changeurlcolor{black}\href{https://doi.org/10.1093/mnras/sts442}{\detokenize{10.1093/mnras/sts442}}}.

\bibitem[{Bahl} and {Baumgardt}(2014)]{2014MNRAS.438.2916B}
{Bahl}, H.; {Baumgardt}, H.
\newblock {A comparison of the distribution of satellite galaxies around
  Andromeda and the results of {\ensuremath{\Lambda}}CDM simulations}.
\newblock {\em \mnras} {\bf 2014}, {\em 438},~2916--2923,
  \href{http://xxx.lanl.gov/abs/1312.3629}{{\normalfont
  [arXiv:astro-ph.GA/1312.3629]}}.
\newblock
  doi:{\changeurlcolor{black}\href{https://doi.org/10.1093/mnras/stt2399}{\detokenize{10.1093/mnras/stt2399}}}.

\bibitem[{Pawlowski} \em{et~al.}(2015){Pawlowski}, {Famaey}, {Merritt}, and
  {Kroupa}]{2015ApJ...815...19P}
{Pawlowski}, M.S.; {Famaey}, B.; {Merritt}, D.; {Kroupa}, P.
\newblock {On the Persistence of Two Small-scale Problems in
  {\ensuremath{\Lambda}}CDM}.
\newblock {\em \apj} {\bf 2015}, {\em 815},~19,
  \href{http://xxx.lanl.gov/abs/1510.08060}{{\normalfont
  [arXiv:astro-ph.GA/1510.08060]}}.
\newblock
  doi:{\changeurlcolor{black}\href{https://doi.org/10.1088/0004-637X/815/1/19}{\detokenize{10.1088/0004-637X/815/1/19}}}.

\bibitem[{Pawlowski}(2018)]{2018MPLA...3330004P}
{Pawlowski}, M.S.
\newblock {The planes of satellite galaxies problem, suggested solutions, and
  open questions}.
\newblock {\em Modern Physics Letters A} {\bf 2018}, {\em 33},~1830004,
  \href{http://xxx.lanl.gov/abs/1802.02579}{{\normalfont
  [arXiv:astro-ph.GA/1802.02579]}}.
\newblock
  doi:{\changeurlcolor{black}\href{https://doi.org/10.1142/S0217732318300045}{\detokenize{10.1142/S0217732318300045}}}.

\bibitem[{Cautun} \em{et~al.}(2015){Cautun}, {Bose}, {Frenk}, {Guo}, {Han},
  {Hellwing}, {Sawala}, and {Wang}]{2015MNRAS.452.3838C}
{Cautun}, M.; {Bose}, S.; {Frenk}, C.S.; {Guo}, Q.; {Han}, J.; {Hellwing},
  W.A.; {Sawala}, T.; {Wang}, W.
\newblock {Planes of satellite galaxies: when exceptions are the rule}.
\newblock {\em \mnras} {\bf 2015}, {\em 452},~3838--3852,
  \href{http://xxx.lanl.gov/abs/1506.04151}{{\normalfont
  [arXiv:astro-ph.GA/1506.04151]}}.
\newblock
  doi:{\changeurlcolor{black}\href{https://doi.org/10.1093/mnras/stv1557}{\detokenize{10.1093/mnras/stv1557}}}.

\bibitem[{Pawlowski} \em{et~al.}(2017){Pawlowski}, {Dabringhausen}, {Famaey},
  {Flores}, {Hammer}, {Hensler}, {Ibata}, {Kroupa}, {Lewis}, {Libeskind},
  {McGaugh}, {Merritt}, {Puech}, and {Yang}]{2017AN....338..854P}
{Pawlowski}, M.S.; {Dabringhausen}, J.; {Famaey}, B.; {Flores}, H.; {Hammer},
  F.; {Hensler}, G.; {Ibata}, R.A.; {Kroupa}, P.; {Lewis}, G.F.; {Libeskind},
  N.I.; {McGaugh}, S.S.; {Merritt}, D.; {Puech}, M.; {Yang}, Y.
\newblock {Considerations on how to investigate planes of satellite galaxies}.
\newblock {\em Astronomische Nachrichten} {\bf 2017}, {\em 338},~854--861,
  \href{http://xxx.lanl.gov/abs/1702.06143}{{\normalfont
  [arXiv:astro-ph.GA/1702.06143]}}.
\newblock
  doi:{\changeurlcolor{black}\href{https://doi.org/10.1002/asna.201713366}{\detokenize{10.1002/asna.201713366}}}.

\bibitem[{Libeskind} \em{et~al.}(2014){Libeskind}, {Knebe}, {Hoffman}, and
  {Gottl{\"o}ber}]{2014MNRAS.443.1274L}
{Libeskind}, N.I.; {Knebe}, A.; {Hoffman}, Y.; {Gottl{\"o}ber}, S.
\newblock {The universal nature of subhalo accretion}.
\newblock {\em \mnras} {\bf 2014}, {\em 443},~1274--1280,
  \href{http://xxx.lanl.gov/abs/1407.0394}{{\normalfont
  [arXiv:astro-ph.CO/1407.0394]}}.
\newblock
  doi:{\changeurlcolor{black}\href{https://doi.org/10.1093/mnras/stu1216}{\detokenize{10.1093/mnras/stu1216}}}.

\bibitem[{Shao} \em{et~al.}(2018){Shao}, {Cautun}, {Frenk}, {Grand},
  {G{\'o}mez}, {Marinacci}, and {Simpson}]{2018MNRAS.476.1796S}
{Shao}, S.; {Cautun}, M.; {Frenk}, C.S.; {Grand}, R.J.J.; {G{\'o}mez}, F.A.;
  {Marinacci}, F.; {Simpson}, C.M.
\newblock {The multiplicity and anisotropy of galactic satellite accretion}.
\newblock {\em \mnras} {\bf 2018}, {\em 476},~1796--1810,
  \href{http://xxx.lanl.gov/abs/1712.05409}{{\normalfont
  [arXiv:astro-ph.GA/1712.05409]}}.
\newblock
  doi:{\changeurlcolor{black}\href{https://doi.org/10.1093/mnras/sty343}{\detokenize{10.1093/mnras/sty343}}}.

\bibitem[{Pawlowski} \em{et~al.}(2011){Pawlowski}, {Kroupa}, and {de
  Boer}]{2011A&A...532A.118P}
{Pawlowski}, M.S.; {Kroupa}, P.; {de Boer}, K.S.
\newblock {Making counter-orbiting tidal debris. The origin of the Milky Way
  disc of satellites?}
\newblock {\em \aap} {\bf 2011}, {\em 532},~A118,
  \href{http://xxx.lanl.gov/abs/1106.2804}{{\normalfont
  [arXiv:astro-ph.CO/1106.2804]}}.
\newblock
  doi:{\changeurlcolor{black}\href{https://doi.org/10.1051/0004-6361/201015021}{\detokenize{10.1051/0004-6361/201015021}}}.

\bibitem[{Milgrom}(1983)]{1983ApJ...270..365M}
{Milgrom}, M.
\newblock {A modification of the Newtonian dynamics as a possible alternative
  to the hidden mass hypothesis.}
\newblock {\em \apj} {\bf 1983}, {\em 270},~365--370.
\newblock
  doi:{\changeurlcolor{black}\href{https://doi.org/10.1086/161130}{\detokenize{10.1086/161130}}}.

\bibitem[{Zhao} \em{et~al.}(2013){Zhao}, {Famaey}, {L{\"u}ghausen}, and
  {Kroupa}]{2013A&A...557L...3Z}
{Zhao}, H.; {Famaey}, B.; {L{\"u}ghausen}, F.; {Kroupa}, P.
\newblock {Local Group timing in Milgromian dynamics. A past Milky
  Way-Andromeda encounter at z > 0.8}.
\newblock {\em \aap} {\bf 2013}, {\em 557},~L3,
  \href{http://xxx.lanl.gov/abs/1306.6628}{{\normalfont
  [arXiv:astro-ph.GA/1306.6628]}}.
\newblock
  doi:{\changeurlcolor{black}\href{https://doi.org/10.1051/0004-6361/201321879}{\detokenize{10.1051/0004-6361/201321879}}}.

\bibitem[{B{\'\i}lek} \em{et~al.}(2018){B{\'\i}lek}, {Thies}, {Kroupa}, and
  {Famaey}]{2018A&A...614A..59B}
{B{\'\i}lek}, M.; {Thies}, I.; {Kroupa}, P.; {Famaey}, B.
\newblock {MOND simulation suggests an origin for some peculiarities in the
  Local Group}.
\newblock {\em \aap} {\bf 2018}, {\em 614},~A59,
  \href{http://xxx.lanl.gov/abs/1712.04938}{{\normalfont
  [arXiv:astro-ph.GA/1712.04938]}}.
\newblock
  doi:{\changeurlcolor{black}\href{https://doi.org/10.1051/0004-6361/201731939}{\detokenize{10.1051/0004-6361/201731939}}}.

\bibitem[{Banik} \em{et~al.}(2018){Banik}, {O'Ryan}, and
  {Zhao}]{2018MNRAS.477.4768B}
{Banik}, I.; {O'Ryan}, D.; {Zhao}, H.
\newblock {Origin of the Local Group satellite planes}.
\newblock {\em \mnras} {\bf 2018}, {\em 477},~4768--4791,
  \href{http://xxx.lanl.gov/abs/1802.00440}{{\normalfont
  [arXiv:astro-ph.GA/1802.00440]}}.
\newblock
  doi:{\changeurlcolor{black}\href{https://doi.org/10.1093/mnras/sty919}{\detokenize{10.1093/mnras/sty919}}}.

\bibitem[{Kroupa}(1997)]{1997NewA....2..139K}
{Kroupa}, P.
\newblock {Dwarf spheroidal satellite galaxies without dark matter}.
\newblock {\em \na} {\bf 1997}, {\em 2},~139--164.
\newblock
  doi:{\changeurlcolor{black}\href{https://doi.org/10.1016/S1384-1076(97)00012-2}{\detokenize{10.1016/S1384-1076(97)00012-2}}}.

\bibitem[{Klessen} and {Kroupa}(1998)]{1998ApJ...498..143K}
{Klessen}, R.S.; {Kroupa}, P.
\newblock {Dwarf Spheroidal Satellite Galaxies without Dark Matter: Results
  from Two Different Numerical Techniques}.
\newblock {\em \apj} {\bf 1998}, {\em 498},~143--155,
  \href{http://xxx.lanl.gov/abs/astro-ph/9711350}{{\normalfont
  [arXiv:astro-ph/astro-ph/9711350]}}.
\newblock
  doi:{\changeurlcolor{black}\href{https://doi.org/10.1086/305540}{\detokenize{10.1086/305540}}}.

\bibitem[{Hammer} \em{et~al.}(2013){Hammer}, {Yang}, {Fouquet}, {Pawlowski},
  {Kroupa}, {Puech}, {Flores}, and {Wang}]{2013MNRAS.431.3543H}
{Hammer}, F.; {Yang}, Y.; {Fouquet}, S.; {Pawlowski}, M.S.; {Kroupa}, P.;
  {Puech}, M.; {Flores}, H.; {Wang}, J.
\newblock {The vast thin plane of M31 corotating dwarfs: an additional fossil
  signature of the M31 merger and of its considerable impact in the whole Local
  Group}.
\newblock {\em \mnras} {\bf 2013}, {\em 431},~3543--3549,
  \href{http://xxx.lanl.gov/abs/1303.1817}{{\normalfont
  [arXiv:astro-ph.CO/1303.1817]}}.
\newblock
  doi:{\changeurlcolor{black}\href{https://doi.org/10.1093/mnras/stt435}{\detokenize{10.1093/mnras/stt435}}}.

\bibitem[{Yang} \em{et~al.}(2014){Yang}, {Hammer}, {Fouquet}, {Flores},
  {Puech}, {Pawlowski}, and {Kroupa}]{2014MNRAS.442.2419Y}
{Yang}, Y.; {Hammer}, F.; {Fouquet}, S.; {Flores}, H.; {Puech}, M.;
  {Pawlowski}, M.S.; {Kroupa}, P.
\newblock {Reproducing properties of MW dSphs as descendants of DM-free TDGs}.
\newblock {\em \mnras} {\bf 2014}, {\em 442},~2419--2433,
  \href{http://xxx.lanl.gov/abs/1405.2071}{{\normalfont
  [arXiv:astro-ph.GA/1405.2071]}}.
\newblock
  doi:{\changeurlcolor{black}\href{https://doi.org/10.1093/mnras/stu931}{\detokenize{10.1093/mnras/stu931}}}.

\bibitem[{Hammer} \em{et~al.}(2019){Hammer}, {Yang}, {Wang}, {Arenou}, {Puech},
  {Flores}, and {Babusiaux}]{2019ApJ...883..171H}
{Hammer}, F.; {Yang}, Y.; {Wang}, J.; {Arenou}, F.; {Puech}, M.; {Flores}, H.;
  {Babusiaux}, C.
\newblock {On the Absence of Dark Matter in Dwarf Galaxies Surrounding the
  Milky Way}.
\newblock {\em \apj} {\bf 2019}, {\em 883},~171,
  \href{http://xxx.lanl.gov/abs/1812.10714}{{\normalfont
  [arXiv:astro-ph.GA/1812.10714]}}.
\newblock
  doi:{\changeurlcolor{black}\href{https://doi.org/10.3847/1538-4357/ab36b6}{\detokenize{10.3847/1538-4357/ab36b6}}}.

\bibitem[{Massari} \em{et~al.}(2018){Massari}, {Breddels}, {Helmi}, {Posti},
  {Brown}, and {Tolstoy}]{2018NatAs...2..156M}
{Massari}, D.; {Breddels}, M.A.; {Helmi}, A.; {Posti}, L.; {Brown}, A.G.A.;
  {Tolstoy}, E.
\newblock {Three-dimensional motions in the Sculptor dwarf galaxy as a glimpse
  of a new era}.
\newblock {\em Nature Astronomy} {\bf 2018}, {\em 2},~156--161,
  \href{http://xxx.lanl.gov/abs/1711.08945}{{\normalfont
  [arXiv:astro-ph.GA/1711.08945]}}.
\newblock
  doi:{\changeurlcolor{black}\href{https://doi.org/10.1038/s41550-017-0322-y}{\detokenize{10.1038/s41550-017-0322-y}}}.

\bibitem[{Massari} \em{et~al.}(2020){Massari}, {Helmi}, {Mucciarelli}, {Sales},
  {Spina}, and {Tolstoy}]{2020A&A...633A..36M}
{Massari}, D.; {Helmi}, A.; {Mucciarelli}, A.; {Sales}, L.V.; {Spina}, L.;
  {Tolstoy}, E.
\newblock {Stellar 3D kinematics in the Draco dwarf spheroidal galaxy}.
\newblock {\em \aap} {\bf 2020}, {\em 633},~A36,
  \href{http://xxx.lanl.gov/abs/1904.04037}{{\normalfont
  [arXiv:astro-ph.GA/1904.04037]}}.
\newblock
  doi:{\changeurlcolor{black}\href{https://doi.org/10.1051/0004-6361/201935613}{\detokenize{10.1051/0004-6361/201935613}}}.

\bibitem[{Li} and {Helmi}(2008)]{2008MNRAS.385.1365L}
{Li}, Y.S.; {Helmi}, A.
\newblock {Infall of substructures on to a Milky Way-like dark halo}.
\newblock {\em \mnras} {\bf 2008}, {\em 385},~1365--1373,
  \href{http://xxx.lanl.gov/abs/0711.2429}{{\normalfont
  [arXiv:astro-ph/0711.2429]}}.
\newblock
  doi:{\changeurlcolor{black}\href{https://doi.org/10.1111/j.1365-2966.2008.12854.x}{\detokenize{10.1111/j.1365-2966.2008.12854.x}}}.

\bibitem[{D'Onghia} and {Lake}(2008)]{2008ApJ...686L..61D}
{D'Onghia}, E.; {Lake}, G.
\newblock {Small Dwarf Galaxies within Larger Dwarfs: Why Some Are Luminous
  while Most Go Dark}.
\newblock {\em \apjl} {\bf 2008}, {\em 686},~L61,
  \href{http://xxx.lanl.gov/abs/0802.0001}{{\normalfont
  [arXiv:astro-ph/0802.0001]}}.
\newblock
  doi:{\changeurlcolor{black}\href{https://doi.org/10.1086/592995}{\detokenize{10.1086/592995}}}.

\bibitem[{Sales} \em{et~al.}(2017){Sales}, {Navarro}, {Kallivayalil}, and
  {Frenk}]{2017MNRAS.465.1879S}
{Sales}, L.V.; {Navarro}, J.F.; {Kallivayalil}, N.; {Frenk}, C.S.
\newblock {Identifying true satellites of the Magellanic Clouds}.
\newblock {\em \mnras} {\bf 2017}, {\em 465},~1879--1888,
  \href{http://xxx.lanl.gov/abs/1605.03574}{{\normalfont
  [arXiv:astro-ph.GA/1605.03574]}}.
\newblock
  doi:{\changeurlcolor{black}\href{https://doi.org/10.1093/mnras/stw2816}{\detokenize{10.1093/mnras/stw2816}}}.

\bibitem[{Erkal} and {Belokurov}(2020)]{2020MNRAS.495.2554E}
{Erkal}, D.; {Belokurov}, V.A.
\newblock {Limit on the LMC mass from a census of its satellites}.
\newblock {\em \mnras} {\bf 2020}, {\em 495},~2554--2563,
  \href{http://xxx.lanl.gov/abs/1907.09484}{{\normalfont
  [arXiv:astro-ph.GA/1907.09484]}}.
\newblock
  doi:{\changeurlcolor{black}\href{https://doi.org/10.1093/mnras/staa1238}{\detokenize{10.1093/mnras/staa1238}}}.

\bibitem[{Nichols} \em{et~al.}(2011){Nichols}, {Colless}, {Colless}, and
  {Bland-Hawthorn}]{2011ApJ...742..110N}
{Nichols}, M.; {Colless}, J.; {Colless}, M.; {Bland-Hawthorn}, J.
\newblock {Accretion of the Magellanic System onto the Galaxy}.
\newblock {\em \apj} {\bf 2011}, {\em 742},~110,
  \href{http://xxx.lanl.gov/abs/1110.2784}{{\normalfont
  [arXiv:astro-ph.CO/1110.2784]}}.
\newblock
  doi:{\changeurlcolor{black}\href{https://doi.org/10.1088/0004-637X/742/2/110}{\detokenize{10.1088/0004-637X/742/2/110}}}.

\bibitem[{Santos-Santos} \em{et~al.}(2021){Santos-Santos}, {Fattahi}, {Sales},
  and {Navarro}]{2021MNRAS.504.4551S}
{Santos-Santos}, I.M.E.; {Fattahi}, A.; {Sales}, L.V.; {Navarro}, J.F.
\newblock {Magellanic satellites in {\ensuremath{\Lambda}}CDM cosmological
  hydrodynamical simulations of the Local Group}.
\newblock {\em \mnras} {\bf 2021}, {\em 504},~4551--4567,
  \href{http://xxx.lanl.gov/abs/2011.13500}{{\normalfont
  [arXiv:astro-ph.GA/2011.13500]}}.
\newblock
  doi:{\changeurlcolor{black}\href{https://doi.org/10.1093/mnras/stab1020}{\detokenize{10.1093/mnras/stab1020}}}.

\bibitem[{Wetzel} \em{et~al.}(2015){Wetzel}, {Deason}, and
  {Garrison-Kimmel}]{2015ApJ...807...49W}
{Wetzel}, A.R.; {Deason}, A.J.; {Garrison-Kimmel}, S.
\newblock {Satellite Dwarf Galaxies in a Hierarchical Universe: Infall
  Histories, Group Preprocessing, and Reionization}.
\newblock {\em \apj} {\bf 2015}, {\em 807},~49,
  \href{http://xxx.lanl.gov/abs/1501.01972}{{\normalfont
  [arXiv:astro-ph.GA/1501.01972]}}.
\newblock
  doi:{\changeurlcolor{black}\href{https://doi.org/10.1088/0004-637X/807/1/49}{\detokenize{10.1088/0004-637X/807/1/49}}}.

\bibitem[{Lynden-Bell} and {Lynden-Bell}(1995)]{1995MNRAS.275..429L}
{Lynden-Bell}, D.; {Lynden-Bell}, R.M.
\newblock {Ghostly streams from the formation of the Galaxy's halo}.
\newblock {\em \mnras} {\bf 1995}, {\em 275},~429--442.
\newblock
  doi:{\changeurlcolor{black}\href{https://doi.org/10.1093/mnras/275.2.429}{\detokenize{10.1093/mnras/275.2.429}}}.

\bibitem[{Torrealba} \em{et~al.}(2016){Torrealba}, {Koposov}, {Belokurov}, and
  {Irwin}]{2016MNRAS.459.2370T}
{Torrealba}, G.; {Koposov}, S.E.; {Belokurov}, V.; {Irwin}, M.
\newblock {The feeble giant. Discovery of a large and diffuse Milky Way dwarf
  galaxy in the constellation of Crater}.
\newblock {\em \mnras} {\bf 2016}, {\em 459},~2370--2378,
  \href{http://xxx.lanl.gov/abs/1601.07178}{{\normalfont
  [arXiv:astro-ph.GA/1601.07178]}}.
\newblock
  doi:{\changeurlcolor{black}\href{https://doi.org/10.1093/mnras/stw733}{\detokenize{10.1093/mnras/stw733}}}.

\bibitem[{Angus} \em{et~al.}(2016){Angus}, {Coppin}, {Gentile}, and
  {Diaferio}]{2016MNRAS.462.3221A}
{Angus}, G.W.; {Coppin}, P.; {Gentile}, G.; {Diaferio}, A.
\newblock {The potential role of NGC 205 in generating Andromeda's vast thin
  corotating plane of satellite galaxies}.
\newblock {\em \mnras} {\bf 2016}, {\em 462},~3221--3242,
  \href{http://xxx.lanl.gov/abs/1608.03763}{{\normalfont
  [arXiv:astro-ph.GA/1608.03763]}}.
\newblock
  doi:{\changeurlcolor{black}\href{https://doi.org/10.1093/mnras/stw1822}{\detokenize{10.1093/mnras/stw1822}}}.

\bibitem[{Libeskind} \em{et~al.}(2016){Libeskind}, {Guo}, {Tempel}, and
  {Ibata}]{2016ApJ...830..121L}
{Libeskind}, N.I.; {Guo}, Q.; {Tempel}, E.; {Ibata}, R.
\newblock {The Lopsided Distribution of Satellite Galaxies}.
\newblock {\em \apj} {\bf 2016}, {\em 830},~121,
  \href{http://xxx.lanl.gov/abs/1606.01516}{{\normalfont
  [arXiv:astro-ph.GA/1606.01516]}}.
\newblock
  doi:{\changeurlcolor{black}\href{https://doi.org/10.3847/0004-637X/830/2/121}{\detokenize{10.3847/0004-637X/830/2/121}}}.

\bibitem[{Pawlowski} \em{et~al.}(2017){Pawlowski}, {Ibata}, and
  {Bullock}]{2017ApJ...850..132P}
{Pawlowski}, M.S.; {Ibata}, R.A.; {Bullock}, J.S.
\newblock {The Lopsidedness of Satellite Galaxy Systems in
  {\ensuremath{\Lambda}}CDM Simulations}.
\newblock {\em \apj} {\bf 2017}, {\em 850},~132,
  \href{http://xxx.lanl.gov/abs/1710.07639}{{\normalfont
  [arXiv:astro-ph.GA/1710.07639]}}.
\newblock
  doi:{\changeurlcolor{black}\href{https://doi.org/10.3847/1538-4357/aa9435}{\detokenize{10.3847/1538-4357/aa9435}}}.

\bibitem[{Gong} \em{et~al.}(2019){Gong}, {Libeskind}, {Tempel}, {Guo},
  {Gottl{\"o}ber}, {Yepes}, {Wang}, {Sorce}, and
  {Pawlowski}]{2019MNRAS.488.3100G}
{Gong}, C.C.; {Libeskind}, N.I.; {Tempel}, E.; {Guo}, Q.; {Gottl{\"o}ber}, S.;
  {Yepes}, G.; {Wang}, P.; {Sorce}, J.; {Pawlowski}, M.
\newblock {The origin of lopsided satellite galaxy distribution in galaxy
  pairs}.
\newblock {\em \mnras} {\bf 2019}, {\em 488},~3100--3108,
  \href{http://xxx.lanl.gov/abs/1906.06128}{{\normalfont
  [arXiv:astro-ph.GA/1906.06128]}}.
\newblock
  doi:{\changeurlcolor{black}\href{https://doi.org/10.1093/mnras/stz1917}{\detokenize{10.1093/mnras/stz1917}}}.

\bibitem[{Brainerd} and {Samuels}(2020)]{2020ApJ...898L..15B}
{Brainerd}, T.G.; {Samuels}, A.
\newblock {Lopsided Satellite Distributions around Isolated Host Galaxies}.
\newblock {\em \apjl} {\bf 2020}, {\em 898},~L15,
  \href{http://xxx.lanl.gov/abs/2007.04703}{{\normalfont
  [arXiv:astro-ph.GA/2007.04703]}}.
\newblock
  doi:{\changeurlcolor{black}\href{https://doi.org/10.3847/2041-8213/aba194}{\detokenize{10.3847/2041-8213/aba194}}}.

\bibitem[{Wang} \em{et~al.}(2021){Wang}, {Libeskind}, {Pawlowski}, {Kang},
  {Wang}, {Guo}, and {Tempel}]{2021ApJ...914...78W}
{Wang}, P.; {Libeskind}, N.I.; {Pawlowski}, M.S.; {Kang}, X.; {Wang}, W.;
  {Guo}, Q.; {Tempel}, E.
\newblock {The Lopsided Distribution of Satellites of Isolated Central
  Galaxies}.
\newblock {\em \apj} {\bf 2021}, {\em 914},~78,
  \href{http://xxx.lanl.gov/abs/2104.12787}{{\normalfont
  [arXiv:astro-ph.GA/2104.12787]}}.
\newblock
  doi:{\changeurlcolor{black}\href{https://doi.org/10.3847/1538-4357/abfc4f}{\detokenize{10.3847/1538-4357/abfc4f}}}.

\bibitem[{Holmberg}(1969)]{1969ArA.....5..305H}
{Holmberg}, E.
\newblock {A study of physical groups of galaxies}.
\newblock {\em Arkiv for Astronomi} {\bf 1969}, {\em 5},~305--343.

\bibitem[{Yang} \em{et~al.}(2006){Yang}, {van den Bosch}, {Mo}, {Mao}, {Kang},
  {Weinmann}, {Guo}, and {Jing}]{2006MNRAS.369.1293Y}
{Yang}, X.; {van den Bosch}, F.C.; {Mo}, H.J.; {Mao}, S.; {Kang}, X.;
  {Weinmann}, S.M.; {Guo}, Y.; {Jing}, Y.P.
\newblock {The alignment between the distribution of satellites and the
  orientation of their central galaxy}.
\newblock {\em \mnras} {\bf 2006}, {\em 369},~1293--1302,
  \href{http://xxx.lanl.gov/abs/astro-ph/0601040}{{\normalfont
  [arXiv:astro-ph/astro-ph/0601040]}}.
\newblock
  doi:{\changeurlcolor{black}\href{https://doi.org/10.1111/j.1365-2966.2006.10373.x}{\detokenize{10.1111/j.1365-2966.2006.10373.x}}}.

\bibitem[{Bovy}(2015)]{2015ApJS..216...29B}
{Bovy}, J.
\newblock {galpy: A python Library for Galactic Dynamics}.
\newblock {\em \apjs} {\bf 2015}, {\em 216},~29,
  \href{http://xxx.lanl.gov/abs/1412.3451}{{\normalfont
  [arXiv:astro-ph.GA/1412.3451]}}.
\newblock
  doi:{\changeurlcolor{black}\href{https://doi.org/10.1088/0067-0049/216/2/29}{\detokenize{10.1088/0067-0049/216/2/29}}}.

\bibitem[{Besla} \em{et~al.}(2012){Besla}, {Kallivayalil}, {Hernquist}, {van
  der Marel}, {Cox}, and {Kere{\v{s}}}]{2012MNRAS.421.2109B}
{Besla}, G.; {Kallivayalil}, N.; {Hernquist}, L.; {van der Marel}, R.P.; {Cox},
  T.J.; {Kere{\v{s}}}, D.
\newblock {The role of dwarf galaxy interactions in shaping the Magellanic
  System and implications for Magellanic Irregulars}.
\newblock {\em \mnras} {\bf 2012}, {\em 421},~2109--2138,
  \href{http://xxx.lanl.gov/abs/1201.1299}{{\normalfont
  [arXiv:astro-ph.GA/1201.1299]}}.
\newblock
  doi:{\changeurlcolor{black}\href{https://doi.org/10.1111/j.1365-2966.2012.20466.x}{\detokenize{10.1111/j.1365-2966.2012.20466.x}}}.

\bibitem[{van den Bergh}(1998)]{1998AJ....116.1688V}
{van den Bergh}, S.
\newblock {The Binary Galaxies NGC 147 and NGC 185}.
\newblock {\em \aj} {\bf 1998}, {\em 116},~1688--1689,
  \href{http://xxx.lanl.gov/abs/astro-ph/9807083}{{\normalfont
  [arXiv:astro-ph/astro-ph/9807083]}}.
\newblock
  doi:{\changeurlcolor{black}\href{https://doi.org/10.1086/300576}{\detokenize{10.1086/300576}}}.

\bibitem[{Geha} \em{et~al.}(2010){Geha}, {van der Marel}, {Guhathakurta},
  {Gilbert}, {Kalirai}, and {Kirby}]{2010ApJ...711..361G}
{Geha}, M.; {van der Marel}, R.P.; {Guhathakurta}, P.; {Gilbert}, K.M.;
  {Kalirai}, J.; {Kirby}, E.N.
\newblock {Local Group Dwarf Elliptical Galaxies. II. Stellar Kinematics to
  Large Radii in NGC 147 and NGC 185}.
\newblock {\em \apj} {\bf 2010}, {\em 711},~361--373,
  \href{http://xxx.lanl.gov/abs/0911.3654}{{\normalfont
  [arXiv:astro-ph.CO/0911.3654]}}.
\newblock
  doi:{\changeurlcolor{black}\href{https://doi.org/10.1088/0004-637X/711/1/361}{\detokenize{10.1088/0004-637X/711/1/361}}}.

\bibitem[{Belokurov} \em{et~al.}(2008){Belokurov}, {Walker}, {Evans}, {Faria},
  {Gilmore}, {Irwin}, {Koposov}, {Mateo}, {Olszewski}, and
  {Zucker}]{2008ApJ...686L..83B}
{Belokurov}, V.; {Walker}, M.G.; {Evans}, N.W.; {Faria}, D.C.; {Gilmore}, G.;
  {Irwin}, M.J.; {Koposov}, S.; {Mateo}, M.; {Olszewski}, E.; {Zucker}, D.B.
\newblock {Leo V: A Companion of a Companion of the Milky Way Galaxy?}
\newblock {\em \apjl} {\bf 2008}, {\em 686},~L83,
  \href{http://xxx.lanl.gov/abs/0807.2831}{{\normalfont
  [arXiv:astro-ph/0807.2831]}}.
\newblock
  doi:{\changeurlcolor{black}\href{https://doi.org/10.1086/592962}{\detokenize{10.1086/592962}}}.

\bibitem[{de Jong} \em{et~al.}(2010){de Jong}, {Martin}, {Rix}, {Smith}, {Jin},
  and {Macci{\`o}}]{2010ApJ...710.1664D}
{de Jong}, J.T.A.; {Martin}, N.F.; {Rix}, H.W.; {Smith}, K.W.; {Jin}, S.;
  {Macci{\`o}}, A.V.
\newblock {The Enigmatic Pair of Dwarf Galaxies Leo IV and Leo V: Coincidence
  or Common Origin?}
\newblock {\em \apj} {\bf 2010}, {\em 710},~1664--1671,
  \href{http://xxx.lanl.gov/abs/0912.3251}{{\normalfont
  [arXiv:astro-ph.CO/0912.3251]}}.
\newblock
  doi:{\changeurlcolor{black}\href{https://doi.org/10.1088/0004-637X/710/2/1664}{\detokenize{10.1088/0004-637X/710/2/1664}}}.

\bibitem[{Bla{\~n}a} \em{et~al.}(2012){Bla{\~n}a}, {Fellhauer}, and
  {Smith}]{2012A&A...542A..61B}
{Bla{\~n}a}, M.; {Fellhauer}, M.; {Smith}, R.
\newblock {Leo IV and V - A possible dwarf galaxy pair?}
\newblock {\em \aap} {\bf 2012}, {\em 542},~A61,
  \href{http://xxx.lanl.gov/abs/1204.1285}{{\normalfont
  [arXiv:astro-ph.GA/1204.1285]}}.
\newblock
  doi:{\changeurlcolor{black}\href{https://doi.org/10.1051/0004-6361/201118442}{\detokenize{10.1051/0004-6361/201118442}}}.

\bibitem[{Fattahi} \em{et~al.}(2013){Fattahi}, {Navarro}, {Starkenburg},
  {Barber}, and {McConnachie}]{2013MNRAS.431L..73F}
{Fattahi}, A.; {Navarro}, J.F.; {Starkenburg}, E.; {Barber}, C.R.;
  {McConnachie}, A.W.
\newblock {Galaxy pairs in the local group.}
\newblock {\em \mnras} {\bf 2013}, {\em 431},~L73--L77,
  \href{http://xxx.lanl.gov/abs/1211.3161}{{\normalfont
  [arXiv:astro-ph.CO/1211.3161]}}.
\newblock
  doi:{\changeurlcolor{black}\href{https://doi.org/10.1093/mnrasl/slt011}{\detokenize{10.1093/mnrasl/slt011}}}.

\bibitem[{Geha} \em{et~al.}(2015){Geha}, {Weisz}, {Grocholski}, {Dolphin}, {van
  der Marel}, and {Guhathakurta}]{2015ApJ...811..114G}
{Geha}, M.; {Weisz}, D.; {Grocholski}, A.; {Dolphin}, A.; {van der Marel},
  R.P.; {Guhathakurta}, P.
\newblock {HST/ACS Direct Ages of the Dwarf Elliptical Galaxies NGC 147 and NGC
  185}.
\newblock {\em \apj} {\bf 2015}, {\em 811},~114,
  \href{http://xxx.lanl.gov/abs/1503.06526}{{\normalfont
  [arXiv:astro-ph.GA/1503.06526]}}.
\newblock
  doi:{\changeurlcolor{black}\href{https://doi.org/10.1088/0004-637X/811/2/114}{\detokenize{10.1088/0004-637X/811/2/114}}}.

\bibitem[{Evslin}(2014)]{2014MNRAS.440.1225E}
{Evslin}, J.
\newblock {Binary satellite galaxies}.
\newblock {\em \mnras} {\bf 2014}, {\em 440},~1225--1232,
  \href{http://xxx.lanl.gov/abs/1310.6896}{{\normalfont
  [arXiv:astro-ph.CO/1310.6896]}}.
\newblock
  doi:{\changeurlcolor{black}\href{https://doi.org/10.1093/mnras/stu340}{\detokenize{10.1093/mnras/stu340}}}.

\bibitem[{Cautun} and {Frenk}(2017)]{2017MNRAS.468L..41C}
{Cautun}, M.; {Frenk}, C.S.
\newblock {The tangential velocity excess of the Milky Way satellites}.
\newblock {\em \mnras} {\bf 2017}, {\em 468},~L41--L45,
  \href{http://xxx.lanl.gov/abs/1612.01529}{{\normalfont
  [arXiv:astro-ph.GA/1612.01529]}}.
\newblock
  doi:{\changeurlcolor{black}\href{https://doi.org/10.1093/mnrasl/slx025}{\detokenize{10.1093/mnrasl/slx025}}}.

\bibitem[{Riley} \em{et~al.}(2019){Riley}, {Fattahi}, {Pace}, {Strigari},
  {Frenk}, {G{\'o}mez}, {Grand}, {Marinacci}, {Navarro}, {Pakmor}, {Simpson},
  and {White}]{2019MNRAS.486.2679R}
{Riley}, A.H.; {Fattahi}, A.; {Pace}, A.B.; {Strigari}, L.E.; {Frenk}, C.S.;
  {G{\'o}mez}, F.A.; {Grand}, R.J.J.; {Marinacci}, F.; {Navarro}, J.F.;
  {Pakmor}, R.; {Simpson}, C.M.; {White}, S.D.M.
\newblock {The velocity anisotropy of the Milky Way satellite system}.
\newblock {\em \mnras} {\bf 2019}, {\em 486},~2679--2694,
  \href{http://xxx.lanl.gov/abs/1810.10645}{{\normalfont
  [arXiv:astro-ph.GA/1810.10645]}}.
\newblock
  doi:{\changeurlcolor{black}\href{https://doi.org/10.1093/mnras/stz973}{\detokenize{10.1093/mnras/stz973}}}.

\bibitem[{Drlica-Wagner} \em{et~al.}(2020){Drlica-Wagner}, {Bechtol}, {Mau},
  {McNanna}, {Nadler}, {Pace}, {Li}, {Pieres}, {Rozo}, {Simon}, {Walker},
  {Wechsler}, {Abbott}, {Allam}, {Annis}, {Bertin}, {Brooks}, {Burke},
  {Rosell}, {Carrasco Kind}, {Carretero}, {Costanzi}, {da Costa}, {De Vicente},
  {Desai}, {Diehl}, {Doel}, {Eifler}, {Everett}, {Flaugher}, {Frieman},
  {Garc{\'\i}a-Bellido}, {Gaztanaga}, {Gruen}, {Gruendl}, {Gschwend},
  {Gutierrez}, {Honscheid}, {James}, {Krause}, {Kuehn}, {Kuropatkin}, {Lahav},
  {Maia}, {Marshall}, {Melchior}, {Menanteau}, {Miquel}, {Palmese}, {Plazas},
  {Sanchez}, {Scarpine}, {Schubnell}, {Serrano}, {Sevilla-Noarbe}, {Smith},
  {Suchyta}, {Tarle}, and {DES Collaboration}]{2020ApJ...893...47D}
{Drlica-Wagner}, A.; {Bechtol}, K.; {Mau}, S.; {McNanna}, M.; {Nadler}, E.O.;
  {Pace}, A.B.; {Li}, T.S.; {Pieres}, A.; {Rozo}, E.; {Simon}, J.D.; {Walker},
  A.R.; {Wechsler}, R.H.; {Abbott}, T.M.C.; {Allam}, S.; {Annis}, J.; {Bertin},
  E.; {Brooks}, D.; {Burke}, D.L.; {Rosell}, A.C.; {Carrasco Kind}, M.;
  {Carretero}, J.; {Costanzi}, M.; {da Costa}, L.N.; {De Vicente}, J.; {Desai},
  S.; {Diehl}, H.T.; {Doel}, P.; {Eifler}, T.F.; {Everett}, S.; {Flaugher}, B.;
  {Frieman}, J.; {Garc{\'\i}a-Bellido}, J.; {Gaztanaga}, E.; {Gruen}, D.;
  {Gruendl}, R.A.; {Gschwend}, J.; {Gutierrez}, G.; {Honscheid}, K.; {James},
  D.J.; {Krause}, E.; {Kuehn}, K.; {Kuropatkin}, N.; {Lahav}, O.; {Maia},
  M.A.G.; {Marshall}, J.L.; {Melchior}, P.; {Menanteau}, F.; {Miquel}, R.;
  {Palmese}, A.; {Plazas}, A.A.; {Sanchez}, E.; {Scarpine}, V.; {Schubnell},
  M.; {Serrano}, S.; {Sevilla-Noarbe}, I.; {Smith}, M.; {Suchyta}, E.; {Tarle},
  G.; {DES Collaboration}.
\newblock {Milky Way Satellite Census. I. The Observational Selection Function
  for Milky Way Satellites in DES Y3 and Pan-STARRS DR1}.
\newblock {\em \apj} {\bf 2020}, {\em 893},~47,
  \href{http://xxx.lanl.gov/abs/1912.03302}{{\normalfont
  [arXiv:astro-ph.GA/1912.03302]}}.
\newblock
  doi:{\changeurlcolor{black}\href{https://doi.org/10.3847/1538-4357/ab7eb9}{\detokenize{10.3847/1538-4357/ab7eb9}}}.

\bibitem[{Hammer} \em{et~al.}(2020){Hammer}, {Yang}, {Arenou}, {Wang}, {Li},
  {Bonifacio}, and {Babusiaux}]{2020ApJ...892....3H}
{Hammer}, F.; {Yang}, Y.; {Arenou}, F.; {Wang}, J.; {Li}, H.; {Bonifacio}, P.;
  {Babusiaux}, C.
\newblock {Orbital Evidences for Dark-matter-free Milky Way Dwarf Spheroidal
  Galaxies}.
\newblock {\em \apj} {\bf 2020}, {\em 892},~3,
  \href{http://xxx.lanl.gov/abs/2002.09493}{{\normalfont
  [arXiv:astro-ph.GA/2002.09493]}}.
\newblock
  doi:{\changeurlcolor{black}\href{https://doi.org/10.3847/1538-4357/ab77be}{\detokenize{10.3847/1538-4357/ab77be}}}.

\bibitem[{Wan} \em{et~al.}(2020){Wan}, {Oliver}, {Lewis}, {Read}, and
  {Collins}]{2020MNRAS.492..456W}
{Wan}, Z.; {Oliver}, W.H.; {Lewis}, G.F.; {Read}, J.I.; {Collins}, M.L.M.
\newblock {On the origin of the asymmetric dwarf galaxy distribution around
  andromeda}.
\newblock {\em \mnras} {\bf 2020}, {\em 492},~456--467,
  \href{http://xxx.lanl.gov/abs/1912.02393}{{\normalfont
  [arXiv:astro-ph.GA/1912.02393]}}.
\newblock
  doi:{\changeurlcolor{black}\href{https://doi.org/10.1093/mnras/stz3477}{\detokenize{10.1093/mnras/stz3477}}}.

\bibitem[{Astropy Collaboration} \em{et~al.}(2018){Astropy Collaboration},
  {Price-Whelan}, {Sip{\H o}cz}, {G{\"u}nther}, {Lim}, {Crawford}, {Conseil},
  {Shupe}, {Craig}, {Dencheva}, {Ginsburg}, {VanderPlas}, {Bradley},
  {P{\'e}rez-Su{\'a}rez}, {de Val-Borro}, {Aldcroft}, {Cruz}, {Robitaille},
  {Tollerud}, {Ardelean}, {Babej}, {Bach}, {Bachetti}, {Bakanov}, {Bamford},
  {Barentsen}, {Barmby}, {Baumbach}, {Berry}, {Biscani}, {Boquien}, {Bostroem},
  {Bouma}, {Brammer}, {Bray}, {Breytenbach}, {Buddelmeijer}, {Burke},
  {Calderone}, {Cano Rodr{\'{\i}}guez}, {Cara}, {Cardoso}, {Cheedella},
  {Copin}, {Corrales}, {Crichton}, {D'Avella}, {Deil}, {Depagne}, {Dietrich},
  {Donath}, {Droettboom}, {Earl}, {Erben}, {Fabbro}, {Ferreira}, {Finethy},
  {Fox}, {Garrison}, {Gibbons}, {Goldstein}, {Gommers}, {Greco}, {Greenfield},
  {Groener}, {Grollier}, {Hagen}, {Hirst}, {Homeier}, {Horton}, {Hosseinzadeh},
  {Hu}, {Hunkeler}, {Ivezi{\'c}}, {Jain}, {Jenness}, {Kanarek}, {Kendrew},
  {Kern}, {Kerzendorf}, {Khvalko}, {King}, {Kirkby}, {Kulkarni}, {Kumar},
  {Lee}, {Lenz}, {Littlefair}, {Ma}, {Macleod}, {Mastropietro}, {McCully},
  {Montagnac}, {Morris}, {Mueller}, {Mumford}, {Muna}, {Murphy}, {Nelson},
  {Nguyen}, {Ninan}, {N{\"o}the}, {Ogaz}, {Oh}, {Parejko}, {Parley}, {Pascual},
  {Patil}, {Patil}, {Plunkett}, {Prochaska}, {Rastogi}, {Reddy Janga},
  {Sabater}, {Sakurikar}, {Seifert}, {Sherbert}, {Sherwood-Taylor}, {Shih},
  {Sick}, {Silbiger}, {Singanamalla}, {Singer}, {Sladen}, {Sooley},
  {Sornarajah}, {Streicher}, {Teuben}, {Thomas}, {Tremblay}, {Turner},
  {Terr{\'o}n}, {van Kerkwijk}, {de la Vega}, {Watkins}, {Weaver}, {Whitmore},
  {Woillez}, {Zabalza}, and {Astropy Contributors}]{2018AJ....156..123A}
{Astropy Collaboration}.; {Price-Whelan}, A.M.; {Sip{\H o}cz}, B.M.;
  {G{\"u}nther}, H.M.; {Lim}, P.L.; {Crawford}, S.M.; {Conseil}, S.; {Shupe},
  D.L.; {Craig}, M.W.; {Dencheva}, N.; {Ginsburg}, A.; {VanderPlas}, J.T.;
  {Bradley}, L.D.; {P{\'e}rez-Su{\'a}rez}, D.; {de Val-Borro}, M.; {Aldcroft},
  T.L.; {Cruz}, K.L.; {Robitaille}, T.P.; {Tollerud}, E.J.; {Ardelean}, C.;
  {Babej}, T.; {Bach}, Y.P.; {Bachetti}, M.; {Bakanov}, A.V.; {Bamford}, S.P.;
  {Barentsen}, G.; {Barmby}, P.; {Baumbach}, A.; {Berry}, K.L.; {Biscani}, F.;
  {Boquien}, M.; {Bostroem}, K.A.; {Bouma}, L.G.; {Brammer}, G.B.; {Bray},
  E.M.; {Breytenbach}, H.; {Buddelmeijer}, H.; {Burke}, D.J.; {Calderone}, G.;
  {Cano Rodr{\'{\i}}guez}, J.L.; {Cara}, M.; {Cardoso}, J.V.M.; {Cheedella},
  S.; {Copin}, Y.; {Corrales}, L.; {Crichton}, D.; {D'Avella}, D.; {Deil}, C.;
  {Depagne}, {\'E}.; {Dietrich}, J.P.; {Donath}, A.; {Droettboom}, M.; {Earl},
  N.; {Erben}, T.; {Fabbro}, S.; {Ferreira}, L.A.; {Finethy}, T.; {Fox}, R.T.;
  {Garrison}, L.H.; {Gibbons}, S.L.J.; {Goldstein}, D.A.; {Gommers}, R.;
  {Greco}, J.P.; {Greenfield}, P.; {Groener}, A.M.; {Grollier}, F.; {Hagen},
  A.; {Hirst}, P.; {Homeier}, D.; {Horton}, A.J.; {Hosseinzadeh}, G.; {Hu}, L.;
  {Hunkeler}, J.S.; {Ivezi{\'c}}, {\v Z}.; {Jain}, A.; {Jenness}, T.;
  {Kanarek}, G.; {Kendrew}, S.; {Kern}, N.S.; {Kerzendorf}, W.E.; {Khvalko},
  A.; {King}, J.; {Kirkby}, D.; {Kulkarni}, A.M.; {Kumar}, A.; {Lee}, A.;
  {Lenz}, D.; {Littlefair}, S.P.; {Ma}, Z.; {Macleod}, D.M.; {Mastropietro},
  M.; {McCully}, C.; {Montagnac}, S.; {Morris}, B.M.; {Mueller}, M.; {Mumford},
  S.J.; {Muna}, D.; {Murphy}, N.A.; {Nelson}, S.; {Nguyen}, G.H.; {Ninan},
  J.P.; {N{\"o}the}, M.; {Ogaz}, S.; {Oh}, S.; {Parejko}, J.K.; {Parley}, N.;
  {Pascual}, S.; {Patil}, R.; {Patil}, A.A.; {Plunkett}, A.L.; {Prochaska},
  J.X.; {Rastogi}, T.; {Reddy Janga}, V.; {Sabater}, J.; {Sakurikar}, P.;
  {Seifert}, M.; {Sherbert}, L.E.; {Sherwood-Taylor}, H.; {Shih}, A.Y.; {Sick},
  J.; {Silbiger}, M.T.; {Singanamalla}, S.; {Singer}, L.P.; {Sladen}, P.H.;
  {Sooley}, K.A.; {Sornarajah}, S.; {Streicher}, O.; {Teuben}, P.; {Thomas},
  S.W.; {Tremblay}, G.R.; {Turner}, J.E.H.; {Terr{\'o}n}, V.; {van Kerkwijk},
  M.H.; {de la Vega}, A.; {Watkins}, L.L.; {Weaver}, B.A.; {Whitmore}, J.B.;
  {Woillez}, J.; {Zabalza}, V.; {Astropy Contributors}.
\newblock {The Astropy Project: Building an Open-science Project and Status of
  the v2.0 Core Package}.
\newblock {\em \aj} {\bf 2018}, {\em 156},~123,
  \href{http://xxx.lanl.gov/abs/1801.02634}{{\normalfont
  [arXiv:astro-ph.IM/1801.02634]}}.
\newblock
  doi:{\changeurlcolor{black}\href{https://doi.org/10.3847/1538-3881/aabc4f}{\detokenize{10.3847/1538-3881/aabc4f}}}.

\bibitem[{Astropy Collaboration} \em{et~al.}(2013){Astropy Collaboration},
  {Robitaille}, {Tollerud}, {Greenfield}, {Droettboom}, {Bray}, {Aldcroft},
  {Davis}, {Ginsburg}, {Price-Whelan}, {Kerzendorf}, {Conley}, {Crighton},
  {Barbary}, {Muna}, {Ferguson}, {Grollier}, {Parikh}, {Nair}, {Unther},
  {Deil}, {Woillez}, {Conseil}, {Kramer}, {Turner}, {Singer}, {Fox}, {Weaver},
  {Zabalza}, {Edwards}, {Azalee Bostroem}, {Burke}, {Casey}, {Crawford},
  {Dencheva}, {Ely}, {Jenness}, {Labrie}, {Lim}, {Pierfederici}, {Pontzen},
  {Ptak}, {Refsdal}, {Servillat}, and {Streicher}]{2013A&A...558A..33A}
{Astropy Collaboration}.; {Robitaille}, T.P.; {Tollerud}, E.J.; {Greenfield},
  P.; {Droettboom}, M.; {Bray}, E.; {Aldcroft}, T.; {Davis}, M.; {Ginsburg},
  A.; {Price-Whelan}, A.M.; {Kerzendorf}, W.E.; {Conley}, A.; {Crighton}, N.;
  {Barbary}, K.; {Muna}, D.; {Ferguson}, H.; {Grollier}, F.; {Parikh}, M.M.;
  {Nair}, P.H.; {Unther}, H.M.; {Deil}, C.; {Woillez}, J.; {Conseil}, S.;
  {Kramer}, R.; {Turner}, J.E.H.; {Singer}, L.; {Fox}, R.; {Weaver}, B.A.;
  {Zabalza}, V.; {Edwards}, Z.I.; {Azalee Bostroem}, K.; {Burke}, D.J.;
  {Casey}, A.R.; {Crawford}, S.M.; {Dencheva}, N.; {Ely}, J.; {Jenness}, T.;
  {Labrie}, K.; {Lim}, P.L.; {Pierfederici}, F.; {Pontzen}, A.; {Ptak}, A.;
  {Refsdal}, B.; {Servillat}, M.; {Streicher}, O.
\newblock {Astropy: A community Python package for astronomy}.
\newblock {\em \aap} {\bf 2013}, {\em 558},~A33,
  \href{http://xxx.lanl.gov/abs/1307.6212}{{\normalfont
  [arXiv:astro-ph.IM/1307.6212]}}.
\newblock
  doi:{\changeurlcolor{black}\href{https://doi.org/10.1051/0004-6361/201322068}{\detokenize{10.1051/0004-6361/201322068}}}.

\bibitem[{Perez} and {Granger}(2007)]{PER-GRA:2007}
{Perez}, F.; {Granger}, B.E.
\newblock {IPython: A System for Interactive Scientific Computing}.
\newblock {\em Computing in Science and Engineering} {\bf 2007}, {\em
  9},~21--29.
\newblock
  doi:{\changeurlcolor{black}\href{https://doi.org/10.1109/MCSE.2007.53}{\detokenize{10.1109/MCSE.2007.53}}}.

\bibitem[Hunter(2007)]{Hunter:2007}
Hunter, J.D.
\newblock Matplotlib: A 2D graphics environment.
\newblock {\em Computing In Science \& Engineering} {\bf 2007}, {\em
  9},~90--95.

\bibitem[{Harris} \em{et~al.}(2020){Harris}, {Millman}, {van der Walt},
  {Gommers}, {Virtanen}, {Cournapeau}, {Wieser}, {Taylor}, {Berg}, {Smith},
  {Kern}, {Picus}, {Hoyer}, {van Kerkwijk}, {Brett}, {Haldane}, {del R{\'\i}o},
  {Wiebe}, {Peterson}, {G{\'e}rard-Marchant}, {Sheppard}, {Reddy}, {Weckesser},
  {Abbasi}, {Gohlke}, and {Oliphant}]{harris2020array}
{Harris}, C.R.; {Millman}, K.J.; {van der Walt}, S.J.; {Gommers}, R.;
  {Virtanen}, P.; {Cournapeau}, D.; {Wieser}, E.; {Taylor}, J.; {Berg}, S.;
  {Smith}, N.J.; {Kern}, R.; {Picus}, M.; {Hoyer}, S.; {van Kerkwijk}, M.H.;
  {Brett}, M.; {Haldane}, A.; {del R{\'\i}o}, J.F.; {Wiebe}, M.; {Peterson},
  P.; {G{\'e}rard-Marchant}, P.; {Sheppard}, K.; {Reddy}, T.; {Weckesser}, W.;
  {Abbasi}, H.; {Gohlke}, C.; {Oliphant}, T.E.
\newblock {Array programming with NumPy}.
\newblock {\em \nat} {\bf 2020}, {\em 585},~357--362,
  \href{http://xxx.lanl.gov/abs/2006.10256}{{\normalfont
  [arXiv:cs.MS/2006.10256]}}.
\newblock
  doi:{\changeurlcolor{black}\href{https://doi.org/10.1038/s41586-020-2649-2}{\detokenize{10.1038/s41586-020-2649-2}}}.

\bibitem[{Virtanen} \em{et~al.}(2020){Virtanen}, {Gommers}, {Oliphant},
  {Haberland}, {Reddy}, {Cournapeau}, {Burovski}, {Peterson}, {Weckesser},
  {Bright}, {van der Walt}, {Brett}, {Wilson}, {Millman}, {Mayorov}, {Nelson},
  {Jones}, {Kern}, {Larson}, {Carey}, {Polat}, {Feng}, {Moore}, {VanderPlas},
  {Laxalde}, {Perktold}, {Cimrman}, {Henriksen}, {Quintero}, {Harris},
  {Archibald}, {Ribeiro}, {Pedregosa}, {van Mulbregt}, and {SciPy 1. 0
  Contributors}]{Virtanen_2020}
{Virtanen}, P.; {Gommers}, R.; {Oliphant}, T.E.; {Haberland}, M.; {Reddy}, T.;
  {Cournapeau}, D.; {Burovski}, E.; {Peterson}, P.; {Weckesser}, W.; {Bright},
  J.; {van der Walt}, S.J.; {Brett}, M.; {Wilson}, J.; {Millman}, K.J.;
  {Mayorov}, N.; {Nelson}, A.R.J.; {Jones}, E.; {Kern}, R.; {Larson}, E.;
  {Carey}, C.J.; {Polat}, {\.I}.; {Feng}, Y.; {Moore}, E.W.; {VanderPlas}, J.;
  {Laxalde}, D.; {Perktold}, J.; {Cimrman}, R.; {Henriksen}, I.; {Quintero},
  E.A.; {Harris}, C.R.; {Archibald}, A.M.; {Ribeiro}, A.H.; {Pedregosa}, F.;
  {van Mulbregt}, P.; {SciPy 1. 0 Contributors}.
\newblock {SciPy 1.0: fundamental algorithms for scientific computing in
  Python}.
\newblock {\em Nature Methods} {\bf 2020}, {\em 17},~261--272,
  \href{http://xxx.lanl.gov/abs/1907.10121}{{\normalfont
  [arXiv:cs.MS/1907.10121]}}.
\newblock
  doi:{\changeurlcolor{black}\href{https://doi.org/10.1038/s41592-019-0686-2}{\detokenize{10.1038/s41592-019-0686-2}}}.

\end{thebibliography}

\end{document}